\documentclass[titlepage,11pt]{article}
\usepackage{amssymb}
\usepackage{amsfonts}
\usepackage{graphicx}
\usepackage{amsmath}
\usepackage{endnotes}
\usepackage{epsfig}
\usepackage{rotating}

\input{tcilatex}
\begin{document}

\title{Average and Quantile Effects in Nonseparable Panel Models\thanks{%
We thank J. Angrist, G. Chamberlain, D. Chetverikov, B. Frandsen, B. Graham, J. Hausman, and
many seminar participants for comments. Brad Larsen and Seongyeon Chang provided capable
research assistance. Parts of this paper were given at the 2007 CEMMAP
Microeconometrics: Measurement Matters Conference, the Shanghai Lecture of
the 2010 World Congress of the Econometric Society, and conferences in
between. We gratefully acknowledge research support from the NSF.}}
\author{Victor Chernozhukov \\
%EndAName
MIT \and Iv\'an Fern\'andez-Val \\
%EndAName
BU \and Jinyong Hahn \\
%EndAName
UCLA \and Whitney Newey \\
%EndAName
MIT}
\maketitle

\begin{abstract}
Nonseparable panel models are important in a variety of economic settings,
including discrete choice. This paper gives identification and estimation
results for nonseparable models under time homogeneity conditions that are
like \textquotedblleft time is randomly assigned\textquotedblright\ or
\textquotedblleft time is an instrument.\textquotedblright\ Partial
identification results for average and quantile effects are given for
discrete regressors, under static or dynamic conditions, in fully
nonparametric and in semiparametric models, with time effects. It is shown
that the usual, linear, fixed-effects estimator is not a consistent
estimator of the identified average effect, and a consistent estimator is
given. A simple estimator of identified quantile treatment effects is given,
providing a solution to the important problem of estimating quantile
treatment effects from panel data. Bounds for overall effects in static and
dynamic models are given. The dynamic bounds provide a partial
identification solution to the important problem of estimating the effect of
state dependence in the presence of unobserved heterogeneity. The impact of $%
T$, the number of time periods, is shown by deriving shrinkage rates for the
identified set as $T$ grows. We also consider semiparametric,
discrete-choice models and find that semiparametric panel bounds can be much
tighter than nonparametric bounds. Computationally-convenient methods for
semiparametric models are presented. We propose a novel inference method
that applies in panel data and other settings and show that it produces
uniformly valid confidence regions in large samples. We give empirical
illustrations.
\end{abstract}

\section{Introduction}

Interesting empirical questions are often formulated in terms of the \textit{%
ceteris paribus} effect of $x$ on $y,$ when observed $x$ is an individual
choice variable partly determined by preferences or technology. Panel data
holds out the hope of controlling for individual preferences or technology
by using multiple observations for a single economic agent. This hope is
particularly difficult to realize with discrete or other nonseparable models
and/or multidimensional individual effects. These models are, by nature, not
additively separable in unobserved individual effects, making them
challenging to identify and estimate. There are some simple solutions, such
as the conditional MLE for the slope parameter of a binary-choice logit
model with an individual location effect. However these are rare and
dependent on specific models or distributions. For example, the slope
parameter of the binary-choice model with a time dummy is identified only
for logit as shown by Chamberlain (2010), and the average treatment effect
is not identified even for logit without a time dummy, as shown below.

A fundamental idea for using panel data to identify the \textit{ceteris
paribus} effect of $x$ on $y$ is to use changes in $x$ over time to estimate
the effect. In order for changes over time in $x$ to correspond to \textit{%
ceteris paribus} effects, the distribution of variables other than $x$ must
not vary over time. This condition is like \textquotedblleft time being
randomly assigned\textquotedblright\ or \textquotedblleft time is an
instrument.\textquotedblright\ In this paper we consider identification via
such time homogeneity conditions. They are also the basis of many previous
panel results, including Chamberlain (1982), Manski (1987), and Honore
(1992). Here we consider the identifying power of time homogeneity for
nonseparable models, i.e. for models that are not additively separable in
unobserved factors. We allow for multidimensional heterogeneity, as
motivated by models where effects of interest, such as price and income
elasticities, are distributed among individuals in unrestricted ways; see
Altonji and Matzkin (2005), Browning and Carro (2007), and Fernandez-Val and
Lee (2010), among others. We also weaken the strict time homogeneity
conditions to allow some time effects.

Models with discrete regressors have many applications and are the subject
of most of this paper. With discrete regressors, time homogeneity only leads
to partial identification of many effects, though some conditional effects
are identified. This paper considers partial identification and estimation
of average and quantile effects, under static or dynamic conditions, in
fully nonparametric and in semiparametric models, with time effects.

For the nonparametric, static model we give simple estimators of the
identified average effect of $x$ on $y$, conditional on $x$ varying over
time. These estimators extend Chamberlain (1982, pp. 10-17) to multiple
regressors with location and scale time effects. We also find that linear,
fixed-effects estimate a variance-weighted average effect instead of the
average effect. For bounded $y$ we move beyond the analysis of identified
effects and give simple estimators of sharp bounds for average effects.
These bounds provide nonparametric, partial-identification estimates of
average effects in important cases, such as binary choice in panel data.

The quantile estimators given here are more novel than the average-effect
estimators. They provide simple estimators of the effect of $x$ on quantiles
of $y,$ conditional on $x$ varying over time, that allow for location and
scale time effects. Estimators of sharp bounds are also provided for the
unconditional, overall quantile effect. The estimators allow for
multidimensional heterogeneity, for example for both location and slope to
vary across individuals in an unrestricted way. In this way we provide a
solution for the important problem of nonparametric quantile regression in
panel data with individual effects, for discrete regressors. Graham, Hahn
and Powell (2009) also consider quantile effects in linear, heterogenous coefficients models, 
but impose conditions which essentially restrict the heterogeneity
to be one-dimensional, and focus on identification of the distribution of
coefficients.

Dynamics is often an important feature of economic models with intertemporal
choice. Here we give a dynamic, nonseparable, panel model that nests the
static one. Simple estimators of bounds on average and quantile effects are
provided. We show that these results provide a partial-identification
solution to the important problem of distinguishing state dependence from
heterogeneity.

This paper shows the impact of the number of time periods $T$ on
identification. We find that the identified set of effects shrinks to a
point exponentially quickly as $T$ grows, when individual effects are
bounded and time period disturbances are not, and that the rate is some
power of $T^{-1}$ more generally. In a nonparametric, dynamic, binary-choice model
we find that the rate is faster the larger the variance of the
period-specific disturbance relative to the variance of the individual
effect.

In numerical examples we find that the nonparametric bounds can be quite
wide, motivating more informative models. Semiparametric models that specify
the distribution of the outcome given regressors and individual effect is an
important class of more informative models. Here we describe both static and
dynamic semiparametric models. When restrictions are imposed on the
heterogeneity, like only some coefficients varying across individuals,
semiparametric models can have substantially tighter bounds than
nonparametric models. We find that in the important binary-logit model with
just a location effect the average effect bounds shrink exponentially
quickly as $T$ grows, in both dynamic and static models, even when the
nonparametric bounds shrink slowly. This result quantifies the gain in
information of a semiparametric model with just a location effect over the
nonparametric model. We also find quite tight bounds for semiparametric
models relative to nonparametric ones in numerical examples.

We show that semiparametric, discrete-choice models have finite dimensional
parameterizations. This reduces bounds calculation and estimation to a
finite-dimensional problem, albeit a large dimensional, highly nonlinear,
and computationally difficult one. To make computation more feasible we use
grids of fixed values for individual effects, so that average choice
probabilities are finite-dimensional, linear combinations. We combine this
with minimum squared distance fitting of data cell probabilities to obtain a
quadratic programming approach for estimating the individual-effect
distributions. This approach is computationally convenient and overcomes
problems with previously proposed methods, as further discussed below. We
also allow the grid to grow in order to approximate the true support points.
It turns out that because the model is finite dimensional there is no need
to limit the number of grid points. Mathematically, a richer fixed grid
simply corresponds to a bigger submodel of the finite-dimensional model.

The semiparametric bounds build on Honor\'{e} and Tamer (2003, 2006) and
Chernozhukov, Hahn, and Newey (2004). Both papers gave results for bounds in
semiparametric, nonlinear, panel-data models. Honore and Tamer (2006)
proposed linear programming, minimum distance, and maximum likelihood
methods for dynamic models. Chernozhukov, Hahn, and Newey (2004) proposed
sieve likelihood estimation of bounds for static models. These approaches
are not very useful for estimation. Plugging in sample frequencies in place
of cell probabilities in the linear-programming algorithm produces empty
identification regions because the frequencies need not satisfy constraints
imposed by the model. Also, the minimum-distance objective function is
computationally difficult, as is sieve maximum likelihood, given the
dimensionality of the individual-effect distributions. Honore and Tamer
(2006) also assumed a fixed known grid for true individual effects, while we
consider an approximation to an unknown grid.

The inferential problem for the semiparametric models is also rather
challenging. The models impose data-dependent constraints that are often
infeasible in finite samples or under misspecification, which produces empty
confidence regions. We overcome these difficulties by projecting these
data-dependent constraints onto the model space using the
quadratic-programming approach mentioned above, thus producing an
always-feasible, data-dependent constraint set. We then suggest linear and
nonlinear programming methods that use these new modified constraints. Our
inference procedures have the appealing justification of targeting the true
model under correct specification and targeting a best approximating model
under incorrect specification. We also develop two novel inferential
procedures, one called the \textit{perturbed bootstrap}, that is described
in the paper, and another called \textit{modified projection,} that is
described in the Supplementary Material. These methods produce uniformly
valid inference in large samples and may be of substantial independent
interest.

We give two empirical illustrations. One is to estimate the effect of unions
on earnings quantiles. There we find that a decline in the union effect as
the quantile increases can be attributed to individual heterogeneity. The
other illustration is to estimate the effects of fertility on women's labor
force participation. There we compare nonparametric and semiparametric
estimates.

Recent research has considered nonseparable panel models with time
homogeneity and continuous regressors. Graham and Powell (2011) give
estimators of the average effect in a linear model with heterogeneous
slopes. Hoderlein and White (2011) give estimators of the average derivative
conditional on equality of regressors across time periods.

Chamberlain (1980, 1984), Altonji and Matzkin (2005), Bester and Hansen
(2008), and others have used control functions for panel data estimation. We
focus instead on time homogeneity with unrestricted dependence between
individual effects and regressors. Bias-corrected, fixed-effects estimation
of semiparametric models has been proposed by Hahn and Kuersteiner (2002),
Alvarez and Arellano (2003), Woutersen (2002), Hahn and Newey (2004), and
Fern\'{a}ndez-Val (2009). These estimators depend on large $T$ for
consistency while we estimate identified effects and bounds for fixed $T$.

Section 2 describes the models and effects we consider. Section 3 discusses
estimation of identified effects. Sections 4 and 5 derive bounds for the
static and dynamic nonparametric models respectively. Section 6 describes
the impact of $T$. Section 7 describes and gives results for semiparametric,
discrete-choice models. Section 8 gives computationally convenient methods
for semiparametric models and numerical examples. Section 9 considers
estimation and inference for semiparametric models. Section 10 gives
empirical examples. The Supplementary Material Chernozhukov et. al. (2012) includes a variety of omitted
discussions and results along with the proofs of results stated in the paper.

\section{The Models and Effects}

The data consist of $n$ observations on $Y_{i}=(Y_{i1},...,Y_{iT})^{\prime }$
and $X_{i}=[X_{i1},...,X_{iT}]^{\prime }$, for a dependent variable $Y_{it}$
and a vector of regressors $X_{it}$. Throughout we assume that the
observations $(Y_{i},X_{i})$\textit{, }$(i=1,...,n)$\textit{, }are
independent and identically distributed. The nonparametric models we
consider satisfy

\bigskip

\textsc{Assumption 1: }\textit{There is a function }$g_{0}(x,\alpha
,\varepsilon )$\textit{\ and vectors }$\alpha _{i}$\textit{\ and }$%
\varepsilon _{it}$\textit{\ of random variables such that}%
\begin{equation*}
Y_{it}=g_{0}(X_{it},\alpha _{i},\varepsilon _{it}),(i=1,...,n;t=1,...,T).
\end{equation*}

The vector $\alpha _{i}$ consists of time invariant individual effects that
often represent individual heterogeneity. The vector $\varepsilon _{it}$
represents period-specific disturbances. Altonji and Matzkin (2005)
considered models satisfying Assumption 1. The invariance of $g_{0}$ over
time in this Assumption does not actually impose any time homogeneity. If
there are no restrictions on $\varepsilon _{it}$ then $t$ could be one of
the components of $\varepsilon _{it},$ allowing the function to vary over
time in a completely general way. The next condition, together with
Assumption 1, imposes time homogeneity on the model.

\bigskip

\textsc{Assumption 2:} $\varepsilon _{it}|X_{i},\alpha _{i}\overset{d}{=}%
\varepsilon _{i1}|X_{i},\alpha _{i}$\textit{, for all }$t.$

\bigskip

This is a static, or \textquotedblleft strictly exogenous\textquotedblright\
time homogeneity condition, where all leads and lags of the regressor are
included in the conditioning variable $X_{i}.$ It requires that the
conditional distribution of $\varepsilon _{it}$ given $X_{i}$ and $\alpha
_{i}$ does not depend on $t,$ but does allow for dependence of $\varepsilon
_{it}$ over time. An equivalent condition is $\tilde{\varepsilon}_{it}|X_{i}%
\overset{d}{=}\tilde{\varepsilon}_{i1}|X_{i}$ for $\tilde{\varepsilon}%
_{it}=(\alpha _{i},\varepsilon _{it}).$ Thus, the time invariant $\alpha
_{i} $ has no distinct role in this model. The condition is just that
whatever the unobserved disturbances are, their conditional distribution
given $X_{i}$ does not depend on $t$.

This seems a basic condition that helps panel data provide information about
the effect of $x$ on $y.$ It is like \textquotedblleft time is randomly
assigned\textquotedblright\ or \textquotedblleft time is an
instrument\textquotedblright\ with the distribution of factors other than $x$
not varying over time, so that changes in $x$ over time can help identify
the effect of $x$ on $y$. Assumption 2 also turns out to be a natural
strengthening of linear model conditions, as shown in Theorem A1 and the
associated discussion in the Supplementary Material.

A dynamic model can be obtained by only including current and lagged $X_{is}$
in the conditioning set for each $t,$ as in the following condition:

\bigskip

\textsc{Assumption 3:} $\varepsilon _{it}|X_{it},...,X_{i1},\alpha _{i}%
\overset{d}{=}\varepsilon _{i1}|X_{i1},\alpha _{i}$,\textit{\ for all }$t.$

\bigskip

This is a \textquotedblleft predetermined\textquotedblright\ version of time
homogeneity that is nested within the static model of Assumptions 1 and 2,
as shown in Theorem A2 of the Supplementary Material. Here the conditional
distribution given only current and lagged regressors must be time
invariant. It also implies that the conditional distribution of $\varepsilon
_{it}$ given current and lagged regressors only depends on $X_{i1}$. Here $%
\varepsilon _{it}$ can be thought of as additional information that is
independent of the past regressors. A conditional-mean version of this
condition arises in rational-expectations models that implies disturbances
have mean zero conditional on past information. Here the stronger
conditional independence restriction is imposed as seems needed for a
nonseparable model. The conditioning on $X_{i1}$ is a way to account for the
initial conditions of this dynamic model. Bhargava and Sargan (1983) adopted
this approach in a linear model as have Honore and Tamer (2006) and Browning
and Carro (2007) in a likelihood setting.

If $X_{it}$ includes lagged $Y_{it}$ then Assumption 3 specifies that the
model is \textquotedblleft dynamically complete,\textquotedblright\ ruling
out $Y_{it}=g_{0}(X_{it},\alpha _{i},\varepsilon _{it})$ as one equation of
a dynamic system. For instance, $X_{it}$ could be $Y_{i,t-1},$ in which case 
$Y_{it}=g_{0}(Y_{it-1},\alpha _{i},\varepsilon _{it})$ is an explicit
nonseparable dynamic model with $\varepsilon _{it}$ being time shocks that
are independent of $Y_{it-1},...,Y_{i1}$. An important example is one where $%
Y_{it}\in \{0,1\}$ is binary, representing state dependence, with $\alpha
_{i}$ representing unobserved heterogeneity. This example is treated in
Section 5.

We will focus in the nonparametric model on two objects, the average
structural function (ASF) of Blundell and Powell (2003) and the quantile
structural function (QSF) of Imbens and Newey (2009). The ASF is%
\begin{equation*}
\mu (x)=E[g_{0}(x,\alpha _{i},\varepsilon _{it})]=\int g_{0}(x,\alpha
,\varepsilon )dF(\alpha ,\varepsilon ),
\end{equation*}%
where throughout the paper $F$ denotes the cumulative distribution function
(CDF) of a random vector that appears as the arguments of $F$. This object
is useful for quantifying the effect of $x$ on the mean of the outcome $%
Y_{it}$. In the treatment-effects literature the average treatment effect
(ATE) of changing $x$ from $x^{b}$ (before) to $x^{a}$ (after) is 
\begin{equation*}
\Delta =\mu (x^{a})-\mu (x^{b}).
\end{equation*}

The QSF $q(\lambda ,x)$ is the $\lambda ^{th}$ quantile of $g_{0}(x,\alpha
_{i},\varepsilon _{it}).$ Under conditions specified below the QSF will
equal the inverse of the CDF of $g_{0}(x,\alpha _{i},\varepsilon _{it})$, 
\begin{equation*}
q(\lambda ,x)=G^{-1}(\lambda ,x),G(y,x)=E[1(g_{0}(x,\alpha _{i},\varepsilon
_{it})\leq y)].
\end{equation*}%
In the treatment-effects literature the $\lambda ^{th}$ quantile treatment
effect (QTE) of changing $x$ from $x^{b}$ to $x^{a}$ is%
\begin{equation*}
\Delta _{\lambda }=q(\lambda ,x^{a})-q(\lambda ,x^{b}),
\end{equation*}%
as in Lehmann (1974). This effect does not give the quantile of the
treatment effect but does quantify the shift in the distribution of $Y_{it}$
that is due to a change in $x.$ It accounts for multidimensional individual
effects that may be correlated with $x$.

The static model implies a conditional-mean model that has been considered
by Chamberlain (1982), Hahn (2001), Wooldridge (2005), and Chernozhukov et.
al. (2007). This conditional-mean model specifies that there is an $\alpha
_{i}$ and $m_{0}(x,\alpha )$ such that $E[Y_{it}|X_{i},\alpha
_{i}]=m_{0}(X_{it},\alpha _{i}).$ A conditional mean ATE, as in Wooldridge
(2005), is $\int [m_{0}(x^{a},\alpha )-m_{0}(x^{b},\alpha )]dF(\alpha )$.
This model and effect differ from those we consider in specifying
conditional-mean restrictions, while we specify conditional distribution
restrictions. In Theorem A3 of the Supplementary Material we show that the
conditional-mean model is implied by Assumptions 1 and 2, or 1 and 3, and
that the conditional mean ATE is equal to the ATE we consider. Thus all
results we give for the ATE, including bounds, apply to the conditional mean
models, such as that of Chernozhukov et. al. (2007).

To help explain the relationship between the conditional-mean model and the
model of our paper, and to illustrate other results, it is useful to
consider examples. Binary choice is a very important model for panel data,
as it has many applications. For this reason we use binary choice as a main
example. The most common model has been one with a scalar individual effect
that is an additive shift to a linear combination of $X_{it}$, where%
\begin{equation*}
Y_{it}=1(X_{it}^{\prime }\beta ^{\ast }+\alpha _{i}\geq \varepsilon _{it}),
\end{equation*}%
for scalar $\varepsilon _{it}$ and an unknown parameter vector $\beta ^{\ast
}$. In this example $g_{0}(x,\alpha ,\varepsilon )=1(x^{\prime }\beta ^{\ast
}+\alpha \geq \varepsilon )$ and the ATE is%
\begin{equation*}
\Delta =\int [1(x^{a\prime }\beta ^{\ast }+\alpha \geq \varepsilon
)-1(x^{b\prime }\beta ^{\ast }+\alpha \geq \varepsilon )]dF(\varepsilon
,\alpha ).
\end{equation*}%
This is an unusual object in the binary choice literature but is equal to a
conditional mean ATE. In particular, if $\varepsilon _{it}$ is independent
of $(X_{i},\alpha _{i})$ with CDF $H(\varepsilon )$ for each $t$. Then $%
E[Y_{it}|X_{i},\alpha _{i}]=\Pr (Y_{it}=1|X_{i},\alpha
_{i})=H(X_{it}^{\prime }\beta ^{\ast }+\alpha _{i})$ and 
\begin{equation*}
\Delta =\int [H(x^{a\prime }\beta ^{\ast }+\alpha )-H(x^{b\prime }\beta
^{\ast }+\alpha )]dF(\alpha ).
\end{equation*}%
Thus the ATE is also the effect of changing $x$ on the choice probabilities
averaged over the individual effect, i.e. the conditional mean ATE.

Our model also includes binary choice with individual-specific slopes as a
special case. Economic motivation for varying slopes is provided by Browning
and Carro (2007, 2009) who point out that with constant slopes the sign of
the treatment effect is the same for every individual and give empirical
examples where varying slopes are important. A general model with varying
slopes is $Y_{it}=1(X_{it}^{\prime }\alpha _{i}\geq \varepsilon _{it})$
where $X_{it}$ now includes a constant and $\varepsilon _{it}$ is
independent of $(X_{i},\alpha _{i})$ with CDF $H(\varepsilon ).$ In this
model%
\begin{equation*}
\Delta =\int [H(x^{a\prime }\alpha )-H(x^{b\prime }\alpha )]dF(\alpha ),
\end{equation*}%
accounting for individual specific slopes. When $X_{it}$ is discrete and
fully saturated (e.g. consists of a full set of dummies, one for every
discrete outcome) this model is actually equivalent to the general static
model. It will be more restrictive when the distribution of $\alpha $ is
restricted in some way, such as having some components of $\alpha $ be
constant. In the semiparametric analysis described below we show how to
impose such restrictions.

Time effects are clearly important in practice but identification of
treatment effects will preclude including $t$ among the regressors $X_{it}$
in the nonparametric model of Assumptions 1 - 3. Identification will be
based on variation over time in $X_{it}$, and if $t$ is a regressor then $%
g_{0}(X_{it},\alpha _{i},\varepsilon _{it})$ has unrestricted variation over
time, precluding identification of the effect of any other regressor. Some
time effects can be allowed for by restricting the way $t$ enters $g_{0}.$
Below we will describe how this is done in semiparametric, discrete-choice
models. With continuous $Y_{it}$ one can allow for location and scale time
effects that are relatively easy to estimate.

\bigskip

\textsc{Assumption 4:} \textit{There is a function }$g_{0}(x,\alpha
,\varepsilon ),$\textit{\ vectors }$\alpha _{i}$\textit{\ and }$\varepsilon
_{it}$\textit{\ of random variables, and constants }$\tau
_{t},s_{t},(t=2,...,T)$\textit{\ such that for }$\tau _{1}=0,$ $s_{1}=1,$ 
\begin{equation*}
Y_{it}=g_{t0}(X_{it},\alpha _{i},\varepsilon _{it}),\text{ }g_{t0}(x,\alpha
,\varepsilon )=\tau _{t}+s_{t}g_{0}(x,\alpha ,\varepsilon
),(i=1,...,n;t=1,...,T).
\end{equation*}

This condition allows the mean and variance of $Y_{it}$ to vary over time in
an unrestricted way. The condition could be generalized to allow for other
time effects, but we leave that to future work. It does not apply to $Y_{it}$
with fixed, discrete support because Assumption 4 does not make sense in
that case. There $t$ must be included \textquotedblleft
inside\textquotedblright\ $g_{0}$, as we do in the semiparametric analysis
described below.

With these time effects the ASF and QSF can depend on $t$. The ASF and QSF
for the first period will be $\mu (x)$ and $q(\lambda ,x)$ as given above,
and for the other periods are%
\begin{equation*}
\mu _{t}(x)=\tau _{t}+s_{t}\mu (x),q_{t}(\lambda ,x)=\tau
_{t}+s_{t}q(\lambda ,x),(t=2,...,T).
\end{equation*}%
Corresponding period-specific and time-averaged ATE and QTE are given by%
\begin{eqnarray}
\mu _{t}(x^{a})-\mu _{t}(x^{b}) &=&s_{t}[\mu (x^{a})-\mu
(x^{b})],q_{t}(\lambda ,x^{a})-q_{t}(\lambda ,x^{b})=s_{t}[q(\lambda
,x^{a})-q(\lambda ,x^{b})],  \label{tqte} \\
&&\left( \frac{\sum_{t=1}^{T}s_{t}}{T}\right) [\mu (x^{a})-\mu
(x^{b})],\left( \frac{\sum_{t=1}^{T}s_{t}}{T}\right) [q(\lambda
,x^{a})-q(\lambda ,x^{b})],  \notag
\end{eqnarray}%
where $s_{1}=1$.

In the rest of this paper we will focus on discrete regressors, imposing the
following condition from here on:

\bigskip

\textsc{Assumption 5:} \textit{The support of }$X_{i}$\textit{\ is finite}$.$

\bigskip

With discrete $X_{it}$ the model can also be written as a multiple
regression with random coefficients, though we find it convenient to use the
notation given here.

\section{Identified Effects in the Nonparametric Static Model}

The analysis of identification in the static model is quite simple. This
simplicity is a virtue, leading to estimators of identified effects and
bounds on unidentified effects that are easy to calculate in a very general
model. For example, this approach gives a simple solution to the important
problem of identification of quantile treatment effects in panel data. The
idea is based on Assumption 2, which states that, conditional on $X_{i},$
the distribution of unobservables does not vary over time. Therefore,
conditional on $X_{i}$ where both $x^{b}$ and $x^{a}$ occur for some time
periods, one can identify effects from the changes in $Y_{it}$ across those
time periods. For the ATE, the identified conditional effects can be
averaged to identify effects conditional on $X_{i}$ being in subsets where
both $x^{b}$ and $x^{a}$ occur for some time period. This idea is a slight
extension of Chamberlain (1982, pp. 10-17) to discrete regressors that are
not binary. For the QTE the distribution functions can be averaged and
inverted to identify corresponding quantile effects. This idea appears to be
novel.

There is a simple approach to allowing for covariates. Suppose $%
x=(x_{1},x_{2}),$ and one is interested in the effect of $x_{1}$ holding $%
x_{2}$ fixed. Then one can take $x^{b}=(x_{1}^{b},x_{2})$ and $%
x^{a}=(x_{1}^{a},x_{2}),$ so that the effect of changing from $x^{b}$ to $%
x^{a}$ is then the effect of interest. Furthermore, one could average these
effects over $x_{2}$ to identify an effect that is averaged over covariates.
We explicitly allow for covariates in the semiparametric models given below.
Because we are already attempting to cover so much ground here, we leave
averaging over covariates in the nonparametric model to future work.

To describe identified effects and their estimators we will focus on the ATE
and QTE conditional on both $x^{a}$ and $x^{b}$ appearing in $X_{i}$ for
some time period. We could also consider effects conditional on smaller
subsets of $X_{i}$ but postpone this until later in order to keep the
exposition relatively simple. We need a little more notation to give a
precise description. Let $1(X_{it}=x)$ denote the indicator function that is
equal to one when $X_{it}=x$ and zero otherwise and let $T_{i}(x)=%
\sum_{t=1}^{T}1(X_{it}=x).$ Here we let the subscript $i$ denote a random
variable that may depend on $X_{i}$ and $Y_{i}$. Let $%
D_{i}=1(T_{i}(x^{a})>0)1(T_{i}(x^{b})>0)$ be the indicator for the event
that $X_{i}$ includes both $x^{a}$ and $x^{b}$ for some time period. Define 
\begin{equation}
\delta =E[g_{0}(x^{a},\alpha _{i},\varepsilon _{i1})-g_{0}(x^{b},\alpha
_{i},\varepsilon _{i1})|D_{i}=1].  \label{ATE cond}
\end{equation}%
This $\delta $ is the ATE for those individuals where both $x^{b}$ and $%
x^{a} $ occur for some time period. This effect may be of interest in many
settings. For example, when $Y_{it}$ is log earnings and $X_{it}\in \{0,1\}$
represents union status, $\delta $ would be the average effect of union
status on earnings for those who changed union status over the time periods
we observe. For a given number of time periods $T,$ this is all one could
hope to identify nonparametrically. However, we may be interested in other
effects too. We might be interested in union effects for those who ever
changed union status at some time. \ This is $\delta $. \ Or we might even
be interested in the effect for those who were ever in a union. Bounds for
such an effect are described below.

A simple estimator of the conditional ATE $\delta $ is%
\begin{equation}
\hat{\delta}=\frac{\sum_{i=1}^{n}D_{i}[\bar{Y}_{i}(x^{a})-\bar{Y}_{i}(x^{b})]%
}{\sum_{i=1}^{n}D_{i}},\bar{Y}_{i}(x)=\left\{ 
\begin{array}{c}
T_{i}(x)^{-1}\sum_{t=1}^{T}1(X_{it}=x)Y_{it},T_{i}(x)>0 \\ 
0,T_{i}(x)=0%
\end{array}%
\right. .  \label{ideff}
\end{equation}%
Consistency of this estimator results from 
\begin{equation*}
E[D_{i}\{\bar{Y}_{i}(x^{a})-\bar{Y}_{i}(x^{b})\}]=E[D_{i}\{g_{0}(x^{a},%
\alpha _{i},\varepsilon _{i1})-g_{0}(x^{b},\alpha _{i},\varepsilon _{i1})\}],
\end{equation*}%
see Lemma A5 of the Supplementary Material. Intuitively, this equation
follows from time being randomly assigned, so that we can estimate the
effect by comparing $Y_{it}$ where $X_{it}=x^{a}$ with $Y_{is}$ where $%
X_{is}=x_{b}$.

Since $\bar{Y}_{i}(x^{a})-\bar{Y}_{i}(x^{b})$ is a difference of means it
can be interpreted as a coefficient of $1(X_{it}=x^{a})$ in a regression of $%
Y_{it}$ on that dummy and on $1(X_{it}=x^{a})+1(X_{it}=x^{b})$. Thus, $\hat{%
\delta}$ is an average of least-squares estimates for each $i$ with $D_{i}=1$%
. From this interpretation we see that $\hat{\delta}$ extends Chamberlain's
(1982, p. 12) estimator to discrete regressors that are not binary. A
consistent estimator of the asymptotic variance of $\sqrt{n}(\hat{\delta}%
-\delta )$ is $n^{-1}\sum_{i=1}^{n}\hat{\psi}_{i}^{2}$ where $\hat{\psi}%
_{i}=nD_{i}[\bar{Y}_{i}(x^{a})-\bar{Y}_{i}(x^{b})-\hat{\delta}%
]/\sum_{i=1}^{n}D_{i}$. For brevity we leave the asymptotic theory to the
Supplementary Material (see Theorem A6) and efficiency results to future
work.

We can also identify and estimate a conditional QTE. Let $G(y,x|D_{i}=1)=\Pr
(g_{0}(x,\alpha _{i},\varepsilon _{i1})\leq y|D_{i}=1)$ denote the CDF of $%
g_{0}(x,\alpha _{i},\varepsilon _{i1})$ conditional on $D_{i}=1$. The QTE
conditional on $D_{i}=1$ is%
\begin{equation*}
\delta _{\lambda }=G^{-1}(\lambda ,x^{a}|D_{i}=1)-G^{-1}(\lambda
,x^{b}|D_{i}=1).
\end{equation*}%
An estimator of this effect can be constructed using a CDF $\Phi (u)$ and a
scalar bandwidth $h$. An estimator of $G(y,x|D_{i}=1)$ is given by%
\begin{equation*}
\hat{G}(y,x|D_{i}=1)=\frac{\sum_{i=1}^{n}D_{i}\bar{G}_{i}(y,x)}{%
\sum_{i=1}^{n}D_{i}},\bar{G}_{i}(y,x)=\left\{ 
\begin{array}{c}
T_{i}(x)^{-1}\sum_{t=1}^{T}1(X_{it}=x)\Phi (\frac{y-Y_{it}}{h}),T_{i}(x)>0,
\\ 
0,T_{i}(x)=0.%
\end{array}%
\right. .
\end{equation*}%
In this estimator the indicator function $1(Y_{it}<y)$ has been replaced by
a smoothed approximation $\Phi (\frac{y-Y_{it}}{h})$, as suggested by Yu and
Jones (1998) for estimating a conditional CDF. An estimator of $\delta
_{\lambda }$ is then 
\begin{equation*}
\hat{\delta}_{\lambda }=\hat{q}_{\lambda }^{a}-\hat{q}_{\lambda }^{b},\hat{q}%
_{\lambda }^{a}=\hat{G}^{-1}(\lambda ,x^{a}|D_{i}=1),\hat{q}_{\lambda }^{b}=%
\hat{G}^{-1}(\lambda ,x^{b}|D_{i}=1).
\end{equation*}%
Note here that we first average, then invert, and then difference. This
estimator solves an important problem of estimating panel quantile effects
and appears to be novel.

A consistent estimator of the asymptotic variance of $\sqrt{n}(\hat{\delta}%
_{\lambda }-\delta _{\lambda })$ is $n^{-1}\sum_{i=1}^{n}\hat{\psi}_{\lambda
i}^{2}$ for 
\begin{equation*}
\hat{\psi}_{\lambda i}= - \frac{nD_{i}}{\sum_{i=1}^{n}D_{i}}\left[ \frac{\bar{G}%
_{i}(\hat{q}^{a},x^{a})-\lambda }{\hat{G}^{\prime }(\hat{q}%
^{a},x^{a}|D_{i}=1)}-\frac{\bar{G}_{i}(\hat{q}^{b},x^{b})-\lambda }{\hat{G}%
^{\prime }(\hat{q}^{b},x^{b}|D_{i}=1)}\right] ,
\end{equation*}%
where $\hat{G}^{\prime }(y,x|D_{i}=1)=\partial \hat{G}(y,x|D_{i}=1)/\partial
y$. Here the denominator terms are actually kernel density estimates. For
this reason one might use different bandwidths $h$ in the numerator and
denominator, with the denominator chosen to be appropriate for density
estimation. Asymptotic theory for this estimator is given in the
Supplementary Material (see Theorem A8). Alternatively, one could simply use
the bootstrap to construct a confidence interval for $\hat{\delta}_{\lambda
}.$

A helpful example is the binary regressor case where $X_{it}\in \{0,1\}.$
Here $X_{it}$ could be thought of as a treatment variable where $X_{it}=1$
for treated and $X_{it}=0$ for untreated. Let $Y_{it}(0)=g_{0}(0,\alpha
_{i},\varepsilon _{it})$ and $Y_{it}(1)=g_{0}(1,\alpha _{i},\varepsilon
_{it})$. Assumption 2 is equivalent to the assumption that the conditional
distribution of $(Y_{it}(0),Y_{it}(1))$ given $X_{i}$ does not vary with $t$%
. This is the key assumption that identifies treatment effects from time
variation in treatment. In this context $\delta
=E[Y_{it}(1)-Y_{it}(0)|D_{i}=1]$ is the ATE for individuals where both
treatment and nontreatment occurs during the observation period. Similarly, $%
\delta _{\lambda }$ is the difference between the $\lambda $ quantile of the
distribution of $Y_{it}(1)$ and the $\lambda $ quantile for $Y_{it}(0)$
conditional on $D_{i}=1$. The ATE and QTE are not identified for those
individuals that either receive treatment in every time period or receive no
treatment in every time period.

In general the usual panel data within (linear fixed effects) estimator is
not a consistent estimator of $\delta .$ This inconsistency results because
the within estimator constrains the slope coefficient to be the same for
each $i$ when the slope is actually varying with $i$. For simplicity we
demonstrate this inconsistency in the binary $X_{it}$ example. The within
estimator $\hat{\delta}_{w}$ is given by 
\begin{equation*}
\hat{\delta}_{w}=\frac{\sum_{i=1}^{n}\sum_{t=1}^{T}(X_{it}-\bar{X}_{i})Y_{it}%
}{\sum_{i=1}^{n}\sum_{t=1}^{T}(X_{it}-\bar{X}_{i})^{2}},\bar{X}%
_{i}=T^{-1} \sum_{t=1}^{T}X_{it}.
\end{equation*}%
Let $\sigma _{i}^{2}=(T-1)^{-1}\sum_{t=1}^{T}(X_{it}-\bar{X}_{i})^{2}$ be the
sample variance over time of $X_{it}$.

\bigskip

\textsc{Theorem 1:} \textit{If Assumptions 1 and 2 are satisfied, } $X_{it} \in \{0,1\}$, $%
E[Y_{it}^{2}]<\infty ,$\textit{\ }$(t=1,...,T)$\textit{, and }$E[D_{i}\sigma
_{i}^{2}]>0$, \textit{then }$\delta =E[D_{i}\{\bar{Y}_{i}(1)-\bar{Y}%
_{i}(0)\}]/E[D_{i}]$ \textit{and} 
\begin{equation}
\hat{\delta}_{w}\overset{p}{\longrightarrow }\delta _{w}=\frac{E[\sigma
_{i}^{2}D_{i}\{\bar{Y}_{i}(1)-\bar{Y}_{i}(0)\}]}{E[\sigma _{i}^{2}D_{i}]}.
\label{lpm}
\end{equation}

Note that the limit of the within estimator is a weighted average of
individual, least-squares estimates $\bar{Y}_{i}(1)-\bar{Y}_{i}(0)$ from
equation (\ref{ideff}). If $T\geq 4$ then the weights $\sigma _{i}^{2}$ vary
over the positive $\sigma _{i}^{2}$ and so the limit $\delta _{w}$ of $\hat{%
\delta}_{w}$ is not the identified conditional ATE $\delta $.

Theorem 1 is different than Yitzhaki (1996) and Angrist (1998), who gave
weighted average interpretations of least squares in other, non-panel
settings. Theorem 1 is also different from Hahn (2001), who found that $\hat{%
\delta}_{w}$ consistently estimates the ATE. Hahn (2001) considered $T=2$
and assumed $X_{i}=(0,1)^{\prime }$. As noted by Hahn (2001), those
conditions are quite special. Theorem 1 is also different from Wooldridge
(2005), who showed that if $b_{i}=E[Y_{it}(1)-Y_{it}(0)|\alpha _{i}]$ is
mean independent of $X_{it}-\bar{X}_{i}$ for each $t$ then linear fixed
effects is a consistent estimator of $\delta $. The problem is that the
mean-independence assumption is very strong when $X_{it}$ is discrete. For
instance, if $T=2$, $X_{i2}-\bar{X}_{i}$ takes on the values $0$ when $%
X_{i}=(1,1)$ or $(0,0)$, $-1/2$ when $X_{i}=(1,0)\,,$ and $1/2$ when $%
X_{i}=(0,1)$. Thus mean independence of $b_{i}$ and $X_{i2}-\bar{X}_{i}$
actually implies that 
\begin{equation*}
E[b_{i}|X_{i}=(1,0)^{\prime }]=E[b_{i}|X_{i}=(0,1)^{\prime
}]=E[b_{i}|X_{i}\in \{(0,0)^{\prime },(1,1)^{\prime }\}].
\end{equation*}%
This is quite close to independence of $b_{i}$ and $X_{i}$, which is not
very interesting if we want to allow the treatment effect to vary with $%
X_{i} $.

The conditional ATE and QTE estimators can easily be modified to accommodate
the time effects of Assumption 4. The changes in $Y_{it}$ over time for
fixed $X_{it}$ can be used to identify and estimate the time effects that
can then be included in the estimation of the ATE and QTE. To describe this
approach, let $\hat{m}_{t}=\sum_{i=1}^{n}1(X_{it}=X_{i1})Y_{it}/%
\sum_{i=1}^{n}1(X_{it}=X_{i1})$ and 
\begin{equation*}
\hat{s}_{t}=\frac{\sum_{i=1}^{n}1(X_{it}=X_{i1})X_{i1}(Y_{it}-\hat{m}_{t})}{%
\sum_{i=1}^{n}1(X_{it}=X_{i1})X_{i1}(Y_{i1}-\hat{m}_{1})},\hat{\tau}_{t}=%
\hat{m}_{t}-\hat{s}_{t}\hat{m}_{1},t=2,...,T.
\end{equation*}%
This $(\hat{\tau}_{t},\hat{s}_{t})^{\prime }$ is an instrumental variables
estimator where the residual is $Y_{it}-\tau _{t}-s_{t}Y_{i1}$, the
instruments are $(1,X_{i1})^{\prime }$, and the estimation is done on the
subsample where $X_{it}=X_{i1}$. These estimators will be consistent and
asymptotically normal as long as $Cov(X_{i1},Y_{i1}|X_{it}=X_{i1})\neq 0$
for each $t=2,...,T.$ One could also use other functions of $X_{i1}$ as
instrumental variables to improve efficiency. We focus on just $X_{i1}$ as
an instrument for simplicity. Graham and Powell (2011) use a similar
approach to identify time effects in a linear model with continuous
regressors.

The time effects are accounted for in ATE and QTE estimation by removing
time location and scale effects from all periods when estimating the first
period effect, and then putting the scale effects back for other periods.
Note first that under Assumption 4 $\delta $ is the conditional ATE for the
first time period. Let $\tilde{Y}_{it}=(Y_{it}-\hat{\mu}_{t})/\hat{s}_{t}$
be the $t^{th}$ period observation with estimated location and scale
removed. Replacing $Y_{it}$ by $\tilde{Y}_{it}$ in the formula for $\hat{%
\delta}$ gives%
\begin{equation*}
\tilde{\delta}=\frac{\sum_{i=1}^{n}D_{i}[\tilde{Y}_{i}(x^{a})-\tilde{Y}%
_{i}(x^{b})]}{\sum_{i=1}^{n}D_{i}},\tilde{Y}_{i}(x)=\left\{ 
\begin{array}{c}
T_{i}(x)^{-1}\sum_{t=1}^{T}1(X_{it}=x)\tilde{Y}_{it},T_{i}(x)>0 \\ 
0,T_{i}(x)=0%
\end{array}%
\right. .
\end{equation*}%
The conditional ATE for the $t^{th}$ time period is given by $s_{t}\delta $
and a time average by $(\sum_{t=1}^{T}s_{t}/T)\delta $ for $s_{1}=1$,
analogously to equation (\ref{tqte}). These can be estimated by $\hat{s}_{t}%
\tilde{\delta}$ and $\bar{s}\tilde{\delta},$ respectively for $\bar{s}%
=\sum_{t=1}^{T}\hat{s}_{t}/T$ and $\hat s_{1}=1.$ These estimators will be
consistent and asymptotically normal. Because of their multistage nature the
bootstrap may provide the easiest approach to carrying out inference on
these estimators, where one resamples from the empirical distribution of $%
(Y_{i},X_{i}),$ $(i=1,...,n)$ to form confidence intervals for the true
parameter. For brevity we omit explicit results.

An analogous approach can be followed to account for time effects in the
QTE. The interpretation of $\delta _{\lambda }$ now becomes QTE for the
first time period conditional on $D_{i}=1$. An estimator of $%
G(y,x|D_{i}=1)=\Pr (g_{0}(x,\alpha _{i},\varepsilon _{i1})\leq y|D_{i}=1)$
that adjusts for time, location and scale is given by%
\begin{equation*}
\tilde{G}(y,x|D_{i}=1)=\frac{\sum_{i=1}^{n}D_{i}\tilde{G}_{i}(y,x)}{%
\sum_{i=1}^{n}D_{i}},\tilde{G}_{i}(y,x)=\left\{ 
\begin{array}{c}
T_{i}(x)^{-1}\sum_{t=1}^{T}1(X_{it}=x)\Phi (\frac{y-\tilde{Y}_{it}}{h}%
),T_{i}(x)>0 \\ 
0,T_{i}(x)=0%
\end{array}%
\right. .
\end{equation*}%
Let $\tilde{q}_{\lambda }^{a}=\tilde{G}^{-1}(\lambda ,x^{a}|D_{i}=1)$ and $%
\tilde{q}_{\lambda }^{b}=\tilde{G}^{-1}(\lambda ,x^{b}|D_{i}=1).$ Estimators
for the conditional QTE for the first period, other periods, and a time
average are given by $\tilde{\delta}_{\lambda }=\tilde{q}_{\lambda }^{a}-%
\tilde{q}_{\lambda }^{b}$, $\hat{s}_{t}\tilde{\delta}_{\lambda
},(t=2,...,T), $ and $\bar{s}\tilde{\delta}_{\lambda }$, respectively. Here
again the bootstrap provides a convenient method for inference. One could
also use quantiles to estimate the time effects, but we avoid that for
simplicity.

\section{Nonparametric Bounds in the Static Model}

When $g_{0}(x,\alpha _{i},\varepsilon _{it})$ is bounded we can estimate
bounds for the ASF and corresponding bounds for the ATE. For the QSF and QTE
we can also estimate bounds without any restriction on $g_{0}$, using the
fact that there are known upper and lower bounds for the indicator function $%
1(g_{0}(x,\alpha _{i},\varepsilon _{it})\leq y).$ The idea of the bounds is
an extension of the estimation of identified effects discussed in the
previous Section. Time homogeneity allows us to use time averages to
estimate the identified parts of the ASF or QSF when $x$ is an element of $%
X_{i},$ i.e. $X_{it}=x$ for some $t$, and apply the lower or upper bounds
when $x$ does not appear in $X_{i}$.

We first describe bounds estimation for the ASF. These bounds depend on
bounds on $g_{0}$ imposed in the following condition:

\bigskip

\textsc{Assumption 6: }$B_{\ell }\leq g_{0}(x,\alpha _{i},\varepsilon
_{it})\leq B_{u}$ \textit{for constants }$B_{\ell }$\textit{\ and }$B_{u}$ 
\textit{and all }$x.$

\bigskip

For example, in the binary-choice model, where $Y_{it}\in \{0,1\}$, upper
and lower bounds are $B_{u}=1$ and $B_{\ell }=0$ respectively. We could
allow $B_{\ell }$\textit{\ }and $B_{u}$ to depend on $x$ and using that
information could tighten the ATE bounds given below. To avoid further
complication we do not allow this.

Let $T_{i}(x)$ and $\bar{Y}_{i}(x)$ be as in Section 3 and $\bar{P}%
(x)=\sum_{i=1}^{n}1(T_{i}(x)=0)/n$ be the sample frequency of $x$ not
occurring in any time period. Estimated lower and upper bounds for $\mu (x)$
are%
\begin{equation*}
\hat{\mu}_{\ell }(x)=n^{-1}\sum_{i=1}^{n}\bar{Y}_{i}(x)+\bar{P}(x)B_{\ell },%
\hat{\mu}_{u}(x)=\hat{\mu}_{\ell }(x)+\bar{P}(x)(B_{u}-B_{\ell }).
\end{equation*}%
Here $\bar{Y}_{i}(x)$ estimates the identified part of the ASF,
corresponding to $T_{i}(x)>0$, and the upper and lower bounds are applied
for observations where $T_{i}(x)=0\,.$ Corresponding estimated lower and
upper bounds for the ATE are $\hat{\Delta}_{\ell }=\hat{\mu}_{\ell }(x^{a})-%
\hat{\mu}_{u}(x^{b})$ and $\hat{\Delta}_{u}=\hat{\mu}_{u}(x^{a})-\hat{\mu}%
_{\ell }(x^{b}).$ The width of these estimated bounds is 
\begin{equation*}
\hat{\Delta}_{u}-\hat{\Delta}_{\ell }=[\bar{P}(x^{a})+\bar{P}%
(x^{b})](B_{u}-B_{\ell }).
\end{equation*}%
For example, for binary choice with a binary regressor, where $B_{u}=1$ and $%
B_{\ell }=0,$ the width of the estimated bounds for the ATE\ is $\bar{P}(0)+%
\bar{P}(1),$ where $\bar{P}(0)$ and $\bar{P}(1)$ are the sample proportions
of $X_{i}$ with $X_{it}=1$ for all $t$ and $X_{it}=0$ for all $t$,
respectively

These estimators will be jointly asymptotically normal under i.i.d. $%
(Y_{i},X_{i})$. The asymptotic variance can be estimated by $\hat{\Sigma}%
=\sum_{i=1}^{n}\hat{\Psi}_{i}\hat{\Psi}_{i}^{\prime }/n$, where%
\begin{equation*}
\hat{\Psi}_{i}=\left( 
\begin{array}{c}
\bar{Y}_{i}(x^{a})-\bar{Y}_{i}(x^{b})+B_{\ell
}1(T_{i}(x^{a})=0)-B_{u}1(T_{i}(x^{b})=0)-\hat{\Delta}_{\ell } \\ 
\bar{Y}_{i}(x^{a})-\bar{Y}_{i}(x^{b})+B_{u}1(T_{i}(x^{a})=0)-B_{\ell
}1(T_{i}(x^{b})=0)-\hat{\Delta}_{u}%
\end{array}%
\right) .
\end{equation*}%
Confidence intervals for the identified set can then be formed using results
of Chernozhukov, Hong, and Tamer (2007) or Beresteanu and Molinari (2008,
pp. 779-781) on estimators of intervals where the upper and lower endpoints
are jointly asymptotically normal.

Turning to the bounds for the QSF, lower and upper estimated bounds for the $%
G(y,x)=\Pr (g_{0}(x,\alpha _{i},\varepsilon _{i1})\leq y)$ are $\hat{G}%
_{\ell }(y,x)=\sum_{i=1}^{n}\bar{G}_{i}(y,x)/n$ and $\hat{G}_{u}(y,x)=\hat{G}%
_{\ell }(y,x)+\bar{P}(x)$ respectively. The idea of these bounds is similar
to the ASF, with a known lower bound of $0$ and upper bound of $1$ for $%
1(g_{0}(x,\alpha _{i},\varepsilon _{i1})\leq y)$. To obtain quantile bounds
we need to invert these functions of $y$. For a strictly
increasing function $G(y)$ with range contained in $[0,1]$ let 
\begin{equation*}
Q(\lambda ,G(\cdot ))=\left\{ 
\begin{array}{l}
-\infty ,\text{ }\lambda \leq \inf_{y}G(y) \\ 
\multicolumn{1}{c}{G^{-1}(\lambda ),\text{ }\inf_{y}G(y)<\lambda
<\sup_{y}G(y)} \\ 
+\infty ,\text{ }\lambda \geq \sup_{y}G(y)%
\end{array}%
\right. .
\end{equation*}%
This is a function with domain $[0,1]$ and range equal to the extended real
line that can be used to invert $\hat{G}_{u}(y,x)$ and $\hat{G}_{\ell
}(y,x). $ Estimators of lower and upper bounds on the QSF are given by 
\begin{equation*}
\hat{q}_{\ell }(\lambda ,x)=Q(\lambda ,\hat{G}_{u}(\cdot ,x)),\hat{q}%
_{u}(\lambda ,x)=Q(\lambda ,\hat{G}_{\ell }(\cdot ,x)).
\end{equation*}%
Corresponding lower and upper bounds for the QTE are $\hat{\Delta}_{\lambda
\ell }=\hat{q}_{\ell }^{a}-\hat{q}_{u}^{b}$ and $\hat{\Delta}_{\lambda u}=%
\hat{q}_{u}^{a}-\hat{q}_{\ell }^{b}$ where $\hat{q}_{\ell }^{a}=\hat{q}%
_{\ell }(\lambda ,x^{a}),$ $\hat{q}_{u}^{a}=\hat{q}_{u}(\lambda ,x^{a}),$ $%
\hat{q}_{\ell }^{b}=\hat{q}_{\ell }(\lambda ,x^{b}),$ and $\hat{q}_{u}^{b}=%
\hat{q}_{u}(\lambda ,x^{b}).$ The width of these bounds depends on the shape
of the empirical distribution of $Y_{it}$ and on $\bar{P}(x).$ The width of
the bounds will be finite when 
\begin{equation}
\max \{\bar{P}(x^{a}),\bar{P}(x^{b})\}<\lambda <\min \{1-\bar{P}(x^{a}),1-%
\bar{P}(x^{b})\},  \label{quant ineq}
\end{equation}%
and otherwise they are infinitely wide.

The bounds will be joint asymptotically normal under the following
regularity condition:

\bigskip

\textsc{Assumption 7: }$\Pr (g_{0}(x,\alpha _{i},\varepsilon _{i1})\leq
y|X_{i})$ \textit{is twice continuously differentiable in }$y$ \textit{with
uniformly bounded derivatives and }$G_{\ell }(y,x)=E[E[1(T_{i}(x)>0)|X_{i}]\Pr
(g_{0}(x,\alpha _{i},\varepsilon _{i1})\leq y|X_{i})]$ \textit{is strictly
increasing in }$y$\textit{\ on the interior of its range for all }$x$. 
\textit{Also }$nh^{4}\longrightarrow 0$ and $nh^{2}\longrightarrow \infty $.

\bigskip

For $\lambda $ satisfying equation (\ref{quant ineq}) the asymptotic
variance can be estimated by $\hat{\Sigma}_{\lambda }=\sum_{i=1}^{n}\hat{\Psi%
}_{\lambda i}\hat{\Psi}_{\lambda i}^{\prime }/n$, where%
\begin{equation*}
\hat{\Psi}_{\lambda i}=\left( 
\begin{array}{c}
\frac{\bar{G}_{i}(\hat{q}_{\ell }^{a},x^{a})+1(T_{i}(x^{a})=0)-\lambda }{%
\hat{G}_{\ell }^{\prime }(\hat{q}_{\ell }^{a},x^{a})}-\frac{\bar{G}_{i}(\hat{%
q}_{u}^{b},x^{b})-\lambda }{\hat{G}_{\ell }^{\prime }(\hat{q}_{u}^{b},x^{b})}
\\ 
\frac{\bar{G}_{i}(\hat{q}_{u}^{a},x^{a})-\lambda }{\hat{G}_{u}^{\prime }(%
\hat{q}_{u}^{a},x^{a})}-\frac{\bar{G}_{i}(\hat{q}_{\ell
}^{b},x^{b})+1(T_{i}(x^{b})=0)-\lambda }{\hat{G}_{\ell }^{\prime }(\hat{q}%
_{\ell }^{b},x^{b})}%
\end{array}%
\right) .
\end{equation*}%
As in estimation of the conditional quantile effect, one might want to use
different bandwidths for numerators and denominators, or just bootstrap to
estimate the asymptotic variance.

Here is a result for both ATE and QTE bounds:

\bigskip

\textsc{Theorem 2:} \textit{Suppose that Assumptions 1, 2, and 5 are
satisfied.\ If Assumption 6 is satisfied} \textit{then there are }$\Delta
_{\ell },$\textit{\ }$\Delta _{u},$\textit{\ and }$\Sigma $ \textit{such
that } \textit{\ }%
\begin{equation*}
\sqrt{n}[(\hat{\Delta}_{\ell },\hat{\Delta}_{u})^{\prime }-(\Delta _{\ell
},\Delta _{u})^{\prime }]\overset{d}{\longrightarrow }N(0,\Sigma ),\hat{%
\Sigma}\overset{p}{\longrightarrow }\Sigma .
\end{equation*}%
\textit{\ where }$\Delta _{\ell }\leq \Delta \leq \Delta _{u}$, \textit{and
these bounds are sharp. If Assumption 7 is satisfied then there are }$\Delta
_{\lambda \ell },$\textit{\ }$\Delta _{\lambda u},$\textit{\ and }$\Sigma
_{\lambda }$ \textit{such that}%
\begin{equation*}
\sqrt{n}[(\hat{\Delta}_{\lambda \ell },\hat{\Delta}_{\lambda u})^{\prime
}-(\Delta _{\lambda \ell },\Delta _{\lambda u})^{\prime }]\overset{d}{%
\longrightarrow }N(0,\Sigma _{\lambda }),\hat{\Sigma}_{\lambda }\overset{p}{%
\longrightarrow }\Sigma _{\lambda }.
\end{equation*}%
\textit{\ where }$\Delta _{\lambda \ell }\leq \Delta _{\lambda }\leq \Delta
_{\lambda u}$\textit{. If }$G_{\ell }(y,x)$ \textit{is also everywhere
strictly increasing in }$y$ \textit{then these bounds are sharp.}

\bigskip

The sharpness conclusion of Theorem 2 for the ATE depends on being able to
let $g_{0}(x,\alpha _{i},\varepsilon _{it})$ take any value between $B_{\ell
}$ and $B_{u}.$ That is not possible for binary choice, where the outcome is
restricted to zero or one. Nevertheless the bounds can still be shown to be
sharp.

Similarly to the treatment-effects literature, we may be interested in the
ATE or QTE, conditional on $X_{i}\in S$ for some set $S$. For example, if $%
X_{it}\in \{0,1\}$ represents treatment then we might be interested in the
effect of treatment conditional on ever treated, i.e. conditional on $%
X_{i}\neq (0,...,0)^{\prime }$. Tighter bounds for such effects can be
formed and in some cases the effects may be identified. These bounds can be
estimated by replacing $1(X_{it}=x)$ by $1(X_{i}\in S)1(X_{it}=x)$ in the
definition of $\bar{Y}_{i}(x)$ and $\bar{G}_{i}(y,x)$, $1(T_{i}(x)=0)$ by $%
1(X_{i}\in S)1(T_{i}(x)=0)$ in the definition of $\bar{P}(x)$, and dividing
through by $\sum_{i=1}^{n}1(X_{i}\in S)/n.$ If $1(X_{i}\in S)\leq D_{i}$ for 
$D_{i}$ from Section 3 the corresponding effects will be identified, and the
upper and lower estimated bounds will be identical.

Time effects can easily be allowed for in quantile-effect bounds by adapting
the approach used earlier. It is not clear that allowing for time effects in
that way makes sense for bounds on the ATE, e.g. for binary choice models
where the support of $Y_{it}$ is fixed. Therefore we focus just on time
effects in quantile bounds. For QTE bounds we can replace $Y_{it}$ by $%
\tilde{Y}_{it}=(Y_{it}-\hat{\mu}_{t})/\hat{s}_{t}$ in the formula for $\hat{G%
}_{\ell }(y,x)$ given above, and interpret $\hat{\Delta}_{\lambda \ell }$
and $\hat{\Delta}_{\lambda u}$ as estimators of the first period bounds.
Estimators of $t^{th}$ period lower and upper bounds for the QTE are then
given by $\hat{s}_{t}\hat{\Delta}_{\lambda \ell }$ and $\hat{s}_{t}\hat{%
\Delta}_{\lambda u}$ respectively. Estimators of time average bounds are $%
\bar{s}\hat{\Delta}_{\lambda \ell }$ and $\bar{s}\hat{\Delta}_{\lambda u},$
where $\bar{s}=\sum_{t=1}^{T}\hat{s}_{t}/T.$ These upper and lower bounds
will be joint asymptotically normal, and their asymptotic variance can be
estimated by the bootstrap.

\section{Nonparametric Bounds in the Dynamic Model}

Analysis of the dynamic model is more challenging than that of the static
one. In the dynamic model of Assumption 3 only the first-period regressor is
common to the conditioning sets for each time period. Consequently location
and scale time effects are not identified, because the conditioning set is
different for every time period. For this reason we do not consider time
effects in the nonparametric dynamic model. Also, the identification and
bounds analysis is limited to objects that are conditional on the first
period or are unconditional. For example, we cannot identify or bound the
ATE conditional on $X_{it}$ changing over time because that event involves
information about all time periods. We can bound unconditional objects and
ones that are conditional on just $X_{i1}.$ These bounds are simple and
novel, for example in providing partial-identification results for the
average effect of state dependence with heterogeneity in both location and
slope when $Y_{it}$ is binary and $X_{it}=Y_{it-1}$.

The model with a binary, lagged dependent variable has $%
Y_{it}=g_{0}(Y_{i,t-1},\alpha _{i},\varepsilon _{it})$, and under Assumption
3, 
\begin{eqnarray*}
\Pr (Y_{it} =1|X_{it},...,X_{i1},\alpha _{i})&=&\int g(Y_{i,t-1},\alpha
_{i},\varepsilon )dF(\varepsilon |\alpha _{i},Y_{i0}) \\
&=&\Pr (Y_{it}=1|Y_{i,t-1},\alpha _{i},Y_{i0}),
\end{eqnarray*}%
where $F(\varepsilon |\alpha _{i},Y_{i0})$ denotes the conditional CDF of $%
\varepsilon _{it}$ given $\alpha _{i}$ and $Y_{i0}.$ Here $\Pr
(Y_{it}=1|Y_{i,t-1},\alpha _{i},Y_{i0})$ does not vary with $t$, and the
model places no other restrictions on $\Pr (Y_{it}=1|Y_{i,t-1},\alpha
_{i},Y_{i0})$. Conditioning on $Y_{i0}$ is present to account correctly for
the initial condition, as in Honore and Tamer (2006) and Browning and Carro
(2007, 2009). The probabilities can be distributed across individuals in any
way at all through the individual effect $\alpha _{i}$. That is we can think
of the four conditional probabilities,%
\begin{equation*}
\Pr (Y_{it}=1|1,\alpha _{i},1),\Pr (Y_{it}=1|0,\alpha _{i},1)\Pr
(Y_{it}=1|1,\alpha _{i},0),\Pr (Y_{it}=1|0,\alpha _{i},0),
\end{equation*}%
as having an unrestricted distribution. Here the ATE is 
\begin{equation*}
\Delta =\int [\Pr (Y_{it}=1|Y_{i,t-1}=1,\alpha ,Y_{0})-\Pr
(Y_{it}=1|Y_{i,t-1}=0,\alpha ,Y_{0})]dF(\alpha ,Y_{0}).
\end{equation*}%
This object quantifies the effect of state dependence in the presence of
individual heterogeneity, an important problem posed by Feller (1943) and
Heckman (1981). The dynamic bounds here provide a simple, estimable,
identified set for this object. This model is considered by Browning and
Carro (2007, 2009), who derive properties of various estimators and
restrictions on $\alpha _{i}$ that lead to identification. We give
nonparametric bounds.

A partition of $X_{i}$ values that preserves the dynamic structure of
Assumption 3 is used to obtain bounds for the ASF and QSF. For each $x$ we
partition $X_{i}$ into realizations where the first occurrence of $x$ is at
time $t$ and the set where $x$ never occurs. This partition is given by $\{%
\mathcal{\bar{X}}(x),\mathcal{X}_{1}(x),...,\mathcal{X}_{T}(x)\}$ where%
\begin{equation*}
\mathcal{X}_{t}(x)=\{X:X_{t}=x,\ X_{s}\neq x\ \forall s<t\},t=1,...,T;%
\mathcal{\bar{X}}(x)=\{X:X_{t}\neq x\ \forall t\}.
\end{equation*}%
Define $\hat{Y}_{i}(x)=\sum_{t=1}^{T}1(X_{i}\in \mathcal{X}_{t}(x))Y_{it}$,
which picks out the $Y_{it}$ for the time period where $x$ first occurs.
Estimated lower and upper ASF bounds are 
\begin{equation*}
\hat{\mu}_{\ell }(x)=n^{-1}\sum_{i=1}^{n}\hat{Y}_{i}(x)+\bar{P}(x)B_{\ell },%
\hat{\mu}_{u}(x)=\hat{\mu}_{\ell }(x)+\bar{P}(x)(B_{u}-B_{\ell })\text{.}
\end{equation*}%
Corresponding lower and upper bounds for $\Delta $ are $\hat{\Delta}_{\ell }=%
\hat{\mu}_{\ell }(x^{a})-\hat{\mu}_{u}(x^{b})$ and $\hat{\Delta}_{u}=\hat{\mu%
}_{u}(x^{a})-\hat{\mu}_{\ell }(x^{b}).$ A joint asymptotic-variance
estimator $\hat{\Sigma}$ can be constructed exactly as for the static case
with $\hat{Y}_{i}(x)$ replacing $\bar{Y}_{i}(x).$

It is interesting to note that the width $\bar{P}(x)(B_{u}-B_{\ell })$ of
the estimated ASF bounds is the same for the dynamic and static models.
Because the static model is a special case of the dynamic one we conjecture
that the bounds for the dynamic model are sharp like the bounds for the
static one, but have not yet been able to show this.

To construct estimated lower and upper bounds for the CDF of $g_{0}(x,\alpha
_{i},\varepsilon _{it})$ let $\hat{G}_{i}(y,x)=\sum_{t=1}^{T}1(X_{i}\in 
\mathcal{X}_{t}(x))\Phi (\frac{y-Y_{it}}{h}).$ The estimated CDF bounds are%
\begin{equation*}
\hat{G}_{\ell }(y,x)=\frac{1}{n}\sum_{i=1}^{n}\hat{G}_{i}(y,x),\hat{G}%
_{u}(y,x)=\hat{G}_{\ell }(y,x)+\bar{P}(x).
\end{equation*}%
Estimated lower and upper bounds for the QSF are then given by 
\begin{equation*}
\hat{q}_{\ell }(\lambda ,x)=Q(\lambda ,\hat{G}_{u}(\cdot ,x)),\hat{q}%
_{u}(\lambda ,x)=Q(\lambda ,\hat{G}_{\ell }(\cdot ,x)).
\end{equation*}%
Corresponding lower and upper bounds for the QTE are $\hat{\Delta}_{\lambda
\ell }=\hat{q}_{\ell }(\lambda ,x^{a})-\hat{q}_{u}(\lambda ,x^{b})$ and $%
\hat{\Delta}_{\lambda u}=\hat{q}_{u}(\lambda ,x^{a})-\hat{q}_{\ell }(\lambda
,x^{b}).$ A joint asymptotic variance estimator $\hat{\Sigma}_{\lambda }$
can be constructed just as for the static case with $\hat{G}_{i}(y,x)$
replacing $\bar{G}_{i}(y,x).$

\bigskip

\textsc{Theorem 3:} \textit{Suppose that Assumptions 1, 3, and 5 are
satisfied.\ If Assumption 6 is satisfied} \textit{then there are }$\Delta
_{\ell },$\textit{\ }$\Delta _{u},$\textit{\ and }$\Sigma $ \textit{such that%
}%
\begin{equation*}
\sqrt{n}[(\hat{\Delta}_{\ell },\hat{\Delta}_{u})^{\prime }-(\Delta _{\ell
},\Delta _{u})^{\prime }]\overset{d}{\longrightarrow }N(0,\Sigma ),\hat{%
\Sigma}\overset{p}{\longrightarrow }\Sigma .
\end{equation*}%
\textit{\ where }$\Delta _{\ell }\leq \Delta \leq \Delta _{u}$\textit{. Also
if Assumption 7 is satisfied with }$X_{i1}$ \textit{%
replacing }$X_i$ \textit{then there are }$\Delta _{\lambda \ell },$%
\textit{\ }$\Delta _{\lambda u},$\textit{\ and }$\Sigma _{\lambda }$ \textit{%
such that}%
\begin{equation*}
\sqrt{n}[(\hat{\Delta}_{\lambda \ell },\hat{\Delta}_{\lambda u})^{\prime
}-(\Delta _{\lambda \ell },\Delta _{\lambda u})^{\prime }]\overset{d}{%
\longrightarrow }N(0,\Sigma _{\lambda }),\hat{\Sigma}_{\lambda }\overset{p}{%
\longrightarrow }\Sigma _{\lambda }.
\end{equation*}%
\textit{\ where }$\Delta _{\lambda \ell }\leq \Delta _{\lambda }\leq \Delta
_{\lambda u}$\textit{. }

\bigskip

Similarly to the static model we may be interested in effects conditional on 
$X_{i1}\in S_{1}$ for some set $S_{1}$. For example, if $X_{it}\in \{0,1\}$
represents treatment then we might be interested in the effect of treatment
conditional on being treated in the first period, i.e. conditional on $%
X_{i1}=1$. Tighter bounds for such effects can be estimated by replacing $%
1(X_{i}\in \mathcal{X}_{t}(x))$ by $1(X_{i1}\in S_{1})1(X_{i}\in \mathcal{X}%
_{t}(x))$ in the definition of $\hat{Y}_{i}(x)$ and $\hat{G}_{i}(y,x)$, $%
1(T_{i}(x)=0)$ by $1(X_{i1}\in S_{1})1(T_{i}(x)=0)$ in the definition of $%
\bar{P}(x)$, and dividing through by $\sum_{i=1}^{n}1(X_{i1}\in S_{1})/n.$

In the binary, lagged-dependent-variable example we have $B_{\ell }=0$ and $%
B_{u}=1$, so the bounds on the ATE are%
\begin{equation*}
\hat{\Delta}_{\ell }=\frac{1}{n}\sum_{i=1}^{n}[\hat{Y}_{i}(1)-\hat{Y}%
_{i}(0)]-\bar{P}(0),\hat{\Delta}_{u}=\hat{\Delta}_{\ell }+\bar{P}(1)+\bar{P}%
(0).
\end{equation*}%
Here $\bar{P}(1)+\bar{P}(0)$ estimates the width of the bounds, providing a
very simple measure of the severity of the problem of identifying state
dependence in the presence of heterogeneity. The bounds will tend to be wide
in short panels but more informative in long ones.

Figure 1 shows the width of corresponding population bounds in a numerical
example based on a dynamic probit model where 
\begin{equation*}
Y_{it}=1(\beta ^{\ast }Y_{i,t-1}+\alpha _{i}\geq \varepsilon
_{it}),\varepsilon _{it}\sim N(0,1),\alpha _{i}\sim N(0,1),\Pr (Y_{i0}=1)=.5.
\end{equation*}%
We consider different DGPs indexed by $\beta ^{\ast }\in \lbrack -2,2]$ and
compute the width of the bounds for $T\in \{2,4,8,16,32,64\}$. The width is
asymmetric with respect to $\beta ^{\ast }=0$ because $\Pr
(X_{i}=(1,...,1)^{\prime })$ grows with $\beta ^{\ast }$, whereas $\Pr
(X_{i}=(0,...,0)^{\prime })$ does not depend on $\beta ^{\ast }$. The width
growing with $\beta ^{\ast }$ may therefore be explained by having fewer switches of $%
Y_{it}$ between one and zero when $\beta ^{\ast }$ is larger. It is
presumably the changes that help identify the ATE. We find that the bounds
can be substantially wide for high values of $\beta ^{\ast }$ even for large 
$T$, consistent with the width of the nonparametric bounds shrinking only at
rate $1/T,$ as shown in the next Section. Semiparametric bounds for this
model that impose the constancy of $\beta ^{\ast }$ across individuals, will
shrink much faster at $T$ grows, as shown in Section 7.

\section{The Impact of $T$}

Increasing $T$ improves identification, shrinking the estimated and
population-identified sets for the objects of interest. The rate at which
the identified set shrinks quantifies this improvement. Here we give rates
for the ASF and, for brevity, leave the quantile results to the
Supplementary Material.

The width of the population bounds for the ASF is $(B_{u}-B_{\ell })\mathcal{%
\bar{P}}(x)$ where%
\begin{equation*}
\mathcal{\bar{P}}(x)=\Pr (X_{i1}\neq x,...,X_{iT}\neq x).
\end{equation*}%
Thus, the rate at which the identified set shrinks, that we will refer to as
the identification rate, is the same as the rate at which $\mathcal{\bar{P}}%
(x)$ shrinks. Factors that determine this rate can be seen when $X_{it}$ is
i.i.d. conditional on $\alpha _{i}$. In that case%
\begin{equation*}
\mathcal{\bar{P}}(x)=E[\Pr (X_{it}\neq x|\alpha _{i})^{T}].
\end{equation*}%
The rate at which $\mathcal{\bar{P}}(x)$ goes to zero will be determined by how much
probability mass of $\Pr (X_{it}\neq x|\alpha _{i})$ is close to one. If $%
\Pr (X_{it}\neq x|\alpha _{i})=1$ with positive probability then $\mathcal{%
\bar{P}}(x)$ does not go to zero. This corresponds to nonidentification of
the ASF, where $x$ does not occur for some individuals as indexed by $\alpha
_{i}$ (see Theorem A11 of the Supplementary Material). On the other hand, if $%
\ \Pr (X_{it}\neq x|\alpha _{i})$ is bounded away from one then the
identified set will shrink exponentially quickly, since $\Pr (X_{it}\neq
x|\alpha _{i})^{T}\leq (1-\varepsilon )^{T}$ for some $\varepsilon >0$. In
between the nonidentified and exponential rate cases there are a range of
rates depending on how much of the distribution of $\Pr (X_{it}\neq x|\alpha
_{i})$ is close to $1$. The following result shows the range of rates.

\bigskip

\textsc{Theorem 4:}\textit{\ Suppose that Assumptions 1, 3, 5, and 6 are
satisfied and }$(X_{i1},X_{i2},...)$\textit{\ is stationary and Markov of
order }$J$\textit{\ conditional on }$\alpha _{i}$\textit{. If for some }$%
\varepsilon >0$\textit{,\ }$\Pr (X_{it}=x|X_{i,t-1},...,X_{i,t-J},\alpha
_{i})\geq \varepsilon $\textit{\ a.s. then }$\mu _{u}(x)-\mu _{\ell }(x)\leq
(B_{u}-B_{\ell })(1-\varepsilon )^{T-J}.$\textit{\ If }$X_{it}$\textit{\ is
i.i.d. conditional on }$\alpha _{i},$\textit{\ }$\Pr (X_{it}\neq x|\alpha
_{i})$\textit{\ is continuously distributed with pdf }$f_{P}(p),$\textit{\
and\ }%
\begin{equation}
f_{P}(p)\leq Cp^{\gamma -1}(1-p)^{v-1},\gamma >0,v>0,\mathit{\ }
\end{equation}%
\textit{then }$\mu _{u}(x)-\mu _{\ell }(x)=O(T^{-v}).$

\bigskip

The upper bound on the rate at which the pdf $f_{P}(p)$ of $\Pr (X_{it}\neq
x|\alpha _{i})$ grows or converges to zero as $p\longrightarrow 1$ provides
an upper bound on the rate at which the identified set shrinks. For example,
if $v=1$ so that $f_{P}(p)$ is bounded as $p\longrightarrow 1,$ then the
identified set shrinks at rate $1/T.$ All of the rates implied by this
result are slower than the exponential rate, reflecting how having $\Pr
(X_{it}\neq x|\alpha _{i})$ close to $1$ affects the rate. Also, $\gamma $
has no effect on the convergence rate because that rate is determined by
closeness of $\Pr (X_{it}\neq x|\alpha _{i})$ to $1$, and not to $0$.

The dynamic, binary-choice model is an example where more explicit
conditions can be given. Suppose $Y_{it}=1(\alpha _{i1}+(\alpha _{i2}-\alpha
_{i1})Y_{i,t-1}\geq \varepsilon _{it})$ and $\varepsilon _{it}$ is i.i.d.
and independent of $\alpha _{i}=(\alpha _{i1}, \alpha _{i2})$ with CDF $H(\varepsilon )$. Here $\Pr
(Y_{it}=1|Y_{i,t-1}=0,\alpha _{i})=H(\alpha _{i1})$ and $\Pr
(Y_{it}=1|Y_{i,t-1}=1,\alpha _{i})=H(\alpha _{i2}).$ Unbounded $\alpha _{i}$
and bounded $\varepsilon _{it}$ will correspond to the unidentified case.
Bounded $\alpha _{i}$ and unbounded $\varepsilon _{it}$ lead to an
exponential convergence rate. The following result covers the in-between
case. Let $f_{\varepsilon }(\varepsilon ),$ $f_{\alpha _{1}}(\alpha )$, and $%
f_{\alpha _{2}}(\alpha )$ denote the pdfs of $\varepsilon _{it},$ $\alpha
_{i1},$ and $\alpha _{i2}$ respectively, all are assumed to be continuously
distributed.

\bigskip

\textsc{Theorem 5:}\textit{\ If }$Y_{it}=1(\alpha _{i1}+(\alpha _{i2}-\alpha
_{i1})Y_{i,t-1}\geq \varepsilon _{it}),$ \textit{where }$\varepsilon
_{it},(t=1,...,T)$ \textit{is i.i.d. and independent of }$(\alpha
_{i1},\alpha _{i2})$\textit{\ and there is }$v,C>0$ \textit{such that for
all }$\varepsilon $%
\begin{equation}
\max_{j=1,2}f_{\alpha _{j}}(\varepsilon )\leq CH(\varepsilon
)^{v-1}[1-H(\varepsilon )]^{v-1}f_{\varepsilon }(\varepsilon ),
\label{density bound}
\end{equation}%
\textit{then }$\Delta _{u}-\Delta _{\ell }=O(T^{-v}).$

\bigskip

Here we see that the identification rate in the nonparametric dynamic model
is related to the tail thickness of the distribution of $\alpha _{i1}$ and $%
\alpha _{i2}$ relative to the distribution of $\varepsilon _{it}$. The
thinner the tail of $f_{\varepsilon }(\varepsilon )$ relative to the tails
of $f_{\alpha _{1}}(\alpha _{1})$ and $f_{\alpha _{2}}(\alpha _{2})$ the
smaller $v$ will need to be to satisfy the inequality in Theorem 5 and the
slower the identification rate will be. In this way the identification rate
is slower the less strong the signal provided by $\varepsilon _{it}$
relative to the individual effects. Here there is no $\gamma $ present
because both left and right tails matter, in order to bound the rate for the
ATE, and not just for the ASF at a particular $x$.

For a specific example consider $\alpha _{i1}$ and $\alpha _{i2}$ as $%
N(0,\sigma _{\alpha }^{2})$ and $\varepsilon _{it}$ as $N(0,\sigma
_{\varepsilon }^{2})$ where $\sigma _{\varepsilon }^{2}\leq \sigma _{\alpha
}^{2}.$ Then for constants $C_{1},$ $C_{2},$ and $v=\sigma _{\varepsilon
}^{2}/\sigma _{\alpha }^{2}$ we have $f_{\alpha _{j}}(\varepsilon
)=C_{1}[f_{\varepsilon }(\varepsilon )]^{v}.$ Also, as is well known for the
Gaussian distribution, $f_{\varepsilon }(\varepsilon )\geq
C_{2}F_{\varepsilon }(\varepsilon )[1-F_{\varepsilon }(\varepsilon )],$
where $F_{\varepsilon }(\varepsilon )$ denotes the CDF of $\varepsilon $. It
follows by $v\leq 1$ that%
\begin{equation*}
f_{\alpha _{j}}(\varepsilon )=C_{1}[f_{\varepsilon }(\varepsilon
)]^{v-1}f_{\varepsilon }(\varepsilon )\leq C_{1}C_{2}^{v-1}F_{\varepsilon
}(\varepsilon )^{v-1}[1-F_{\varepsilon }(\varepsilon )]^{v-1}f_{\varepsilon
}(\varepsilon ).
\end{equation*}%
Thus equation (\ref{density bound}) is satisfied with $v=\sigma
_{\varepsilon }^{2}/\sigma _{\alpha }^{2}$ so that 
\begin{equation*}
\Delta _{u}-\Delta _{\ell }=O(T^{-\sigma _{\varepsilon }^{2}/\sigma _{\alpha
}^{2}}).
\end{equation*}%
Hence the width of the bounds shrinks at a rate no larger than $T^{-1}$ and
the rate is slower the smaller $\sigma _{\varepsilon }^{2}/\sigma _{\alpha
}^{2}$ is. It can also be shown that convergence is faster than $T^{-1}$
when $\sigma _{\varepsilon }^{2}>\sigma _{\alpha }^{2}$ and increases with $%
\sigma _{\varepsilon }^{2}/\sigma _{\alpha }^{2}$. Thus we see that the
stronger the signal provided by $\varepsilon $ relative to that provided by $%
\alpha ,$ in the sense that the higher $\sigma _{\varepsilon }^{2}$ is
relative to $\sigma _{\alpha }^{2}$, the faster will be the identification
rate.

One can obtain analogous results in a static model. If $X_{it}=1(\alpha
_{i}\geq \eta _{it})$ is a binary regressor where $\eta _{it}$ is i.i.d.
over time then the identification rate will be $T^{-v}$ when the inequality
in Theorem 5 is satisfied with the pdf $f_{\eta }(\eta )$ of $\eta _{it}$
replacing the pdf $f_{\varepsilon }(\varepsilon ).$ If $\alpha _{i}$ and $%
\eta _{it}$ are distributed as $N(0,\sigma _{\alpha }^{2})$ and $\eta _{it}$
as $N(0,\sigma _{\eta }^{2})$ respectively with $\sigma _{\eta }^{2}\leq
\sigma _{\alpha }^{2}$, then the identified set shrinks at rate $T^{-\sigma
_{\eta }^{2}/\sigma _{\alpha }^{2}}$. For brevity we omit the details.

\section{Semiparametric Multinomial Choice Models}

The nonparametric bounds are informative but may be quite wide for small $T$%
. They can be tightened by imposing additional structure on the model. One
way to do this is to specify a parametric model for the conditional
distribution of $Y_{i}$ given values for $(X_{i},\alpha _{i}).$ We focus
here on multinomial choice models. In those models $Y_{i}$ is one of a
finite number of outcomes, denoted here by $\{Y^{1},...,Y^{J}\}.$ The
parametric part of the model are the known conditional probabilities $%
\mathcal{L}_{j}^{k}(\alpha ,\beta )$ of $Y_{i}=Y^{j}$\textit{\ }given $%
\alpha _{i}$ and $X_{i}\in \mathcal{X}^{k},(k=1,...,K),$ where $\beta $ is a
parameter vector with true value $\beta ^{\ast }$, and $\mathcal{X}^{k}$ is
the set of $X_{i}$ values being conditioned on. Formulating the model in
this way allows for $X_{i}$ that are lagged dependent variables. The
nonparametric part of the model will be the unknown CDF's $F_{k}^{\ast
}(\alpha ),(k=1,...,K)$ of $\alpha _{i}$ conditional on $X_{i}$ in each $%
\mathcal{X}^{k}.$ The model then satisfies

\bigskip

\textsc{Assumption 8:} $\Pr (Y_{i}=Y^{j}|X_{i}\in \mathcal{X}^{k})=\int 
\mathcal{L}_{j}^{k}(\alpha ,\beta ^{\ast })dF_{k}^{\ast }(\alpha
),(j=1,...,J;k=1,...,K)$\textit{.}

\bigskip

Some examples may be helpful. An important example is a binary choice model
where $Y_{it}\in \{0,1\}$, $\alpha $ is a scalar location individual effect, 
$\Pr (Y_{it}=1|X_{i},\alpha _{i},\beta ^{\ast })=H(X_{it}^{\prime }\beta
^{\ast }+\alpha _{i})$ for a CDF $H(\varepsilon ),$ and $Y_{i1},...,Y_{iT}$
are mutually independent conditional on $X_{i}$ and $\alpha _{i}$. In this
case we would let $\mathcal{X}^{k}$ be a singleton given by the $k^{th}$
value $X^{k}$ in the finite support of $X_{i}$ and%
\begin{equation}
\mathcal{L}_{j}^{k}(\alpha ,\beta )=\prod_{t=1}^{T}H(X_{t}^{k\prime }\beta
+\alpha )^{Y_{t}^{j}}[1-H(X_{t}^{k\prime }\beta +\alpha )]^{1-Y_{t}^{j}}.
\label{semistat}
\end{equation}%
Time effects can be included in this model by specifying that some
components of $X_{t}^{k}$ only depend on $t.$ This model can also be
generalized to allow for some slopes to vary across individuals by
specifying that 
\begin{equation}
\mathcal{L}_{j}^{k}\left( \alpha ,\beta \right)
=\prod_{t=1}^{T}H(z_{t}^{\prime }\beta _{1}+X_{t1}^{k\prime }\beta
_{2}+X_{t2}^{k\prime }\alpha )^{Y_{t}^{j}}[1-H(z_{t}^{\prime }\beta
_{1}+X_{t1}^{k\prime }\beta _{2}+X_{t2}^{k\prime }\alpha )]^{1-Y_{t}^{j}}.
\label{semi}
\end{equation}%
This model allows the coefficients of $X_{t2}^{k}$ to vary with individuals,
which will include a location effect when some element of $X_{t2}^{k}$ does
not vary with $t$ or $k.$

This set up also allows for dynamic models. For example, consider a binary
choice model with a lagged dependent variable where $\Pr
(Y_{it}=1|Y_{i,t-1},...,Y_{i0},\alpha _{i},\beta ^{\ast })=H(Y_{i,t-1}\beta
^{\ast }+\alpha _{i}).$ Here $X_{i}=(Y_{i,T-1},...,Y_{i0})$ and we take $%
K=2, $ with $\mathcal{X}^{k}=\{X_{i}:X_{i1}=Y_{i0}=k-1\}.$ The parametric
part of the model is%
\begin{eqnarray}
\mathcal{L}_{j}^{k}\left( \alpha ,\beta \right)
&=&\prod_{t=2}^{T}H(Y_{t-1}^{j}\beta +\alpha
)^{Y_{t}^{j}}[1-H(Y_{t-1}^{j}\beta +\alpha )]^{1-Y_{t}^{j}}  \label{semidyn}
\\
&&\times H((k-1)\beta +\alpha )^{Y_{1}^{j}}[1-H((k-1)\beta +\alpha
)]^{1-Y_{1}^{j}}.  \notag
\end{eqnarray}%
This model could be generalized to allow individual specific coefficients
for the dynamic effect, time effects, and other covariates, including the
model of Browning and Carro (2009). For brevity we omit this generalization.

The ATE and its bounds can be decomposed into a weighted average of
conditional ATE and corresponding bounds, weighted by the identified $\Pr
(X_{i}\in \mathcal{X}^{k})$. The semiparametric model may restrict the
conditional bounds so we focus first on them. We will assume that a
conditional ATE takes the form%
\begin{equation*}
\Delta ^{k}=\int \Delta (\alpha ,\beta ^{\ast })dF_{k}^{\ast }(\alpha ),
\end{equation*}%
where $\Delta (\alpha ,\beta )$ denotes a treatment effect conditional on $%
\alpha $. For example, in the model of equation (\ref{semistat}) we could
take $\Delta (\alpha ,\beta )=H(x^{a\prime }\beta +\alpha )-H(x^{b\prime
}\beta +\alpha ),$ in which case%
\begin{equation*}
\Delta ^{k}=\int [H(x^{a\prime }\beta ^{\ast }+\alpha )-H(x^{b\prime }\beta
^{\ast }+\alpha )]dF_{k}^{\ast }(\alpha )
\end{equation*}%
is the ATE conditional on $X_{i}=X^{k}$. One could also consider the ASF
conditional on $X_{i}=X^{k},$ that would be $\int H(x^{\prime }\beta ^{\ast
}+\alpha )dF_{k}^{\ast }(\alpha )$ in this example.

Neither $\Delta ^{k}$ nor $\beta ^{\ast }$ need be identified. Instead,
there may be sets of $\beta ^{\ast }$ and ATE values that are consistent
with the distribution of the data. To describe the identified sets let $%
\mathcal{P}=(\mathcal{P}_{1}^{1},...,\mathcal{P}_{J}^{1},...,\mathcal{P}%
_{J}^{K})^{\prime }$ denote the vector of population choice probabilities
with $\mathcal{P}_{j}^{k}=\Pr (Y_{i}=Y^{j}|X_{i}\in \mathcal{X}^{k})$ and 
\begin{equation*}
\mathcal{F}_{k}(\beta ,\mathcal{P})=\{F_{k}:\mathcal{P}_{j}^{k}=\int 
\mathcal{L}_{j}^{k}\left( \alpha ,\beta \right) dF_{k}(\alpha ),j=1,...,J\},
\end{equation*}%
where $\mathcal{F}_{k}(\beta ,\mathcal{P})$ may be empty. The identified set
for $\beta ^{\ast }$ is%
\begin{equation*}
B=\{\beta \text{ s.t. }\mathcal{F}_{k}(\beta ,\mathcal{P})\neq \varnothing
,\forall k=1,...,K\}.
\end{equation*}%
That is, $B$ is the set where there exist individual effect distributions
such that integrals of model probabilities equal population choice
probabilities. Sharp upper and lower bounds $\Delta _{u}^{k}$ and $\Delta
_{\ell }^{k}$ for $\Delta ^{k}$ are given by%
\begin{equation}
\Delta _{u}^{k}=\sup_{\beta \in B,F_{k}\in \mathcal{F}_{k}(\beta ,\mathcal{P}%
)}\int \Delta (\alpha ,\beta )dF_{k}\left( \alpha \right) ,\text{ }\Delta
_{\ell }^{k}=\inf_{\beta \in B,F_{k}\in \mathcal{F}_{k}(\beta ,\mathcal{P}%
)}\int \Delta (\alpha ,\beta )dF_{k}\left( \alpha \right) .  \label{sate}
\end{equation}%
This characterization of bounds for the ATE extends that of Honore and Tamer
(2006) from a finite dimensional $F_{k}$, where $\alpha $ is restricted to a
known fixed grid, to infinite-dimensional $F_{k}$ where any distribution for 
$\alpha $ is allowed.

For purposes of comparison with the nonparametric results we consider models
without trends, where the semiparametric models in equations (\ref{semistat}%
) and (\ref{semidyn}) are nested in the nonparametric static or dynamic
model. In those models $\Delta ^{k}$ will be identified if it is also
identified in the nonparametric model. In the static case $\Delta ^{k}$ is
nonparametrically identified if $X_{t}^{k}$ takes on the values $x^{b}$ and $%
x^{a}$ for some time periods$.$ This follows similarly to the identification
of the conditional effect $\delta $ in Section 3. Therefore, in static
models obtaining a smaller identified set by imposing the restrictions of a
semiparametric model is limited to those $\Delta ^{k}$ where at least one of 
$x^{b}$ or $x^{a}$ does not appear in any time period. In what follows we
focus on these $\Delta ^{k}$.

When slopes vary across individuals the semiparametric bounds may be no
tighter than the nonparametric ones. To illustrate consider a binary-choice
model with a single binary regressor $X_{it},$ where $Y_{it}=1((\alpha
_{i2}-\alpha _{i1})X_{it}+\alpha _{i1}>\varepsilon _{it}),$ $\varepsilon
_{it}$ is independent of $(X_{i},\alpha _{i2},\alpha _{i1}),$ and $%
\varepsilon _{it}$ has known CDF $H(\varepsilon )$ that is strictly
increasing on the entire real line. The joint distribution of $H(\alpha
_{i1})$ and $H(\alpha _{i2})$ conditional on $X_{i}=X^{k}$ is entirely
unrestricted. Therefore when $X^{k}=(0,...,0)^{\prime }$ the fact that $%
E[H(\alpha _{i1})|X_{i}=X^{k}]=E[Y_{it}|X_{i}=X^{k}]$ for every every $t$,
and so is identified gives no information about $E[H(\alpha
_{i2})|X_{i}=X^{k}].$ Thus, $E[H(\alpha _{i2})|X_{i}=X^{k}]$ can be anything
in the unit interval. \ Therefore, the width of the bound for $\Delta
^{k}=E[H(\alpha _{i2})-H(\alpha _{i1})|X_{i}=X^{k}]$ will be equal to the
width in the nonparametric case, $\Delta _{u}^{k}-\Delta _{\ell }^{k}=1$.
More generally, in the panel binary choice model of equation (\ref{semi}),
when there are no time effects, every coefficient of $X_{it}$ varies across
individuals, and $X_{it}$ is fully saturated (e.g. is a complete set of
dummies, one for every possible value of $X_{it}$), the semiparametric
bounds will equal the nonparametric ones.

In the binary-regressor case the width of the overall bound on the ATE is
given by 
\begin{equation}
\Delta _{u}-\Delta _{\ell }=\mathcal{\bar{P}}(0)(\Delta _{u}^{1}-\Delta
_{\ell }^{1})+\mathcal{\bar{P}}(1)(\Delta _{u}^{2}-\Delta _{\ell }^{2}).
\label{semi width}
\end{equation}%
where we assume $X^{1}=(0,...,0)^{\prime }$ and\textit{\ }$%
X^{2}=(1,...,1)^{\prime }$. The semiparametric bounds will be smaller than
the nonparametric bounds if and only if $\Delta _{u}^{1}-\Delta _{\ell }^{1}$
or $\Delta _{u}^{2}-\Delta _{\ell }^{2}$ are smaller than the nonparametric
values of $1.$ This decomposition also shows that the semiparametric
identification rate will be determined by the nonparametric rate, which
governs how fast $\mathcal{\bar{P}}(0)$ and $\mathcal{\bar{P}}(1)$ shrink,
and the rate that the conditional bounds converge. When the slope does not
vary across individuals it turns out that the conditional bounds can
converge very rapidly. The following result shows this in static and
dynamic, binary-choice logit models with binary regressors.

\bigskip

\textsc{Theorem 6: }\textit{Suppose that }$H(v)=e^{v}/(1+e^{v}),$ $\Delta
(\beta ,\alpha )=H(\beta +\alpha )-H(\alpha ),$\textit{\ and either equation
(\ref{semistat}) is satisfied with, }$X_{it}\in \{0,1\}$, \textit{and }$%
X^{1}=(0,...,0)^{\prime }$ \textit{and }$X^{2}=(1,...,1)^{\prime },$ \textit{%
or equation (\ref{semidyn}) is satisfied with }$k\in \{1,2\}$. \textit{Then
there are }$C>0$ \textit{and }$1>\varepsilon >0$\textit{\ such that}%
\begin{equation*}
\Delta _{u}^{k}-\Delta _{\ell }^{k}\leq C(1-\varepsilon )^{T},k=1,2.
\end{equation*}

\bigskip

This fast rate occurs because $T$ conditional moments of a one-to-one
transformation of $\alpha _{i}$ are identified from probabilities of various 
$Y$ values, and these moments lead to a fast approximation of the
conditional ATE. For example, $\Pr (Y_{i}=(1,...,1)^{\prime
}|X_{i}=X^{1})=E[H(\alpha _{i})^{T}|X_{i}=X^{1}]$, and other conditional
moments of $H(\alpha _{i})$ can be similarly identified. For the logit $%
H(\alpha )$, identification of these moments leads to fast approximation of $%
\Delta ^{1}=E[H(\beta ^{\ast }+\alpha _{i})-H(\alpha _{i})|X_{i}=X^{1}]$ and
hence to fast shrinkage of the conditional bound.

From equation (\ref{semi width}) we see that the semiparametric
identification rate in this example will be at least exponential, and may be
even faster, depending on the nonparametric rate. This result illustrates
how imposing a single, additive individual effect can speed up the
identification rate. We expect that this type of improvement will extend
beyond the logit model with binary regressors.

\section{Computation of Semiparametric Bounds}

In this section we discuss computation of population bounds, give examples,
and present theoretical results. A challenge for computation and for
estimation is the dimensionality of the unknown parameters and the nonlinearity
of the probabilities in those parameters. A useful feature of multinomial
panel models is that they are finite dimensional, in spite of the presence
of distributions. The following lemma shows that one only need consider
discrete distributions with $J$ unknown support points in the specification
of the likelihood and the bounds for the ATE. Let \textit{$\Upsilon $}
denote the set of possible values for the individual effect and $\mathbb{B}$
the set of parameters for $\beta .$

\bigskip

\textsc{Lemma 7: }\textit{If Assumptions 5 and 8 are satisfied and }$%
\mathcal{L}_{j}^{k}\left( \alpha ,\beta \right) $ \textit{is a measurable
function of }$\alpha $ \textit{for each }$\beta \in \mathbb{B},$ \textit{%
then for each }$\beta $\textit{\ and every CDF }$F_{k}$\textit{\ on $%
\Upsilon $ there is a discrete distribution }$F_{k}^{J}$\textit{\ with no
more than }$J$\textit{\ support points such that }$\int \mathcal{L}%
_{j}^{k}\left( \alpha ,\beta \right) dF_{k}^{J}(\alpha )=\int \mathcal{L}%
_{j}^{k}(\alpha ,\beta )dF_{k}(\alpha )$\textit{\ }$(j=1,...,J).$ \textit{%
If, in addition, }$\Delta (\alpha ,\beta )$ \textit{is bounded for each }$%
\beta $ \textit{then }$\Delta _{u}^{k}$ \textit{and }$\Delta _{\ell }^{k}$%
\textit{\ are not affected by restricting attention to }$F_{k}\in \mathcal{F}%
_{k}(\beta )$\textit{\ that are discrete with no more than }$J$ \textit{%
support points.}

\bigskip

Thus, no matter what the dimension of $\alpha $ is, the multinomial panel
model is finite dimensional, with the number of parameters given by $\dim
(\beta )+(2J-1)^{K}.$ Another implication of this result is that the
distribution of the individual effect is generally not identified in
multinomial models. For example, if the true distribution $F_{k}^{\ast }$
were continuous then Lemma 7 would imply that there is a discrete
distribution that gives exactly the same likelihood. The proof of this
result is similar to Lindsay's (1983) result that the maximum likelihood
estimator of a mixture model has a finite support. It is interesting that
the model takes a discrete mixture form, although the finite-dimensional
nature of the model is expected because the data have finite support.

Although the individual-effect distribution can be taken to be finite
dimensional, the dimension can be large, and the probabilities depend
nonlinearly on the support points for the individual effect. We overcome
this challenge by using an approximation with a fixed but large number of
support points for the individual effects. This approximation makes
approximate probabilities and the ATE linear in parameters, simplifying
computation. Honore and Tamer (2006) used a similar approach, but assumed
that the true distribution of individual effects had known support points.
We explicitly allow for approximation of unknown support points.

To describe how the approximation can be used to calculate the identified
set, let $M$ denote a number of support points for the individual effect and 
\textit{$\Upsilon _{M}\mathcal{=}$}$(\bar{\alpha}_{1M},...,\bar{\alpha}%
_{MM})^{\prime }$ be a grid of fixed values for the individual effect. Also
let $\pi =(\pi ^{1\prime },...,\pi ^{K\prime })^{\prime }$ denote a $%
MK\times 1$ vector of possible probabilities, with each $\pi ^{k}$ an
element of the $M$ dimensional unit simplex $\mathcal{S}_{M}$. Approximate
model probabilities are 
\begin{equation*}
P_{j}^{k}(\beta ,\pi ,M)=\sum_{m=1}^{M}\pi _{m}^{k}\mathcal{L}_{j}^{k}\left( 
\bar{\alpha}_{mM},\beta \right) \text{.}
\end{equation*}%
Consider the function%
\begin{equation*}
T_{\lambda }(\beta ,\pi ,M)=\sum_{j,k}w_{j}^{k}\left[ \mathcal{P}%
_{j}^{k}-P_{j}^{k}(\beta ,\pi ,M)\right] ^{2}+\lambda _{M}\pi ^{\prime }\pi ,
\end{equation*}%
where $w_{j}^{k}$ are positive weights, such as the chi-square ones $%
\mathcal{P}^{k}/\mathcal{P}_{j}^{k},$ for $\mathcal{P}^{k}=\Pr (X_{i}\in 
\mathcal{X}^{k})$, and $\lambda _{M}>0$ is a penalty multiplier that
controls the impact of the penalty term $\lambda _{M}\pi ^{\prime }\pi $.
This term is present to help regularize the objective function and ensures a
nonsingular Hessian matrix. Let $\tilde{T}_{\lambda }(\beta ,M)=\min_{\pi
\in \mathcal{S}_{M}^{K}}T_{\lambda }(\beta ,\pi ,M)$ and let $\epsilon
_{M}>0 $ be a positive scalar. We approximate the identified set for $\beta $
by 
\begin{equation*}
B(M)=\{\beta :\tilde{T}_{\lambda }(\beta ,M)\leq \epsilon _{M}\},\epsilon
_{M}>0.
\end{equation*}%
The use of $\epsilon _{M}$ here in allowing a range of values of the
objective function is analogous to Manski and Tamer's (2002) estimation
method. A positive $\epsilon _{M}$ ensures that the set sequence $%
(B(M))_{M=1}^{\infty }$ is lower hemi-continuous and that $B(M)$ need not be
smaller than the identified set, even though the individual effect
distributions are restricted by fixing their support points for each $M.$

We calculate the identified set by letting $M$ grow and $\lambda _{M}$ and $%
\epsilon _{M}$ shrink until there is little change in $B(M)$. Calculation of 
$\tilde{T}_{\lambda }(\beta ,M)$ is straightforward because it is the
minimum of a quadratic function. In practice we have found that $B(M)$
changes little as $M$ increases even when $M$ is quite small. As $M$ grows
and $\epsilon _{M}$ shrinks the set $B(M)$ will converge to the identified
set under conditions given below.

For the ATE bounds, note 
\begin{equation*}
D^{k}(M)=\{\sum_{m=1}^{M}\pi _{m}^{k}\Delta (\bar{\alpha}_{mM},\beta
):T_{\lambda }(\beta ,\pi ,M)\leq \epsilon _{M}\}
\end{equation*}%
is the set of possible conditional ATE (given $X \in \mathcal{X}^{k})$ that are consistent
with $\tilde{T}_{\lambda }(\beta ,M)\leq \epsilon _{M}$. Approximate lower
and upper bounds are%
\begin{equation*}
\Delta _{\ell }^{k}(M)=\min D^{k}(M),\Delta _{u}^{k}(M)=\max D^{k}(M).
\end{equation*}%
As $M$ grows and $\epsilon _{M}$ shrinks these bounds will converge to $%
\Delta _{\ell }^{k}$ and $\Delta _{u}^{k}$ respectively, under conditions
given below.\ 

Computation of these ATE bounds is challenging because it requires searching
over a large dimensional set of possible $\pi $. In practice we start with a
smaller set of probabilities and then try others. Specifically, let $\tilde{%
\pi}(\beta )\in \arg \min_{\pi \in \mathcal{S}_{M}^{K}}T_{\lambda }(\beta
,\pi ,M),$ $\tilde{S}^{k}(\beta )=\{\pi ^{k}:P_{j}^{k}(\beta ,\pi ,M)=$ $%
P_{j}^{k}(\beta ,\tilde{\pi}(\beta ),M),$ $j=1,...,J\},$ and 
\begin{equation*}
\tilde{\Delta}_{\ell }^{k}(M)=\min_{\beta \in B(M),\pi ^{k}\in \tilde{S}%
^{k}(\beta )}\sum_{m=1}^{M}\pi _{m}^{k}\Delta (\bar{\alpha}_{mM},\beta ),%
\text{ }\tilde{\Delta}_{u}^{k}(M)=\max_{\beta \in B(M),\pi ^{k}\in \tilde{S}%
^{k}(\beta )}\sum_{m=1}^{M}\pi _{m}^{k}\Delta (\bar{\alpha}_{mM},\beta ).
\end{equation*}%
For each $\beta $ these bounds are easy to calculate by linear programming.
We have done so and then checked to see if other values $\pi $ violate these
bounds. We have not found this to be so for values of $M$ that we use to
compute $\beta $. We conjecture that these bounds also converge to the
population bounds as $M\longrightarrow \infty $ although we have not yet
been able to prove this (because we have not been able to show that the ATE
bounds are continuous in the true probabilities).

We carry out some numerical calculations for the probit model where 
\begin{equation*}
Y_{it}=1(\beta ^{\ast }X_{it}+\alpha _{i}\geq \varepsilon _{it}),\varepsilon
_{it}\sim N(0,1),X_{it}=1(\alpha _{i}\geq \eta _{it}),\eta _{it}\sim
N(0,1),\alpha _{i}\sim N(0,1).
\end{equation*}%
We consider different DGPs indexed by $\beta ^{\ast }\in \lbrack -2,2]$ and $%
T\in \{2,3\}$. Figures 2 and 3 show nonparametric bounds for ATEs and
semiparametric bounds for $\beta ^{\ast }$ and ATEs for $T=2$ and $T=3$,
respectively. The semiparametric bounds are obtained using the computational
algorithm described above with $M=100$ and $\lambda _{M}=1.3\times 10^{-8}$.
The elements of the fixed grid $\Upsilon _{M}$ are located at the
percentiles of the standard normal distribution. We find that $\beta ^{\ast }
$ is not identified for $T=2$, extending the result of Chamberlain (2010) to
this example without time dummy. This result also holds for $T=3$, although
it is difficult to appreciate in the figure because the identified set $B$
is very small. The nonparametric bounds for the ATEs (NP-bounds) can be very
wide, even when we impose monotonicity (NPM-bounds) as described in the
Supplementary Material. The semiparametric bounds for the ATEs (SP-bounds)
are tighter than the nonparametric bounds and shrink very fast with $T$. In
the Supplementary Material we report similar results for the logit,
including nonidentification of the ATEs, except that $\beta ^{\ast }$ is
identified, as is well known. Honore and Tamer (2006) also found tight
bounds for the coefficient of a dynamic model.

To show that the approximate sets converge to the identified set as $M$
grows we impose some conditions. Let $d(\alpha ,\tilde{\alpha})$ denote a
metric on the set \textit{$\Upsilon $ }of possible values for $\alpha $.

\bigskip

\textsc{Assumption 9:} \textit{(i) $\Upsilon $ is a compact metric space
with metric }$d(\alpha ,\tilde{\alpha})$\textit{; ii) }$\eta
(M)=\sup_{\alpha \in \Upsilon }\min_{\tilde{\alpha}\in \Upsilon
_{M}}d(\alpha ,\tilde{\alpha})$\textit{\ }$\longrightarrow 0$ as $%
M\longrightarrow \infty ;$ \textit{(iii) }$\mathbb{B}$ \textit{is a compact
subset of }$\Re ^{b}$\textit{; (iv) there is }$C$ \textit{such that for all }%
$(\alpha ,\beta ),(\tilde{\alpha},\tilde{\beta})\in $\textit{$\Upsilon $}$%
\times \mathbb{B}$, $\left\vert \mathcal{L}_{j}^{k}\left( \tilde{\alpha},%
\tilde{\beta}\right) -\mathcal{L}_{j}^{k}\left( \alpha ,\beta \right)
\right\vert \leq C[d(\tilde{\alpha},\alpha )+\left\Vert \tilde{\beta}-\beta
\right\Vert ];$ \textit{and v) }$\Delta (\alpha ,\beta )$ \textit{is
continuous on $\Upsilon $}$\times \mathbb{B}$\textit{.}

\bigskip

Although condition (i) seems restrictive, unbounded individual effects may
be allowed if $\Upsilon $\textit{\ }is chosen appropriately. For example, in
the binary-choice model of equation (\ref{semistat}) this condition will be
satisfied if \textit{$\Upsilon $} is taken to be a two-point
compactification of the real line and $d(\alpha ,\tilde{\alpha})$ is
specified appropriately, as shown in the following result.

\bigskip

\textsc{Lemma 8: }\textit{If Assumptions 5 and 8 and equation (\ref{semistat}%
) are satisfied, where }$H(v)$\textit{\ is strictly monotonic on }$\Re $ 
\textit{with bounded continuous derivative, and }$\mathbb{B}$\textit{\ is a
compact subset of }$\Re ^{b},$\textit{\ then there is a metric }$d(\alpha ,%
\tilde{\alpha})$\textit{\ and for each }$M$\textit{\ there is }$\Upsilon
_{M}=\{\bar{\alpha}_{1M},...,\bar{\alpha}_{MM}\}$\textit{\ such that
Assumption 9 is satisfied with }$\eta (M)=1/(M-1).$\textit{\ }

\bigskip

For the convergence results for the identified set we use the Hausdorff set
metric,%
\begin{equation*}
d_{H}(A,B)=\max \{\sup_{a\in A}\inf_{b\in B}d(a,b),\sup_{b\in B}\inf_{a\in
A}d(a,b)\}.
\end{equation*}%
\textit{\ }

\bigskip

\textsc{Theorem 9: } \textit{If Assumptions 5, 8, and 9 are satisfied, }$%
\epsilon _{M}\longrightarrow 0$, \textit{and} $\left( \eta (M)+\lambda
_{M}\right) /\epsilon _{M}\longrightarrow 0$\textit{\ then as }$%
M\longrightarrow \infty ,$%
\begin{equation*}
d_{H}(B(M),B)\longrightarrow 0,\Delta _{\ell }^{k}(M)\longrightarrow \Delta
_{\ell }^{k},\Delta _{u}^{k}(M)\longrightarrow \Delta _{u}^{k}.
\end{equation*}

\section{Estimation and Inference}

Under Assumptions 5 and 8 the complete description of the data-generating
process is provided by the parameter vector $(${$P_{X}^{\prime }$}$,${$P$}$%
^{\prime })^{\prime },$ where $P_{X}=(P^{k},k=1,...,K)^{\prime }$ and $%
P=(P_{j}^{k},j=1,...,J,k=1,...,K)^{\prime }.$ The true value of the
parameter vector is $\Pi =({\mathcal{P}}_{X}^{\prime },\mathcal{P}^{\prime
})^{\prime }$, where $\mathcal{P}_{X}=(\mathcal{P}^{k},k=1,...,K)^{\prime }$
and $\mathcal{P}=(\mathcal{P}_{j}^{k},j=1,...,J,k=1,...,K)^{\prime },$ and
the empirical estimate is $\hat{\Pi}=({\hat{P}}_{X}^{\prime },\hat{P}%
^{\prime })^{\prime }$, where $\hat{P}_{X}=(\hat{P}^{k},k=1,...,K)^{\prime }$
and $\hat{P}=(\hat{P}_{j}^{k},j=1,...,J,k=1,...,K)^{\prime }.$

The estimation method is like the computational one in using
linear-in-parameters approximations to the probabilities. Here we describe
the estimation method and give a consistency result, and in the
Supplementary Material we provide the implementation details. We follow
 the same steps as the computational one except that we use
estimated weights $\hat{w}_{j}^{k}$ and estimated probabilities $\hat{P}%
_{j}^{k}$. Let $\hat{M}$ be a choice of $M$ that may depend on the data and
sample size, and 
\begin{equation*}
\hat{T}_{\lambda }(\beta ,\pi )=\sum_{j,k}\hat{w}_{j}^{k}\left[ \hat{P}%
_{j}^{k}-P_{j}^{k}(\beta ,\pi ,\hat{M})\right] ^{2}+\lambda _{n}\pi ^{\prime
}\pi .
\end{equation*}%
Let $\hat{T}_{\lambda }(\beta )=\min_{\pi \in \mathcal{S}_{M}^{K}}\hat{T}%
_{\lambda }(\beta ,\pi )$ and $\epsilon _{n}$ $>0$ be a positive scalar. We
estimate the identified set for $\beta $ by 
\begin{equation*}
\hat{B}=\{\beta \in \mathbb{B}:\hat{T}_{\lambda }(\beta )\leq \epsilon
_{n}\},
\end{equation*}%
where $\mathbb{B}$ is the parameter space and $\epsilon _{n}$ is a cut-off
parameter that shrinks to zero with the sample size, as in Manski and Tamer
(2002) and Chernozhukov, Hong, and Tamer (2007). The ATE bounds can be
estimated by 
\begin{equation*}
\hat{\Delta}_{\ell }^{k}=\min \hat{D}^{k},\hat{\Delta}_{u}^{k}=\max \hat{D}%
^{k},\hat{D}^{k}=\{\sum_{m=1}^{M}\pi _{m}^{k}\Delta (\bar{\alpha}_{mM},\beta
):\hat{T}_{\lambda }(\beta ,\pi )\leq \epsilon _{n}\}.
\end{equation*}

This approach to estimation (and computation) can be easily modified to
handle the case where the distribution of the individual effect is
restricted to be the same across some values of $k$. Such a modification
could be implemented by imposing equality of $\pi _{m}^{k}$ across those
values of $k.$ An example would be a model where the distribution of $\alpha
_{i}$ did not depend on some component of $X_{it}.$ That restriction could
be imposed setting $\pi _{m}^{k}$ to be equal across $k$ where the other
components of $X_{it}$ do not vary. Or in a case with a lagged dependent
variable we could restrict the distribution of $\alpha $ to only depend on
the initial condition by imposing equality of $\pi _{m}^{k}$ across all $k$
where $Y_{i0}$ takes on a particular value.

The following is a consistency result.

\bigskip

\textsc{Theorem 10: } \textit{If Assumptions 5, 8, and 9 are satisfied, }$%
\hat{w}_{j}^{k}\overset{p}{\longrightarrow }w_{j}^{k}>0,$ $\hat{P}_{j}^{k}%
\overset{p}{\longrightarrow }\mathcal{P}_{j}^{k}$, $\epsilon
_{n}\longrightarrow 0$, \textit{and }$\left( n^{-1}+\eta (\hat{M})+\lambda
_{n}\right) /\epsilon _{n}\overset{p}{\longrightarrow }0$,\textit{\ then} $%
d_{H}(\hat{B},B)\overset{p}{\longrightarrow }0,\hat{\Delta}_{\ell }^{k}%
\overset{p}{\longrightarrow }\Delta _{\ell }^{k},\hat{\Delta}_{u}^{k}\overset%
{p}{\longrightarrow }\Delta _{u}^{k}.$

\bigskip

It is interesting to note that no upper limit is placed on $M$ in this
result or in Theorem 9. The reason for this is that the model is finite
dimensional, so there is no need for such a limit. Mathematically, a richer,
fixed grid simply corresponds to a bigger submodel of the finite-dimensional
model.

Turning now to the inference for the semiparametric models, we note that it
is rather challenging. The estimators of parameters and ATE are obtained by
nonlinear programming subject to data-dependent constraints that are
modified to respect the constraints of the model. The distributions of these
highly-complex estimators are not tractable, and are also non-regular in the
sense that the limit versions of these distributions do not vary with
perturbations of the DGP in a continuous\ fashion. This implies that the
usual bootstrap is not consistent. To overcome all of these difficulties we
will rely on a variation of the bootstrap, which we call the perturbed
bootstrap. We also give an alternative inference method based on a modified
projection in the Supplementary Material.

The usual bootstrap computes the critical value -- the $\alpha $-quantile of
the distribution of a test statistic -- given a consistently-estimated
data-generating process (DGP). If this critical value is not a continuous
function of the DGP, the usual bootstrap fails to consistently estimate the
critical value. We instead consider the perturbed bootstrap, where we
compute a set of critical values generated by suitable perturbations of the
estimated DGP and then take the most conservative critical value in the set.
If the perturbations cover at least one DGP that gives a more conservative
critical value than the true DGP does, then this approach yields a valid
inference procedure.

The approach outlined above is most closely related to the Monte-Carlo
inference approach of Dufour (2006); see also Romano and Wolf (2000) for a
finite-sample inference procedure for the mean that has a similar spirit. In
the set-identified context, this approach was first applied in the MIT
thesis work of Rytchkov (2007); see also Chernozhukov (2007).

We consider the problem of performing inference on a real parameter $\theta
^{\ast }$. For example, $\theta ^{\ast }$ can be an upper (or lower) bound
on the conditional ATE $\Delta ^{k}$ such as 
\begin{equation*}
\theta ^{\ast }(P)=\max_{\beta \in B^{\ast }(P),F_{k}\in \mathcal{F}%
_{k}(\beta ,P^{\ast }(P))}\int \Delta (\alpha ,\beta )dF_{k}\left( \alpha
\right) ,\ 
\end{equation*}%
where $P^{\ast }$ denotes the projection of $P$ onto the model space $\Xi
=\{P:\exists \beta \in \mathbb{B}$ with $\mathcal{F}_{k}(\beta ,P)\neq
\varnothing ,\forall k=1,...,K\}$, i.e. 
\begin{equation*}
P^{\ast }(P)=\arg \min_{\tilde{P}\in \Xi }W({\tilde{P}},P),\ \ W({\tilde{P}}%
,P)=n\sum_{j,k}\hat{P}^{k}\frac{(P_{j}^{k}-{\tilde{P}}_{j}^{k})^{2}}{\tilde{P%
}_{j}^{k}},
\end{equation*}%
and $B^{\ast }(P)$ is the corresponding projection for the identified set of
the parameter, i.e. 
\begin{equation*}
B^{\ast }(P)=\left\{ \beta \in \mathbb{B}:\exists \tilde{P}\in P^{\ast }(P)%
\text{ with }\mathcal{F}_{k}(\beta ,\tilde{P})\neq \varnothing
,k=1,...,K\right\} .
\end{equation*}%
Alternatively, $\theta ^{\ast }$ can be an upper (or lower) bound on a
scalar functional $c^{\prime }\beta ^{\ast }$ of the parameter $\beta ^{\ast
}$. Then we define 
\begin{equation*}
\theta ^{\ast }(P)=\max_{\beta \in B^{\ast }(P)}c^{\prime }\beta .
\end{equation*}%
In both cases we project $P$ onto the model space in order to address the
problem of infeasibility of constraints defining the parameters of interest
under misspecification or sampling error. Under misspecification, we
interpret our inference as targeting the parameters of interest in a best
approximating model; see the Supplementary Material on the modified
projection method for further details. Under correct specification, our inference targets
the parameters of interest in the true model.

In order to perform inference on the true value $\theta ^{\ast }=\theta
^{\ast }(\mathcal{P})$ of the parameter, we use the statistic 
\begin{equation*}
S_{n}=\hat{\theta}-\theta ^{\ast },
\end{equation*}%
where $\hat{\theta}=\theta ^{\ast }(\hat{P})$. Let $G_{n}(s,P)$ denote the
distribution function of $S_{n}(P)=\hat{\theta}-\theta ^{\ast }(P)$, when
the data follow the DGP $P$. The goal is to estimate the distribution of the
statistic $S_{n}$ under the true DGP $P=\mathcal{P}$, that is, to estimate $%
G_{n}(s,\mathcal{P})$.

The method proceeds by constructing a confidence region $CR_{1-\gamma }(%
\mathcal{P})$ that contains the true DGP $\mathcal{P}$ with probability $%
1-\gamma $, close to one. For efficiency purposes, we also want the
confidence region to be an efficient estimator of $\mathcal{P}$, in the
sense that as $n\rightarrow \infty $, $d_{H}(CR_{1-\gamma }(\mathcal{P}),%
\mathcal{P})=O_{p}(n^{-1/2}),$ where $d_{H}$ is the Hausdorff distance
between sets. Specifically, in our case we use 
\begin{equation}
CR_{1-\gamma }(\mathcal{P})=\{P\in S_{J}^{K}:W(P,\hat{P})\leq c_{1-\gamma
}(\chi _{K(J-1)}^{2})\},  \label{coverage}
\end{equation}%
where $c_{1-\gamma }(\chi _{K(J-1)}^{2})$ is the $(1-\gamma )$-quantile of
the $\chi _{K(J-1)}^{2}$ distribution and $W$ is the goodness-of-fit
statistic: 
\begin{equation*}
W(P,\hat{P})=n\sum_{j,k}\hat{P}^{k}\frac{\left( \hat{P}_{j}^{k}-P_{j}^{k}%
\right) ^{2}}{P_{j}^{k}}.
\end{equation*}%
Then we define the estimates of the lower and upper bounds on the quantiles of $%
G_{n}(s,\mathcal{P})$ as 
\begin{equation}
\underline{G}_{n}^{-1}(\alpha ,\mathcal{P})/\overline{G}_{n}^{-1}(\alpha ,%
\mathcal{P})=\inf /\sup_{P\in CR_{1-\gamma }(\mathcal{P})}G_{n}^{-1}(\alpha
,P),
\end{equation}%
where $G_{n}^{-1}(\alpha ,P)=\inf \{s:G_{n}(s,P)\geq \alpha \}$ is the $%
\alpha $-quantile of the distribution function $G_{n}(s,P)$. Then we
construct a $(1-\alpha -\gamma )\cdot 100\%$ confidence region for the
parameter of interest as 
\begin{equation*}
CR_{1-\alpha -\gamma }(\theta ^{\ast })=\left[ \underline{\theta },\overline{%
\theta }\right]
\end{equation*}%
where, for $\alpha =\alpha _{1}+\alpha _{2}$, 
\begin{equation*}
\underline{\theta }=\hat{\theta}-\overline{G}_{n}^{-1}(1-\alpha _{1},%
\mathcal{P}),\ \overline{\theta }=\hat{\theta}-\underline{G}_{n}^{-1}(\alpha
_{2},\mathcal{P}).
\end{equation*}%
This formulation allows for both one-sided intervals (either $\alpha _{1}=0$
or $\alpha _{2}=0$) or two-sided intervals ($\alpha _{1}=\alpha _{2}=\alpha
/2)$.

For the inference results we condition on the observed distribution of $X$
and thus set $P_{X}=\mathcal{P}_{X}=\hat{P}_{X}.$ We make the following
assumption about the data-generating process.

\bigskip

\textsc{Assumption 10: }\ $\Pi \in $\thinspace $\mathbb{P=}%
\{(P_{X},P):P^{k}>\varepsilon ,P_{j}^{k}>\varepsilon ;j=1,...,J,k=1,...,K\}$
for some $\varepsilon >0$.

\bigskip

The following theorem shows that this method delivers (uniformly) valid
inference on the parameter of interest.

\bigskip

\textsc{Theorem 11:} \textit{If Assumptions 5, 8, and 9 are satisfied then
for any sequence of data-generating process }$\Pi = \Pi_n$ \textit{satisfying Assumption 10},%
\begin{equation*}
\lim_{n\rightarrow \infty }\text{Pr}_{\Pi }(\theta ^{\ast }\in \left[
\underline{\theta },\overline{\theta }\right] )\geq 1-\alpha -\gamma .
\end{equation*}

\bigskip

In practice, we use the following simulation approach to compute the
confidence intervals.

\bigskip

\textsc{Algorithm: Perturbed Bootstrap}
\begin{enumerate}
\item \textit{Draw a potential DGP $P_{r}=(P_{r1}^{\prime
},...,P_{rK}^{\prime }),$ where $P_{rk}\sim \mathcal{M}(n\hat{P}^{k},(\hat{P}%
_{1}^{k},...,\hat{P}_{J}^{k}))/(n\hat{P}^{k})$ and $\mathcal{M}$ denotes the
multinomial distribution. }

\item \textit{Keep $P_{r}$ if it passes the chi-square goodness-of-fit test
at the $\gamma $ level in equation (\ref{coverage}), using $K(J-1)$ degrees
of freedom, and proceed to the next step. Otherwise reject, and repeat step
1. }

\item \textit{Estimate the distribution $G_{n}(s,P_{r})$ of $S_{n}(P_{r})$
by simulation under the DGP $P_{r}$. }

\item \textit{Repeat steps 1 to 3 for $r=1,...,R$, obtaining $%
\{G_{n}(s,P_{r})$, $r=1,...,R\}.$}

\item \textit{Let $\hat{\underline{G}}_{n}^{-1}(\alpha ,\mathcal{P})/\hat{%
\overline{G}}_{n}^{-1}(\alpha ,\mathcal{P})=\min /\max \{G_{n}^{-1}(\alpha
,P_{1}),...,G_{n}^{-1}(\alpha ,P_{R})\},$ and construct a $1-\alpha -\gamma $
confidence region for the parameter of interest as $CR_{1-\alpha -\gamma
}(\theta ^{\ast })=\left[ \underline{\theta },\overline{\theta }\right] $,
where $\underline{\theta }=\hat{\theta}-\hat{\overline{G}}_{n}^{-1}(1-\alpha
_{1},\mathcal{P}),$ $\overline{\theta }=\hat{\theta}-\hat{\underline{G}}%
_{n}^{-1}(\alpha _{2},\mathcal{P})$, and $\alpha _{1}+\alpha _{2}=\alpha .$ }
\end{enumerate}

\bigskip

\section{Empirical Examples}

We illustrate the estimation and inference results with two empirical
examples. One estimates identified effects and calculates bounds for the
effect of unions on earnings quantiles. The other compares nonparametric and
semiparametric bounds for the effect of fertility on women's labor force
participation.

\subsection{Union Premium}

We revisit the empirical question of how unions impact wage structure using
panel data. Our major contribution here is to estimate the effect without
imposing the assumption that unobserved heterogeneity is some additive term
that can be simply differenced out. In our model unobserved heterogeneity
can have an almost unrestricted impact on the structural/causal response
functions, with the time homogeneity serving as the only restriction.

Our analysis is motivated by previous empirical studies that find
differences in unobservables between union and nonunion workers. For
instance, in an influential study, Chamberlain (1982) finds strong evidence
of heterogeneity bias in the estimation of the union effect by comparing
estimates of cross-sectional models and panel data models with additive
heterogeneity. This finding demonstrates the important need of controlling
for unobserved heterogeneity. Also, Angrist and Newey (1991) reject the
hypothesis that the unobserved heterogeneity acts solely in an additive
fashion, motivating the need to control for more general unobserved
heterogeneity. Card (1996) found differences in the union and selection
effect across skill levels. Here we account fully for differences across
individuals in the union effect while allowing correlation of that effect
with union status, thus accounting for selection. Recently Frandsen (2011)
focused on quantile union effects using a regression-discontinuity design
that estimates union effects for those near a union election discontinuity
rather than for those whose union status changes. We find a flatter quantile
profile than he does, consistent with his theoretical results that suggest a
flatter profile away from the discontinuity.

We use data from the National Longitudinal Survey (Youth Sample). The sample
consists of full-time, young, working males, 20 to 29 years old in 1986,
followed over the period 1986 to 1993. We exclude individuals who failed to
provide sufficient information for each year, were in the active armed
forces or were students any year, or who reported too high (more than \$500
per hour) or too low (less than \$1 per hour) wages. The final sample
includes 2,065 men followed over 8 years. We use the union membership and
the log-hourly wage rate in 1980 dollars as the covariate and the outcome
variables. The union membership variable reflects whether or not the
individual had his wage set by a collective bargaining agreement. Vella and
Verbeek (1998) also used data from the NLSY for different years and found
evidence of important union effect heterogeneity with a random effects model.

We begin by imposing the stationarity condition that income with and without
union membership has the same distribution in each time period but also will
allow for location and scale time effects. It turns out that time effects
are not important in this data. Some covariates are also allowed for since
time-invariant covariates are absorbed in the individual effects.
Insensitivity to time effects also suggests that time-varying covariates may
not be important though a fuller exploration would be useful. For brevity we
focus on the case without covariates.

In our analysis, we focus on estimating the union quantile effect for the
subpopulations of workers that ever became unionized within the sample (47\%
of the sample) or that were unionized in the first year (20\% of the
sample). For these subpopulations, the union effect is not point-identified,
since there are 13\% of the ever-unionized workers that always stayed
unionized between 1986 and 1993, and there are 32\% of the workers unionized
in 1986 that remained unionized until 1993. However, we hope to construct
informative bounds on the union effect. We consider both a static model that
allows for the union membership decisions to be strictly exogenous with
respect to wage-setting decisions, and a dynamic model that allows for the
union-membership decisions to be only predetermined with respect to
wage-setting decisions. We shall also report the estimates of the union
effect for the subpopulation of workers who change their union status at
least once within the sample. For this subpopulation, the effect is
point-identified in the static model, that is, the bounds on the union
effect collapse to a point. We shall not estimate the union effect for the
entire population of workers, since the bounds are completely uninformative
in this case. This happens because more than half of the workers are never
unionized within the sample (see Table 1).

All the results are reported in Table 1 and Figure 4. Table 1 assesses the
plausibility of the time-homogeneity assumption by comparing moments and
quantiles of the cross-sectional distributions of log-wages across years for
workers that do not change union status. Under time homogeneity, these cross
sectional distributions should remain time invariant in the static model. In
the table we observe distributional changes across years, but most of the
variation can be captured by additive location effects for both
always-unionized and never-unionized workers.

Panels A and B of fig. 4 present the estimates of the union effect in the
static model for the subpopulation of workers who change their union status
at least once within the sample. In panel A we compare our panel data
estimates of quantile effects that control for individual heterogeneity with
pooled estimates that do not control for individual heterogeneity. In the
pooled estimates, we see that the quantile effect of union membership is
positive but declines sharply at the upper end of the distribution, which
agrees with previous cross-sectional findings (Chamberlain, 1994). A common
explanation for this phenomenon is that the high-skill workers at the lower
end of the earning distribution tend to join the union, whereas the
high-skill workers at the high end of the earning distribution tend not to
join the union. The estimated quantile effect in the cross-section therefore
captures this selection effect of unobserved skills. In the panel-data
estimates, which control for unobserved skills, we see that the quantile
effects of union membership become very flat across the quantile indices.
Thus, by controlling for individual heterogeneity, we have eliminated the
selection effect. Panel B shows that the results are not sensitive to the
inclusion of location and scale effects.

Panel C presents estimated bounds on the union effect for the subpopulation
of workers that ever became unionized within the sample using the static
model with time effects. The bounds are informative, and show that the
effect is positive for most of the quantile indices. The panel also shows
bounds obtained using the assumption of monotonic and positive union effect
on earnings described in the Supplementary Material. These bounds are also
informative, and in fact are substantially tighter than the bounds obtained
without the monotonicity assumption. Panel D presents similar bounds on the
union effect for the subpopulation of workers unionized in the first period
using the dynamic model. The bounds in this case are not informative, even
after imposing monotonicity.

All the panels include 90\% uniform confidence bands for the quantile union
effects constructed by bootstrap with 200 repetitions. These bands allow us
to make visual simultaneous inference on the entire quantile functions. For
example, we cannot reject that the identified union effect is constant and
positive for all the quantiles. For the ever unionized, the quantile union
effect is positive for a large range of quantiles.

\subsection{Female Labor Force Participation}

For an application of the semiparametric bounds we consider a binary choice
panel model of female labor force participation. We focus on the
relationship between participation and the presence of young children in the
household. Other studies that estimate similar models of participation in
panel data include Heckman and MaCurdy (1980, 1982), Chamberlain (1984),
Hyslop (1999), Chay and Hyslop (2000), Carrasco (2001), Carro (2007), and
Fern\'{a}ndez-Val (2009).

The empirical analysis is based on a sample of married women from the
National Longitudinal Survey of Youth 1979 (NLSY79). The sample consists of
1,587 married women. Only women continuously married, not students or in the
active forces, and with complete information on the relevant variables in
the entire sample period are selected from the survey. Descriptive
statistics for the sample are shown in Table 2. The labor force
participation variable ($LFP$) is an indicator that takes the value one if
the woman's employment status is \textquotedblleft in the labor
force\textquotedblright\ according to the CPS definition, and zero
otherwise. The fertility variable ($kids$) indicates whether the woman has
any children younger than 3 years. We focus on very young, preschool
children as most empirical studies find that their presences have the
strongest impact on the mother's participation decision. $LFP$ is stable
across the years considered, whereas $kids$ is decreasing. The proportion of
women that change fertility status grows steadily with the number of time
periods of the panel, but there are still $49\%$ of the women in the sample
for which the effect of fertility is not identified after 3 periods.

The empirical specification we use is similar to Chamberlain (1984). In
particular, we estimate the following equation 
\begin{equation*}
LFP_{it}=\mathbf{1}\left\{ \beta ^{\ast }\cdot kids_{it}+\alpha
_{i} \geq \epsilon _{it} \right\} ,
\end{equation*}%
where $\alpha _{i}$ is an individual-specific effect. The parameters of
interest are $\beta ^{\ast }$ and the ATE of fertility on participation. We
compute nonparametric and semiparametric probit and logit bounds for these
parameters. We also obtain linear and nonlinear fixed effects estimates,
together with large-$T$ analytical bias corrected estimates and conditional
fixed effects logit estimates.\footnote{%
The analytical corrections use the estimators of the bias based on expected
quantities in Fern\'{a}ndez-Val (2009).} The nonparametric bounds impose
monotonicity on the effects. For the semiparametric bounds, we use the
method described in Section 9 with penalty $\lambda _{n}=1/(n\log n)$ and
iterate the quadratic program 3 times with initial weights $\hat{w}_{j}^{k}=%
\hat{P}^{k}$. This iteration makes the estimates insensitive to the penalty
and weighting. We search over discrete distributions with $\hat{M}=23$
support points at $\{-\infty ,-4,-3.6,...,3.6,4,\infty \}$ for the parameter 
$\beta ^{\ast }$, and with $\hat{M}=163$ support points at $\{-\infty
,-8,-7.9,...,7.9,8,\infty \}$ for the ATE. The estimates are based on panels
of 2 and 3 time periods, both of them starting in 1990.

Table 3 reports estimates and 95\% confidence regions for the parameters of
interest. The confidence regions for the nonparametric bounds are
constructed using the normal approximation $(95\%\ N)$ and nonparametric
bootstrap with 200 repetitions $(95\%\ B)$. The confidence regions for the
semiparametric bounds are obtained using the procedures described in Section
9 and the Supplementary Material. For the perturbed bootstrap method $%
(95\%\ PB)$ we use $R=100$, $\gamma =.01$, $\alpha _{1}=\alpha _{2}=.02,$
and 200 simulations from each DGP to approximate the distribution of the
statistic. For the modified projection method $(95\%\ MP)$, the confidence
interval for $\mathcal{P}$ in the first stage is approximated by 5,000 DGPs
drawn from the empirical multinomial distributions that pass the
goodness-of-fit test. Together the modified projection and the perturbed
bootstrap took several days to compute on a personal computer. We also
include confidence intervals obtained by a canonical projection method $%
(95\%\ CP)$ less robust to model misspecification than the modified
projection method, that intersects a nonparametric confidence interval for $%
\mathcal{P}$ with the space of probabilities compatible with the
semiparametric model $\Xi $: 
\begin{equation*}
CR_{1-\alpha }(\mathcal{P})=\left\{ P\in \Xi :W(P,\hat{P})\leq c_{1-\alpha
}(\chi _{K(J-1)}^{2})\right\} .
\end{equation*}%
For the fixed-effects estimators, the confidence regions are based on the
asymptotic normal approximation. The semiparametric estimates are shown for $%
\epsilon _{n}=0$, i.e., for the solution that gives the minimum value in the
quadratic problem.

Overall, we find that the nonparametric bound estimates and confidence
regions are too wide to provide informative evidence about the relationship
between participation and fertility. The semiparametric bounds offer a good
compromise between producing more informative results without adding too
much structure to the model. Thus, these estimates are always inside the
confidence regions of the nonparametric model and do not suffer important
efficiency losses relative to the fixed-effects estimates.
Another salient feature of the results is that the misspecification problem
of the canonical projection method clearly arises in this application. Thus,
this procedure gives empty confidence regions for the panel with 3 periods.
The perturbed bootstrap and modified projection methods produce similar
(non-empty) confidence regions for the model parameters and ATEs.

The semiparametric intervals for the ATE cover the -9.6\% estimate of
Chamberlain (1984) for the expected effect of having an additional young
child on the participation probability. He obtained this estimate from a
correlated, random-coefficient probit model, a richer specification that
includes education and fertility covariates, and a different sample from the
PSID.

% FIGURE %%%%%%%%%%%%%%%%%%%%%%%%%%%%%%%%%%%%%%%%%%%%

\begin{figure*}

\begin{center}

\centering\epsfig{figure=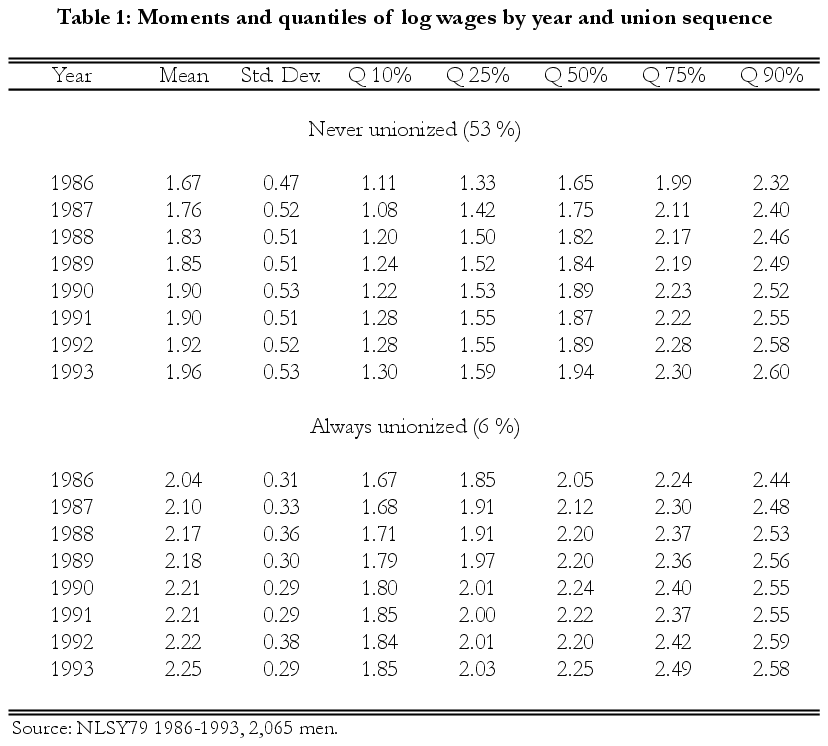,width=8in,height=10in}

\end{center}

\end{figure*}

% figure %%%%%%%%%%%%%%%%%%%%%%%%%%%%%%%%%%%%%%%%%%%%

% FIGURE %%%%%%%%%%%%%%%%%%%%%%%%%%%%%%%%%%%%%%%%%%%%

\begin{figure*}

\begin{center}

\centering\epsfig{figure=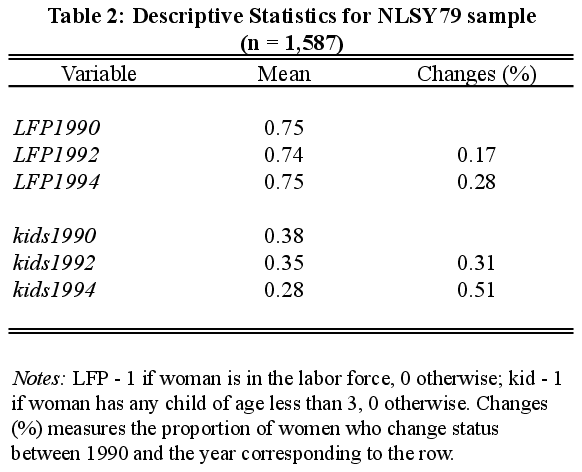,width=8in,height=10in}

\end{center}

\end{figure*}

% figure %%%%%%%%%%%%%%%%%%%%%%%%%%%%%%%%%%%%%%%%%%%%

% FIGURE %%%%%%%%%%%%%%%%%%%%%%%%%%%%%%%%%%%%%%%%%%%%

\begin{figure*}

\begin{center}

\centering\epsfig{figure=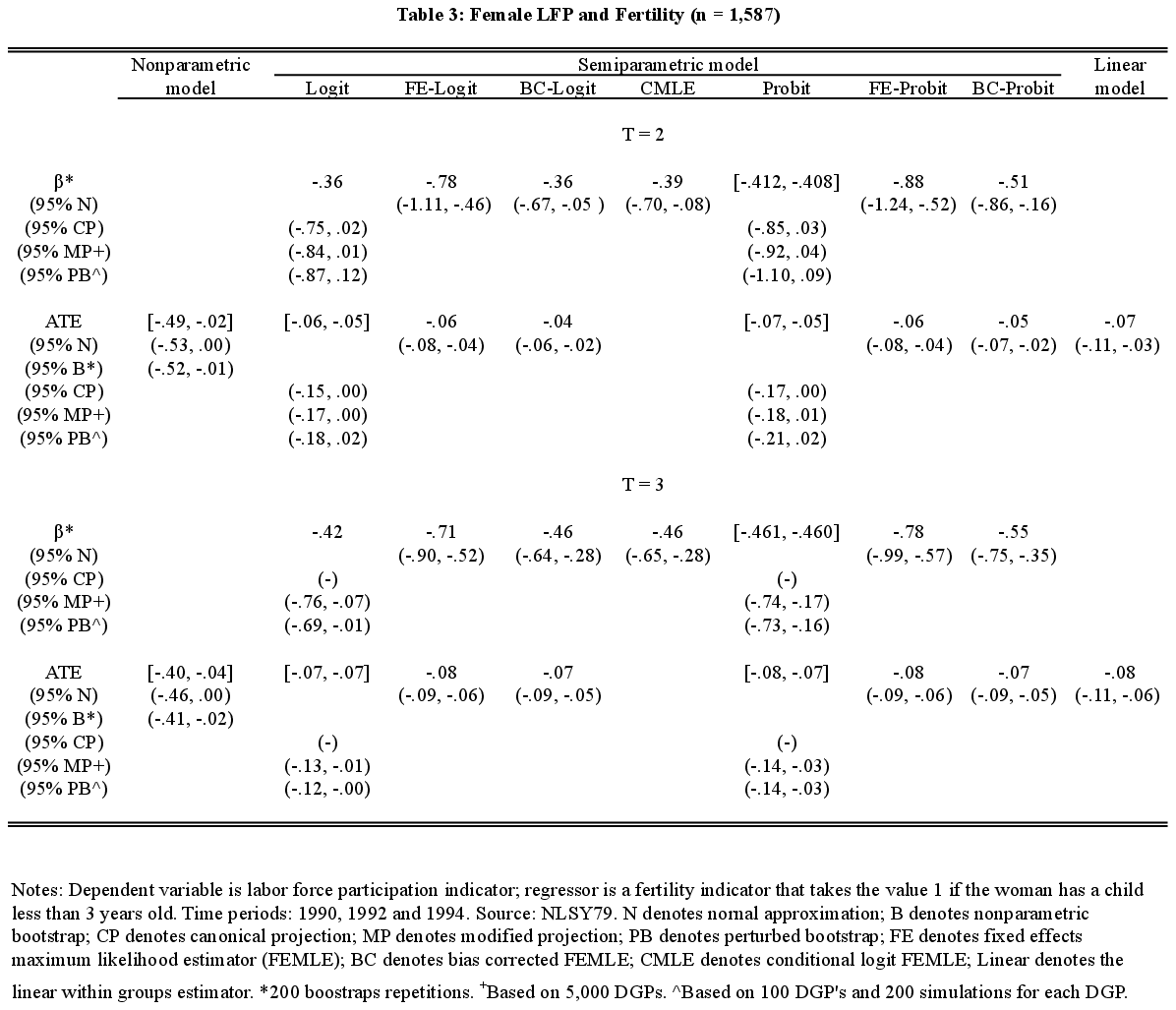,width=7.5in,height=9in}

\end{center}

\end{figure*}

% figure %%%%%%%%%%%%%%%%%%%%%%%%%%%%%%%%%%%%%%%%%%%%

% FIGURE %%%%%%%%%%%%%%%%%%%%%%%%%%%%%%%%%%%%%%%%%%%%

\begin{figure}

\begin{center}

\centering\epsfig{figure=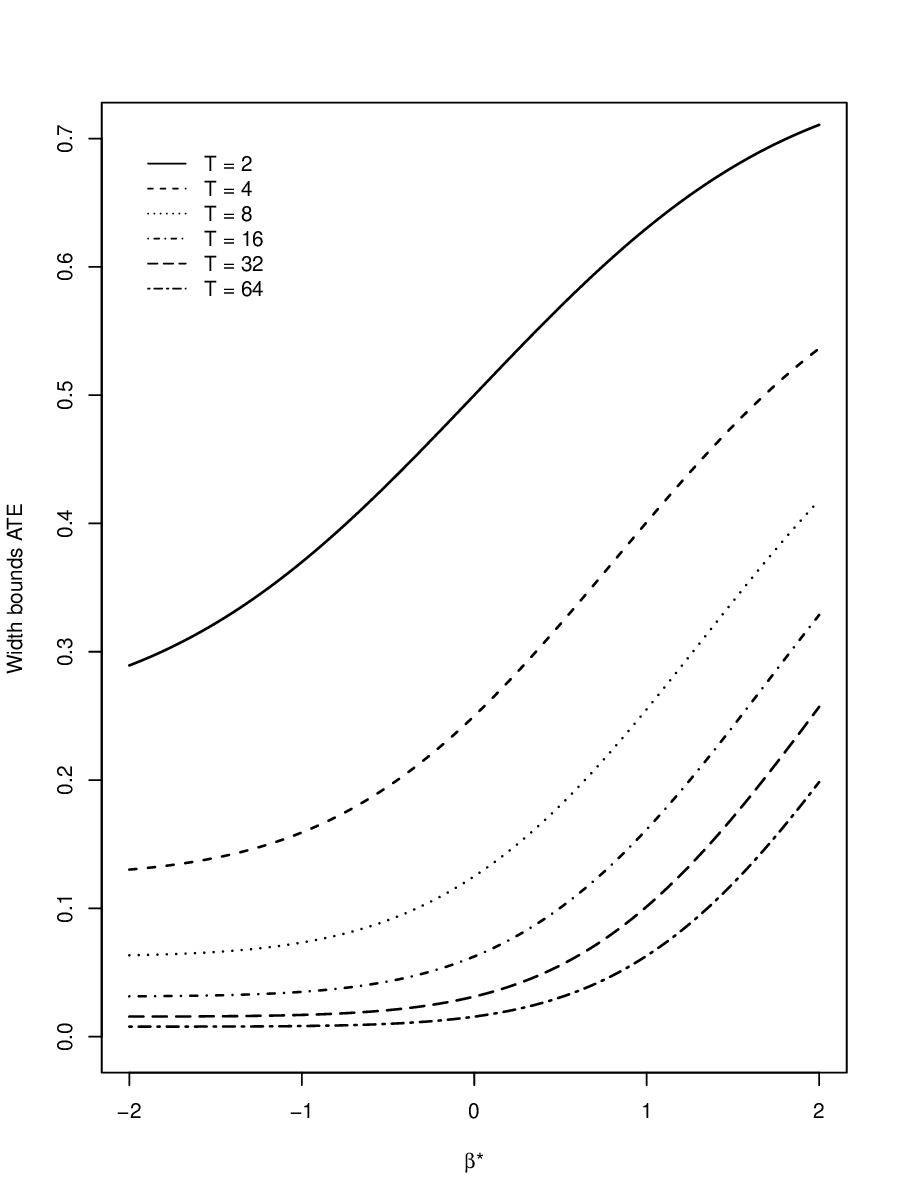,width=6in,height=8in}

\caption{\label{fig: width_bounds_dynamic_probit} Width of
nonparametric bounds for the ATE in dynamic binary choice probit
models with $ Y_{it} = 1(\beta^* Y_{i,t-1} + \alpha_i \geq
\varepsilon_{it})$, $\varepsilon_{it} \sim  N(0,1)$, $\alpha_i \sim
N(0,1)$, $\Pr(Y_{i0} = 1) = .5$, $\beta^* \in [-2, 2]$, and $T \in
\{2, 4, 8, 16, 32, 64 \}$.}

\end{center}

\end{figure}

% figure %%%%%%%%%%%%%%%%%%%%%%%%%%%%%%%%%%%%%%%%%%%%

% FIGURE %%%%%%%%%%%%%%%%%%%%%%%%%%%%%%%%%%%%%%%%%%%%

\begin{figure}

\begin{center}

\centering\epsfig{figure=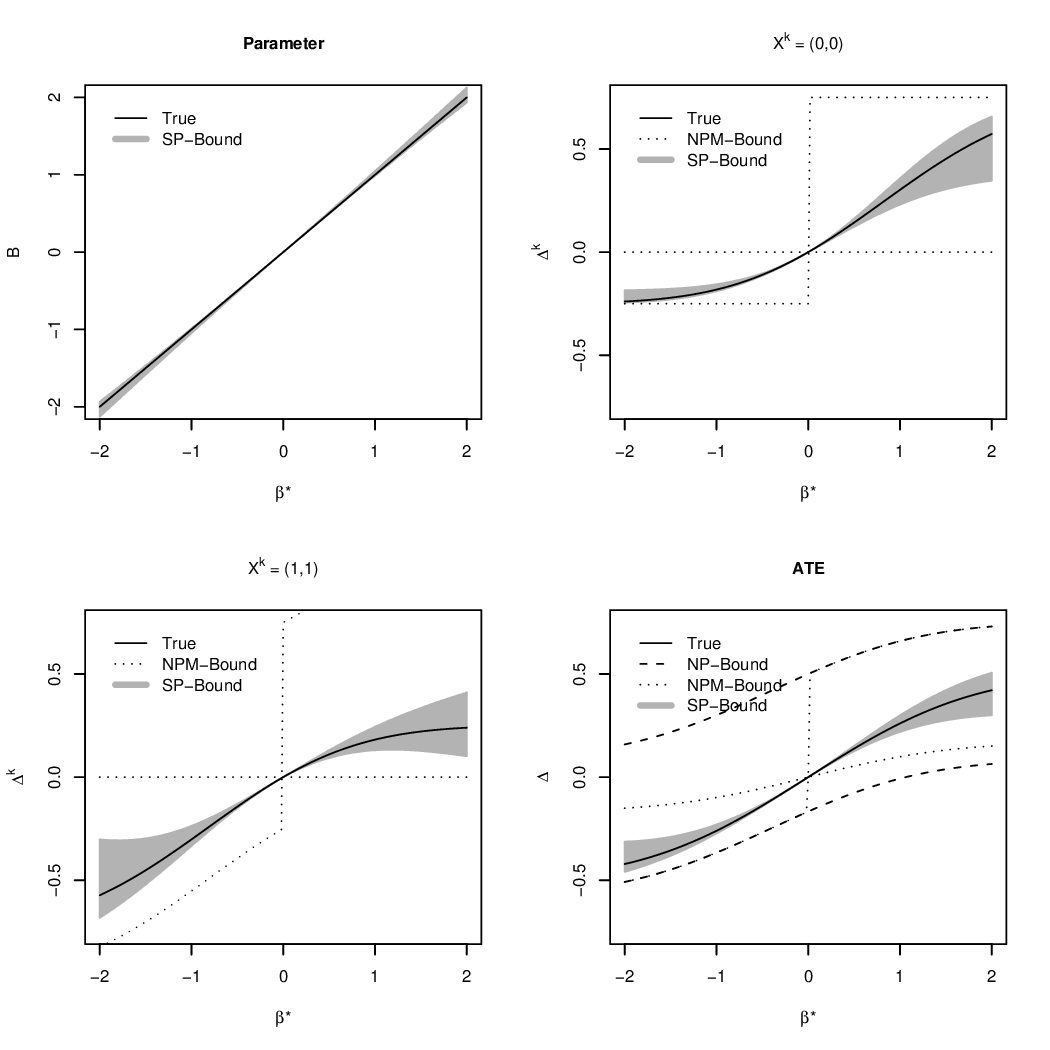,width=6.5in,height=6.5in}

\caption{\label{fig: probit-QP-T2} Identified set for parameter and
ATEs in binary choice probit models with $ Y_{it} = 1(\beta^{\ast}
X_{it} + \alpha_i \geq \varepsilon_{it})$, $\varepsilon_{it} \sim
N(0,1)$, $X_{it} = 1(\alpha_i \geq \eta_{it})$, $\eta_{it} \sim
N(0,1)$, $\alpha_i \sim N(0,1)$, $\beta^{\ast} \in [-2, 2]$, and $T
= 2$.}

\end{center}

\end{figure}

% figure %%%%%%%%%%%%%%%%%%%%%%%%%%%%%%%%%%%%%%%%%%%%

% FIGURE %%%%%%%%%%%%%%%%%%%%%%%%%%%%%%%%%%%%%%%%%%%%

\begin{figure}

\begin{center}

\centering\epsfig{figure=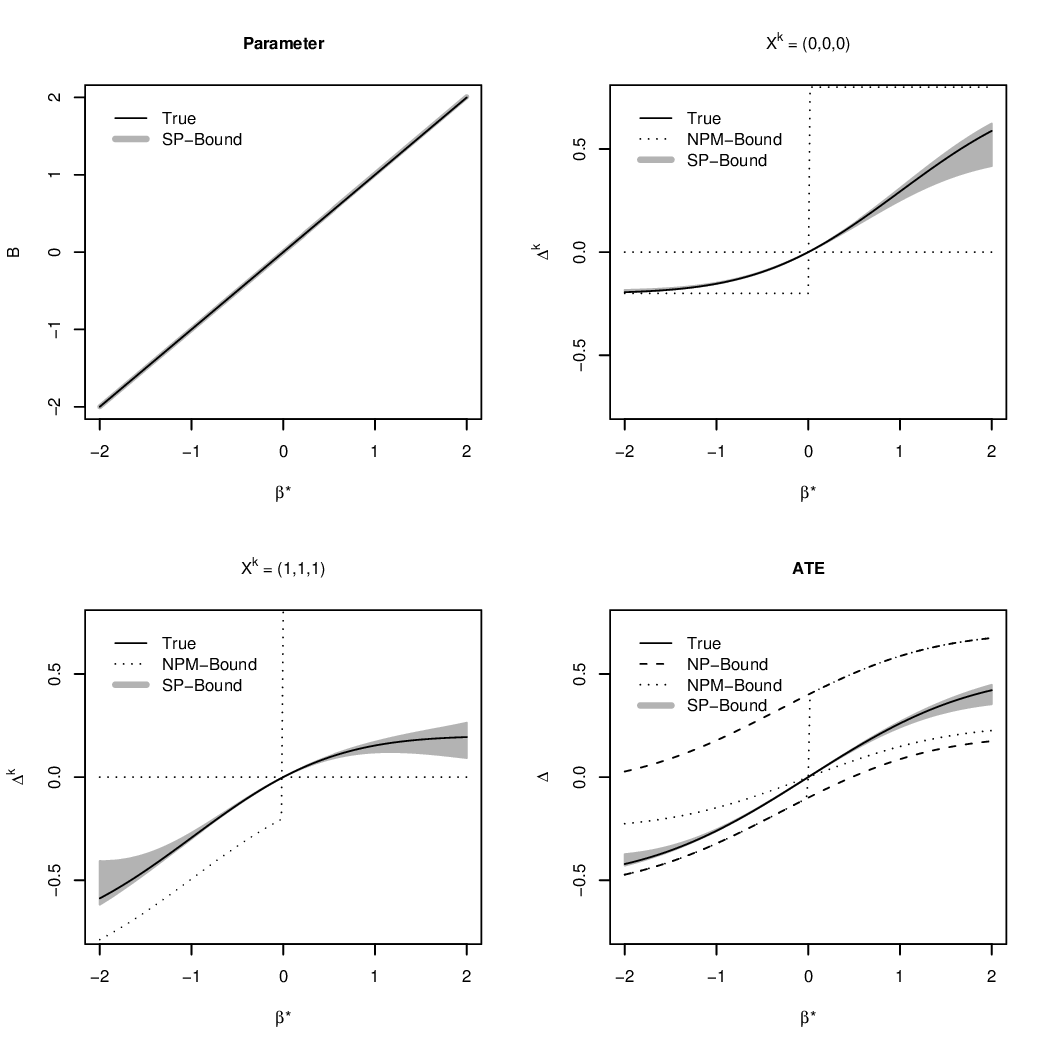,width=6.5in,height=6.5in}

\caption{\label{fig: probit-QP-T3} Identified set for parameter and
ATEs in binary choice probit models with $ Y_{it} = 1(\beta^{\ast}
X_{it} + \alpha_i \geq \varepsilon_{it})$, $\varepsilon_{it} \sim
N(0,1)$, $X_{it} = 1(\alpha_i \geq \eta_{it})$, $\eta_{it} \sim
N(0,1)$, $\alpha_i \sim N(0,1)$, $\beta^{\ast} \in [-2, 2]$, and $T
= 3$.}

\end{center}

\end{figure}

% figure %%%%%%%%%%%%%%%%%%%%%%%%%%%%%%%%%%%%%%%%%%%%

% FIGURE %%%%%%%%%%%%%%%%%%%%%%%%%%%%%%%%%%%%%%%%%%%%

\begin{figure}

\begin{center}

\centering\epsfig{figure=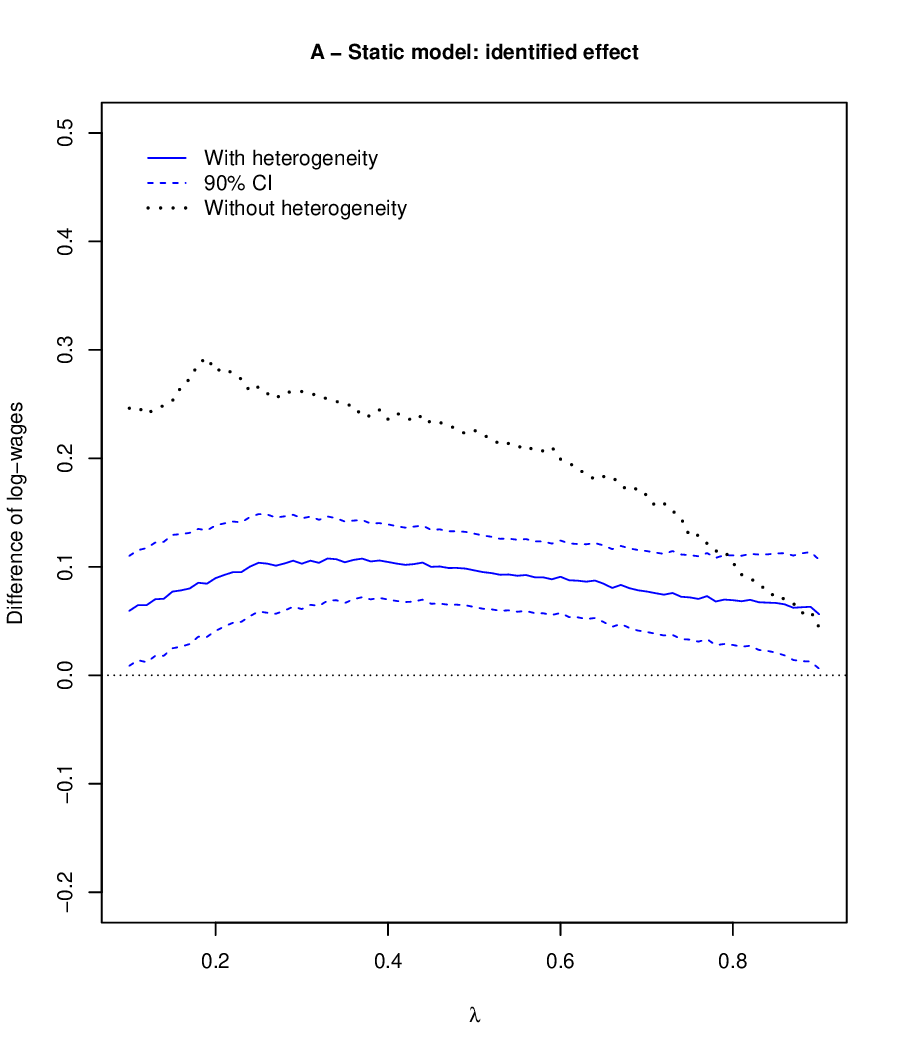,width=.48\textwidth,height=.48\textwidth}
\epsfig{figure=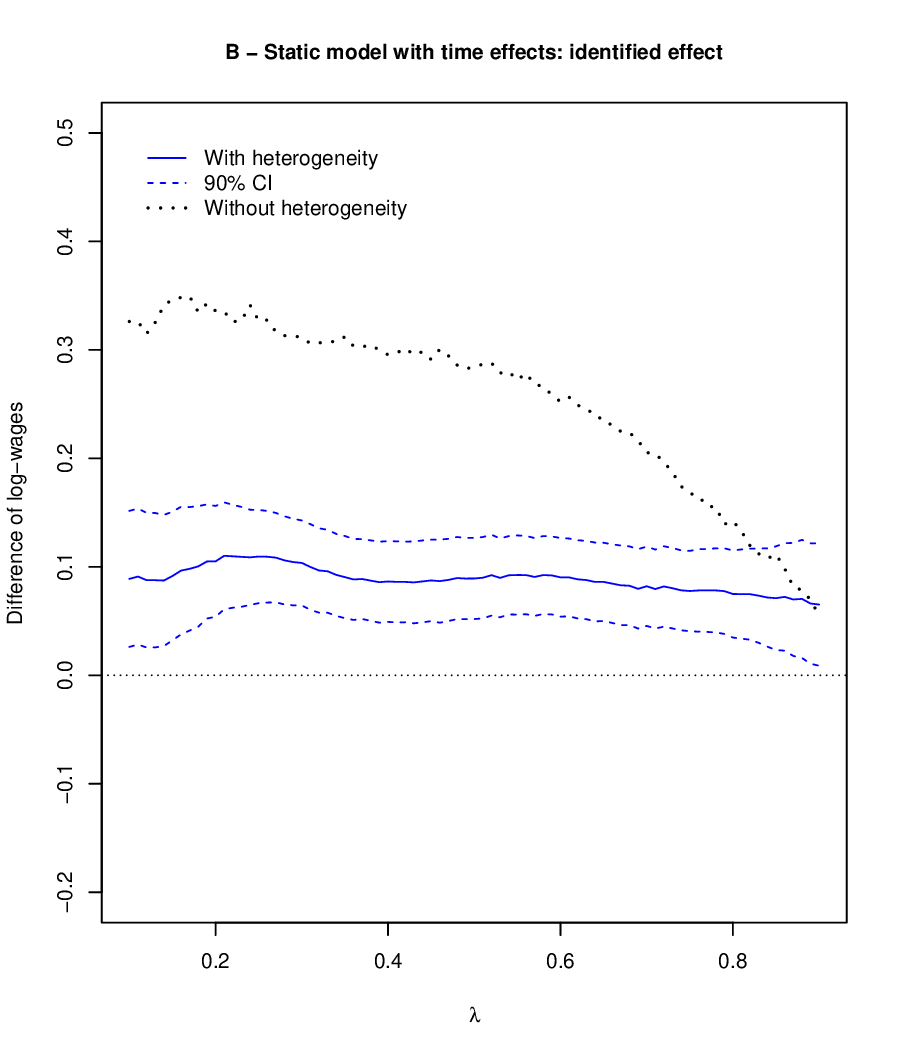,width=.48\textwidth,height=.48\textwidth}
\epsfig{figure=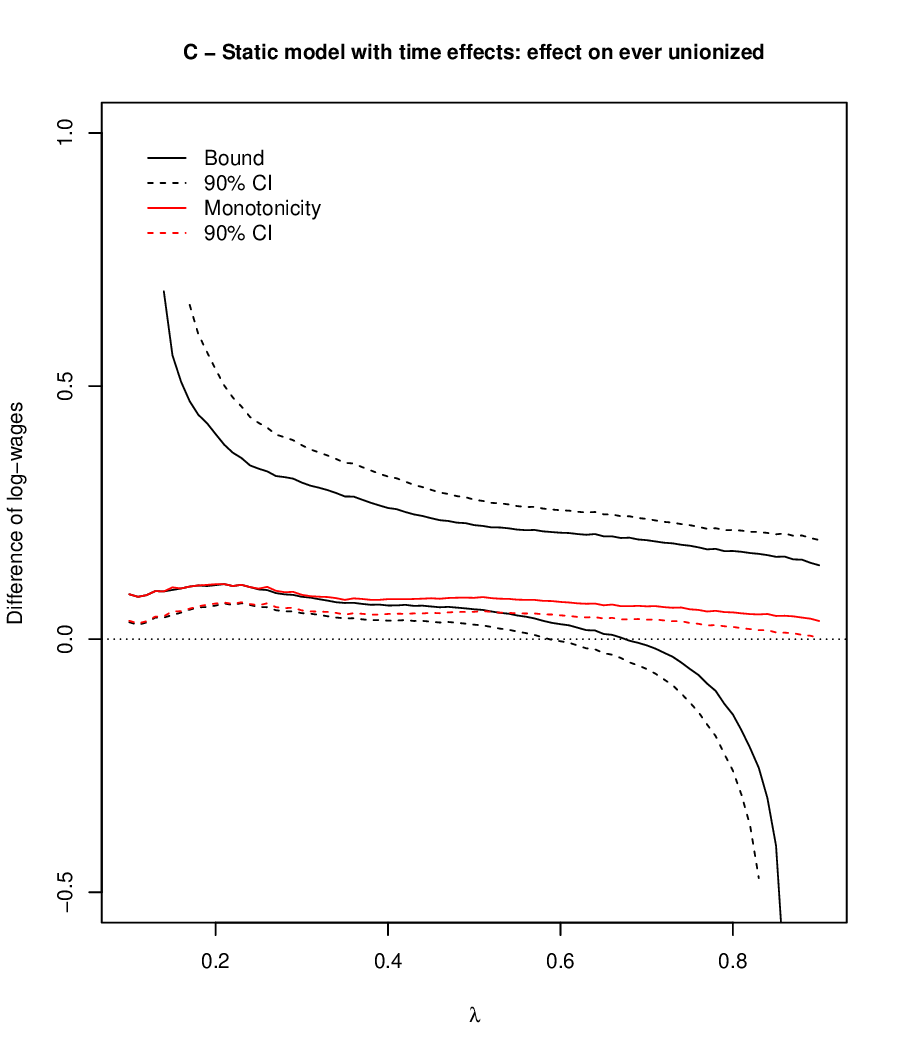,width=.48\textwidth,height=.48\textwidth}
\epsfig{figure=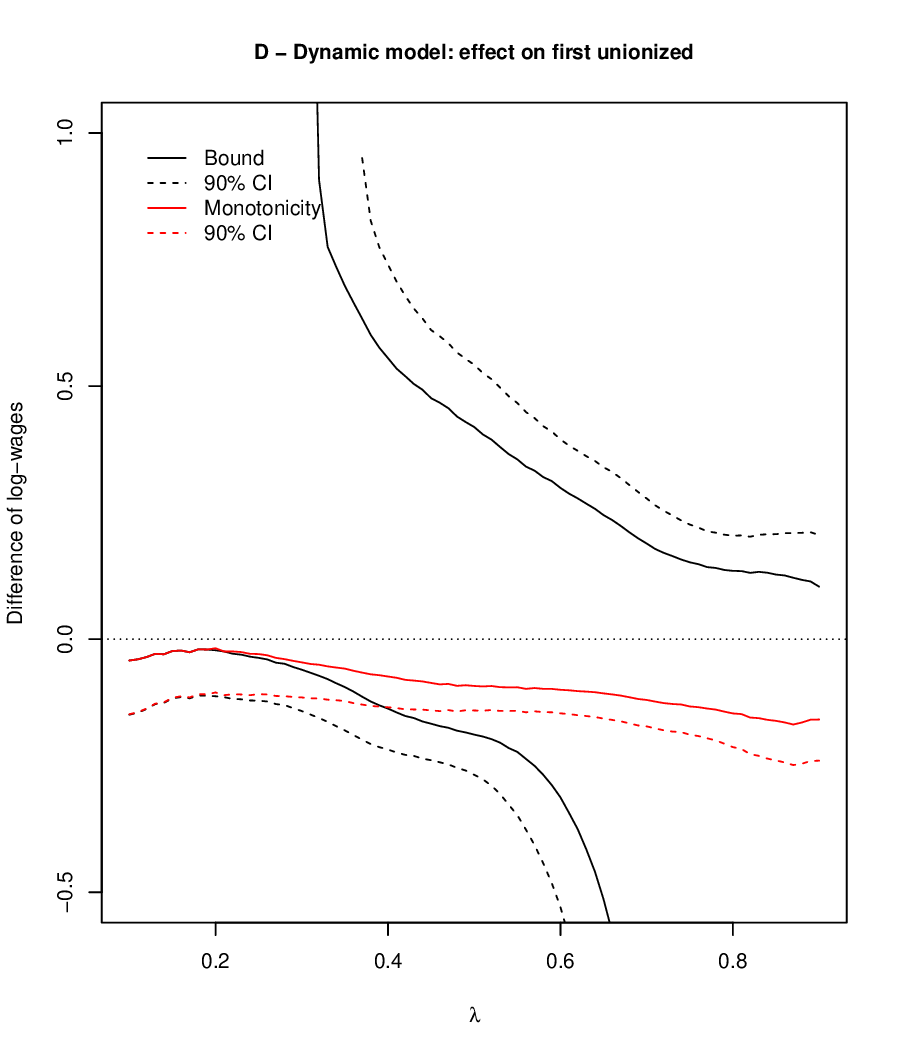,width=.48\textwidth,height=.48\textwidth}

\caption{\label{Fig: union} Quantile union effects for male workers. Panel A displays point and interval estimates of the identified quantile union
effects in the static model with and without accounting for individual heterogeneity. Panel B displays point and interval estimates of the identified quantile union
effects in the static model with location and scale time effects, averaged across time periods with and without accounting for individual heterogeneity. Panel C displays point and interval estimates of the bounds for the quantile effect on the ever unionized in the static model with time effects, with and without imposing monotonicity. Panel D displays point and interval estimates of the bounds for the quantile effect on the unionized in the first period in the dynamic model, with and without imposing monotonicity. Estimates based on NLSY79 for the years 1986--1993. 90\% confidence intervals obtained by bootstrap with 200 repetitions.}

\end{center}

\end{figure}

% figure %%%%%%%%%%%%%%%%%%%%%%%%%%%%%%%%%%%%%%%%%%%%

\clearpage

\begin{center}
\Large{Supplemental Material for  Average and Quantile Effects in
Nonseparable Panel Models}

\bigskip

\normalsize{
Victor Chernozhukov,  Iv\'an Fern\'andez-Val, 
Jinyong Hahn, and Whitney Newey}
\end{center}

\bigskip
\bigskip

\renewcommand{\thesection}{A\arabic{section}}
\setcounter{section}{0}

\section{Introduction}

In this supplemental material we provide omitted discussions, results, and
proofs by Section in the same order they are referred to in the paper. Let
w.p.a.1 denote "with probability approaching one" and $C$ denote a generic
constant that may be different in different uses.

\section{Supplements to Section 2}

We begin with the omitted discussion and results referred to in Section 2 of
the paper. These concern the general, nonseparable model of Assumptions 1 -
3 and apply whether or not the regressors are discrete.

\subsection{Time homogeneity in the linear model}

We will first show that Assumption 2 is a natural generalization of the
following linear model:%
\begin{equation}
Y_{it}=X_{it}^{\prime }\beta _{0}+\alpha _{i}+\varepsilon
_{it},E[X_{is}\varepsilon _{it}]=0\text{ for all }s\text{ and }t.
\label{std lin mod}
\end{equation}%
This is a standard linear model that leads to consistency of the within and
other estimators. Let $\bar{E}(\cdot |X_{i})$ denote the linear projection
on $vec(X_{i}),$ as in Chamberlain (1982).

\bigskip

\textsc{Theorem A1:} \textit{Suppose that }$Y_{i}$\textit{\ and }$X_{i}$%
\textit{\ have finite second moments. Then equation (\ref{std lin mod}) is
satisfied if and only if there is }$\tilde{\varepsilon}_{it}$\textit{\ with} 
\begin{equation}
Y_{it}=X_{it}^{\prime }\beta _{0}+\tilde{\varepsilon}_{it}\text{, }\bar{E}(%
\tilde{\varepsilon}_{it}|X_{i})=\bar{E}(\tilde{\varepsilon}%
_{i1}|X_{i}),(t=2,...,T).  \label{lin proj}
\end{equation}

\bigskip

Proof: If eq. (\textit{\ref{std lin mod}}) is satisfied let $\tilde{%
\varepsilon}_{it}=\alpha _{i}+\varepsilon _{it}$. By orthogonality of $%
\varepsilon _{it}$ with $X_{is}$ for all $s$ and $t$ we have $\bar{E}%
(\varepsilon _{it}|X_{i})=0$ for all $t$, so that 
\begin{equation*}
\bar{E}(\tilde{\varepsilon}_{it}|X_{i})=\bar{E}(\alpha _{i}|X_{i})+\bar{E}%
(\varepsilon _{it}|X_{i})=\bar{E}(\alpha _{i}|X_{i})=\bar{E}(\alpha
_{i}|X_{i})+\bar{E}(\varepsilon _{i1}|X_{i})=\bar{E}(\tilde{\varepsilon}%
_{i1}|X_{i}).
\end{equation*}%
Now suppose eq. (\ref{lin proj}) is satisfied. Let $\alpha _{i}=\bar{E}[%
\tilde{\varepsilon}_{i1}|X_{i}]$ and $\varepsilon _{it}=\tilde{\varepsilon}%
_{it}-\alpha _{i}$. Then $Y_{it}=X_{it}^{\prime }\beta _{0}+\alpha
_{i}+\varepsilon _{it}$ by construction and 
\begin{equation*}
E[X_{is}\varepsilon _{it}]=E[X_{is}(\tilde{\varepsilon}_{it}-\bar{E}[\tilde{%
\varepsilon}_{i1}|X_{i}])]=E[X_{is}(\tilde{\varepsilon}_{it}-\bar{E}[\tilde{%
\varepsilon}_{it}|X_{i}])]=0,
\end{equation*}%
where the second equality follows by $\bar{E}(\tilde{\varepsilon}%
_{it}|X_{i})=\bar{E}(\tilde{\varepsilon}_{i1}|X_{i})$ and the third quality
by orthogonality of each element of $X_{i}$ with the projection residual.
Q.E.D.

\bigskip

This result shows that the standard linear model of equation (\ref{std lin
mod}) is equivalent to the model of equation (\ref{lin proj}). The second
model is one that satisfies a time homogeneity condition analogous to
Assumption 2. In equation (\ref{lin proj}) the linear projection of the
disturbance on the elements of $X_{i}$ is time invariant. What Assumption 2
does is strengthen this to time invariance of the conditional distribution.
This strengthening seems like a natural thing to do when moving from a
linear model to a nonlinear, nonseparable model.

\subsection{Relationship between static and dynamic models}

We next show that the static model is nested within the dynamic model.

\bigskip

\textsc{Theorem A2:} \textit{If Assumptions 1 and 2 are satisfied then
Assumptions 1 and 3 are satisfied.}

\bigskip

Proof: Note that Assumptions 1 and 2 allow some flexibility in the
definition of $\alpha _{i},$ because Assumption 1 just specifies that there
exists $\alpha _{i}$ with $Y_{it}=g_{0}(X_{it},\alpha _{i},\varepsilon
_{it}).$ This equation continues to hold if more variables are added to $%
\alpha _{i}$. Furthermore, we can add any function of $X_{i}$ to $\alpha
_{i} $ without changing Assumption 2. Let $\tilde{\alpha}_{i}=(\alpha
_{i},X_{i})$. Then Assumptions 1 and 2 are also satisfied for this $\tilde{%
\alpha}_{i}$. Furthermore, since $X_{it},...,X_{i1}$ are included in $\tilde{%
\alpha}$ and Assumption 2 for the original $\alpha _{i}$ implies that $%
\varepsilon _{it}|\tilde{\alpha}_{i}\overset{d}{=}\varepsilon _{i1}|\tilde{%
\alpha}_{i}$ we have%
\begin{equation*}
\varepsilon _{it}|X_{it},...,X_{i1},\tilde{\alpha}_{i}\overset{d}{=}%
\varepsilon _{it}|\tilde{\alpha}_{i}\overset{d}{=}\varepsilon _{i1}|\tilde{%
\alpha}_{i}\overset{d}{=}\varepsilon _{i1}|X_{i1},\tilde{\alpha}_{i}.
\end{equation*}%
Thus we see that Assumptions 1 and 2 imply existence of $\alpha _{i}=\tilde{%
\alpha}_{i}$ such that Assumptions 1 and 3 are also satisfied. That is,
Assumptions 1 and 2 imply Assumptions 1 and 3. Q.E.D.

\subsection{Relationship between nonseparable models and conditional mean
models}

Next we show that the nonseparable models given here imply conditional mean
models where the ATE is also the conditional mean ATE.

\bigskip

\textsc{Theorem A3:} \textit{Suppose that Assumption 1 is satisfied\ and }$%
E[|g_{0}(x,\alpha _{i},\varepsilon _{it})|]<\infty $\textit{\ for all }$x.$ 
\textit{If Assumption 2 is satisfied then for }$\tilde{\alpha}_{i}=X_{i}$%
\textit{\ and }$m_{0}(x,\tilde \alpha )=$\textit{\ }$\int g_{0}(x,\alpha
,\varepsilon )dF(\alpha ,\varepsilon |\tilde{\alpha}),$%
\begin{equation*}
E[Y_{it}|X_{i},\tilde{\alpha}_{i}]=m_{0}(X_{it},\tilde{\alpha}_{i}),\mu
(x)=\int m_{0}(x,\tilde{\alpha})dF(\tilde{\alpha}).
\end{equation*}%
\textit{If Assumption 3 is satisfied then for }$\tilde{\alpha}=(\alpha
,X_{1})$ \textit{and }$m_{0}(x,\tilde{\alpha})=$\textit{\ }$\int
g_{0}(x,\alpha ,\varepsilon )dF(\varepsilon |\tilde{\alpha})$,%
\begin{equation*}
E[Y_{it}|X_{it},...,X_{i1},\tilde{\alpha}_{i}]=m_{0}(X_{it},\tilde{\alpha}%
_{i}),\mu (x)=\int m_{0}(x,\tilde{\alpha})F(d\tilde{\alpha}).
\end{equation*}%
Proof: By Assumption 2, for $\tilde{\alpha}=X$ and $m_{0}(x,\tilde \alpha
)=\int g_{0}(x,\alpha ,\varepsilon )dF(\alpha ,\varepsilon |X)$ we have 
\begin{eqnarray*}
E[Y_{it}|X_{i},\tilde{\alpha}_{i}] &=&E[g_{0}(X_{it},\alpha _{i},\varepsilon
_{it})|X_{i}]=\int g_{0}(X_{it},\alpha ,\varepsilon )dF(\alpha ,\varepsilon |%
\tilde{\alpha}_{i})=m_{0}(X_{it},\tilde{\alpha}_{i}), \\
\int m_{0}(x,\tilde{\alpha})dF(\tilde{\alpha}) &=&\int g_{0}(x,\alpha
,\varepsilon )dF(\alpha ,\varepsilon |\tilde{\alpha})dF(\tilde{\alpha})=\mu
(x).
\end{eqnarray*}%
Similarly, Assumption 3 implies, for $\tilde{\alpha}_{i}=(\alpha
_{i},X_{1i}) $,%
\begin{eqnarray*}
E[Y_{it}|X_{it},...,X_{i1},\tilde{\alpha}_{i}] &=&\int g_{0}(X_{it},\alpha
_{i},\varepsilon )dF(\varepsilon |X_{it},...,X_{i1},\alpha _{i}) \\
&=&\int g_{0}(X_{it},\alpha _{i},\varepsilon )dF(\varepsilon |\alpha
_{i},X_{i1})=m_{0}(X_{it},\tilde{\alpha}_{i}), \\
\int m_{0}(x,\tilde{\alpha})dF(\tilde{\alpha}) &=&\int g_{0}(x,\alpha
,\varepsilon )dF(\varepsilon |\alpha ,X_{1})dF(\alpha ,X_{1}) \\
&=&\int g_{0}(x,\alpha ,\varepsilon )dF(\varepsilon ,\alpha ,X_{1})=\mu
(x).Q.E.D.
\end{eqnarray*}

It may be helpful to explain this result and relate it to Chamberlain
(1982). First, it should be noted that Assumptions 1 and 2 only assume the
existence of some $\alpha _{i}$ such that the conditions are satisfied.
Thus, we are free to choose $\alpha _{i}$ in whatever way is convenient. A
convenient choice for Theorem A3 turns out to be $\tilde{\alpha}_{i}=X_{i}$,
where we use the $\tilde{\alpha}_{i}$ notation to distinguish this time
invariant effect from the one in Assumptions 1 and 2. Note then that the
first conclusion implies that for $m_{0}(x,X)=\int g(x,\alpha ,\varepsilon
)dF(\alpha ,\varepsilon |X),$%
\begin{equation}
E[Y_{it}|X_{i}]=m_{0}(X_{it},X_{i})\text{.}  \label{nonlin Chamb}
\end{equation}%
This statement has no content for any one time period, because the effect of 
$X_{it}$ in the first argument of $m(X_{it},X_{i})$ is indistinguishable
from the effect of $X_{it}$ that appears in the second argument. However,
for multiple time periods it does have content, because $m_{0}(x,X)$ is time
invariant. Equation (\ref{nonlin Chamb}) implies that the effect of changing 
$X_{it}$ on $E[Y_{it}|X_{i}]$ will be different than the effect on $%
E[Y_{is}|X_{i}]$ for $s\neq t$. Furthermore, this form leads directly to
identification of conditional mean ATE conditioned on $X_{i}$. For any $%
X_{i} $ where $X_{it}=x^{b}$ and $X_{is}=x^{a}$ for some $t$ and $s$, 
\begin{equation*}
E[Y_{is}-Y_{it}|X_{i}]=m_{0}(x^{a},X_{i})-m_{0}(x^{b},X_{i}),
\end{equation*}%
that is a conditional mean ATE given $X_{i}.$

It may also help to think of $m(X_{it},X_{i})$ as a nonlinear version of
Chamberlain's (1982) multivariate regression for panel data. In the linear
model of equation (\ref{std lin mod}), for $\bar{E}[\alpha _{i}|X_{i}]=\pi
^{\prime }vec(X_{i})$ we have 
\begin{equation*}
\bar{E}[Y_{it}|X_{i}]=X_{it}^{\prime }\beta _{0}+\pi ^{\prime }vec(X_{i})=%
\bar{m}(X_{it},X_{i})\text{, }\bar{m}(x,X)=x^{\prime }\beta _{0}+\pi
^{\prime }vec(X).
\end{equation*}%
For a single time period $\beta _{0}$ is indistinguishable from coefficients
in $\pi $, but multiple time periods can be used to identify $\beta _{0}$
from these regressions. Equation (\ref{nonlin Chamb}) is like this except it
is jointly nonlinear in its first and second arguments.

\section{Supplements to Section 3}

\subsection{Auxiliary results}

We turn now to identification and estimation with discrete regressors in the
static case. Here we use the idea that ``time is an instrument" or ``time is
randomly assigned." This allows us to vary the time period so as to match $x$
with $X_{it}$ and achieve identification.

The following Lemma applies this idea to obtain specific results. Let $%
g_{it}(x)=g_{0}(x,\alpha _{i},\varepsilon _{it})$.

\bigskip

\textsc{Lemma A4:} \textit{If Assumptions 1 and 2 are satisfied then}%
\begin{equation*}
E[\bar{G}_{i}(y,x)|X_{i}]=1(T_{i}(x)>0)E[\Phi (\frac{y-g_{i1}(x)}{h})|X_{i}].
\end{equation*}%
\textit{If in addition }$E[|g_{0}(x,\alpha _{i},\varepsilon _{it})|]<\infty $%
\textit{\ for all }$x$\textit{\ then} 
\begin{equation*}
E[\bar{Y}_{i}(x)|X_{i}]=1(T_{i}(x)>0)E[g_{i1}(x)|X_{i}].
\end{equation*}%
\textit{.}

Proof: By Assumptions 1 and 2,%
\begin{eqnarray*}
E[1(X_{it} &=&x)\Phi (\frac{y-Y_{it}}{h})|X_{i}]=E[1(X_{it}=x)\Phi (\frac{%
y-g_{it}(x)}{h})|X_{i}] \\
&=&1(X_{it}=x)E[\Phi (\frac{y-g_{it}(x)}{h})|X_{i}]=1(X_{it}=x)E[\Phi (\frac{%
y-g_{i1}(x)}{h})|X_{i}].
\end{eqnarray*}%
Therefore, we have%
\begin{eqnarray*}
E[\bar{G}_{i}(y,x)|X_{i}]
&=&1(T_{i}(x)>0)T_{i}(x)^{-1}\sum_{t=1}^{T}E[1(X_{it}=x)\Phi (\frac{y-Y_{it}%
}{h})|X_{i}] \\
&=&1(T_{i}(x)>0)T_{i}(x)^{-1}\sum_{t=1}^{T}1(X_{it}=x)E[\Phi (\frac{%
y-g_{i1}(x)}{h})|X_{i}] \\
&=&1(T_{i}(x)>0)E[\Phi (\frac{y-g_{i1}(x)}{h})|X_{i}].
\end{eqnarray*}%
We also have%
\begin{eqnarray*}
E[1(X_{it}
&=&x)Y_{it}|X_{i}]=E[1(X_{it}=x)g_{it}(x)|X_{i}]=1(X_{it}=x)E[g_{it}(x)|X_{i}]
\\
&=&1(X_{it}=x)E[g_{i1}(x)|X_{i}]
\end{eqnarray*}%
so the second conclusion follows similarly to the first. Q.E.D.

\bigskip

We can use the previous result to show how $\delta $ is identified.

\bigskip

\textsc{Lemma A5:} \textit{If Assumptions 1 and 2 are satisfied, }$%
E[|g_{0}(x,\alpha _{i},\varepsilon _{it})|]<\infty $\textit{\ for all }$x,$\ 
\textit{and }$\mathit{\Pr }(D_{i}=1)>0$ \textit{then }$\delta =E[D_{i}\{\bar{%
Y}_{i}(x^{a})-\bar{Y}_{i}(x^{b})\}]/E[D_{i}].$

\bigskip

Proof: Note that $D_{i}=D_{i}1(T_{i}(x^{b})>0)=D_{i}1(T_{i}(x^{a})>0).$
Therefore, by Lemma A4 
\begin{eqnarray*}
E[D_{i}\{\bar{Y}_{i}(x^{a})-\bar{Y}_{i}(x^{b})\}|X_{i}] &=&D_{i}E[\bar{Y}%
_{i}(x^{a})|X_{i}]-D_{i}E[\bar{Y}_{i}(x^{b})|X_{i}] \\
&=&D_{i}1(T_{i}(x^{a})>0)E[g_{i1}(x^{a})|X_{i}]-D_{i}1(T_{i}(x^{b})>0)E[g_{i1}(x^{b})|X_{i}]
\\
&=&D_{i}E[g_{i1}(x^{a})-g_{i1}(x^{b})|X_{i}]=E[D_{i}\left\{
g_{i1}(x^{a})-g_{i1}(x^{b})\right\} |X_{i}]
\end{eqnarray*}%
The conclusion then follows by iterated expectations. \textit{Q.E.D.}

\bigskip

The asymptotic normality of $\hat{\delta}$ and consistency of the asymptotic
variance estimator are simple applications of standard theory, as in the
following result, that forms a prototype for the asymptotic normality of the
nonparametric ATE bounds. Let $P=E[D_{i}].$

\bigskip

\textsc{Theorem A6:} \textit{If Assumptions 1 and 2 are satisfied, }$%
E[|g_{0}(x,\alpha _{i},\varepsilon _{it})|^{2}]<\infty $\textit{\ for all }$%
x,$ \textit{and}\ $\mathit{\Pr }(D_{i}=1)>0,$ \textit{then }$\sqrt{n}(\hat{%
\delta}-\delta )\overset{d}{\longrightarrow }N(0,V)$ \textit{and} $%
\sum_{i=1}^{n}\hat{\psi}_{i}^{2}/n\overset{p}{\longrightarrow }V,$ \textit{%
where }$V=E[\psi _{i}^{2}]$\textit{\ and} $\psi _{i}=P^{-1}D_{i}\left[ \bar{Y%
}_{i}(x^{a})-\bar{Y}_{i}(x^{b})-\delta \right] .$

\bigskip

Proof: Let $d_{i}=D_{i}\{\bar{Y}_{i}(x^{a})-\bar{Y}_{i}(x^{b})\}$ so that $%
\hat{\delta}=\bar{d}/\bar{D}$. By the central limit theorem (CLT), $\bar{d}$
and $\bar{D}$ are root-$n$ consistent for $\mu _{d}=E[d_{i}]$ and $P$. Then
by $P>0$ and $\delta =\mu _{d}/P,$%
\begin{eqnarray*}
\sqrt{n}(\hat{\delta}-\delta ) &=&\sqrt{n}(\frac{\bar{d}}{\bar{D}}-\frac{\mu
_{d}}{P})=\sqrt{n}\bar{D}^{-1}[\bar{d}-\mu _{d}-\delta (\bar{D}-P)] \\
&=&\sqrt{n}P^{-1}[\bar{d}-\mu _{d}-\delta (\bar{D}-P)]+o_{p}(1)=%
\sum_{i=1}^{n}\psi _{i}/\sqrt{n}+o_{p}(1).
\end{eqnarray*}%
The first conclusion then follows by the CLT. For the second conclusion note
that%
\begin{equation*}
\sum_{i}(\hat{\psi}_{i}-\psi _{i})^{2}/n\leq C(\bar{D}^{-1}-P^{-1})^{2}%
\sum_{i}d_{i}^{2}/n+C(\bar{D}^{-1}\hat{\delta}-P^{-1}\delta
)^{2}\sum_{i}D_{i}^{2}/n\overset{p}{\longrightarrow }0.
\end{equation*}%
Therefore, the second conclusion follows by a standard argument. Q.E.D.

\bigskip

We now give an intermediate result that is useful for showing asymptotic
normality for the estimator of the identified quantile treatment effect.
This will also serve as a prototype for the proofs of Theorems 2 and 3 in
the body of the paper. Let $\hat{G}_{1}(y,x)=\hat{G}(y,x|D_{i}=1)$, $%
G_{1}(y,x)=G(y,x|D_{i}=1),$ $G_{i}(y,x)=1(T_{i}(x)>0)T_{i}(x)^{-1}%
\sum_{t=1}^{T}1(X_{it}=x)1(Y_{it}\leq y),$ and $G_1'(y,x) = \partial G_1(y,x)/\partial y.$

\bigskip

\textsc{Lemma A7:} \textit{If Assumption 7 is satisfied with }$G_{\ell
}(y,x) $\textit{\ replaced by }$G_{1}(y,x)$\textit{\ then for any }$%
0<\lambda <1$\textit{\ and any }$x$\textit{, there exists }$\hat{q}_{\lambda
}$\textit{\ with }$\hat{G}_{1}(\hat{q}_{\lambda },x)=\lambda $ \textit{%
satisfying}%
\begin{equation*}
\sqrt{n}(\hat{q}_{\lambda }-q_{\lambda })= - G_{1}^{\prime }(q_{\lambda
},x)^{-1}\frac{1}{\sqrt{n}}P^{-1}\sum_{i}D_{i}\left[ G_{i}(q_{\lambda
},x)-\lambda \right] +o_{p}(1).
\end{equation*}

Proof: Note that $\hat{G}_{1}(y,x)$ is strictly monotonic increasing in $y$
and converges to $0$ and $1$ as $y$ goes to $-\infty $ and $\infty $
respectively. Therefore there is a unique $\hat{q}_{\lambda }$ such that $%
\hat{G}_{1}(\hat{q}_{\lambda },x)=\lambda $. Also, by $G_{1}(y,x)$ strictly
monotonic in $y$ there is a unique $q_{\lambda }$ solving $G_{1}(q_{\lambda
},x)=\lambda $. By $G_{1}(y,x)$ strictly monotonic and continuous, it
follows that for all $\varepsilon >0$ small enough,%
\begin{equation*}
0<G_{1}(q_{\lambda }-\varepsilon ,x)<G_{1}(q_{\lambda },x)=\lambda .
\end{equation*}%
By $\hat{G}_{1}(q_{\lambda }-\varepsilon ,x)\overset{p}{\longrightarrow }%
G_{1}(q_{\lambda }-\varepsilon ,x)$ it follows that w.p.a.1, for all $y\leq
q_{\lambda }-\varepsilon $%
\begin{equation*}
\hat{G}_{1}(y,x)\leq \hat{G}_{1}(q_{\lambda }-\varepsilon
,x)<G_{1}(q_{\lambda },x)=\lambda .
\end{equation*}%
Thus, it follows that $\hat{q}_{\lambda }\geq q_{\lambda }-\varepsilon $
w.p.a.1. Similarly it follows that $\hat{q}_{\lambda }\leq q_{\lambda
}+\varepsilon $ w.p.a.1. Since $\varepsilon $ is arbitrary, we have $\hat{q}%
_{\lambda }\overset{p}{\longrightarrow }q_{\lambda }$.

Next, note that $G_{1}(y,x)$ is differentiable in $y$ by Assumption 7, so
that $g_{i1}(x)$ is continuously distributed conditional on $D_{i}=1.$ Thus, 
$g_{it}(x)$ is also continuously distributed conditional on $D_{i}=1$ by
Assumption 2. It follows that as $h\longrightarrow 0$, $\Phi(\frac{y-g_{it}(x)}{%
h})\longrightarrow 1(g_{it}(x)\leq y)$ with probability one. By the
dominated convergence theorem this convergence is also in mean-square. Recall that 
\begin{equation*}
G_{i}(y,x)=\left\{ 
\begin{array}{c}
T_{i}(x)^{-1}\sum_{t=1}^{T}1(X_{it}=x)1(Y_{it}\leq y),T_{i}(x)>0, \\ 
0,T_{i}(x)=0.%
\end{array}%
\right.
\end{equation*}%
We have $\bar{G}_{i}(y,x)\longrightarrow G_{i}(y,x)$ in mean square, so that%
\begin{eqnarray*}
&&\sum_{i=1}^{n}[D_{i}\bar{G}_{i}(y,x)-D_{i}G_{i}(y,x)]/n\overset{p}{%
\longrightarrow }0, \\
&&\sum_{i=1}^{n}\{D_{i}\bar{G}_{i}(y,x)-E[D_{i}\bar{G}%
_{i}(y,x)]-D_{i}G_{i}(y,x)+E[D_{i}G_{i}(y,x)]\}/\sqrt{n}\overset{p}{%
\longrightarrow }0.
\end{eqnarray*}%
Let $W_{i}=g_{0}(x,\alpha _{i},\varepsilon _{i1})$ and $f(w)$ and $F(w)$
denote the pdf and CDF of $W_{i}$ conditional on $D_{i}=1$ and $P=E[D_{i}].$
Note that $\Phi (\frac{y-w}{h})F(w)$ converges to zero as $w\longrightarrow
\infty $ and as $w\longrightarrow -\infty $. Therefore, integration by parts
gives%
\begin{eqnarray*}
E[\bar{G}_{i}(y,x)|D_{i} &=&1]=\int \Phi (\frac{y-w}{h})f(w)dw=h^{-1}\int
\phi (\frac{y-w}{h})F(w)dw \\
&=&\int \phi (u)F(y-hu)du=F(y)+(h^{2}/2)\int \phi (u)F^{\prime \prime }(y-%
\bar{h}u)u^{2}du \\
&=&F(y)+o(h^{2})=G_{1}(y,x)+o(h^{2}),
\end{eqnarray*}%
where the fifth equality follows by an expansion 
\begin{equation*}
F(y-hu)=F(y)-F^{\prime }(y)hu+F^{\prime \prime }(y-\bar{h}u)h^{2}u^{2}/2,
\end{equation*}%
and $\bar{h}$ can depend on $u.$ Therefore it follows by $%
E[D_{i}G_{i}(q_{\lambda },x)]=PG_{1}(q_{\lambda },x)=P\lambda $ that%
\begin{eqnarray*}
\sum_{i=1}^{n}D_{i}[\bar{G}_{i}(q_{\lambda },x)-\lambda ]/\sqrt{n}
&=&\sum_{i=1}^{n}\{D_{i}\bar{G}_{i}(q_{\lambda },x)-E[D_{i}\bar{G}%
_{i}(q_{\lambda },x)]\}/\sqrt{n} \\
&&+\sqrt{n}\{E[D_{i}\bar{G}_{i}(q_{\lambda },x)]-\lambda P\}-\lambda
\sum_{i=1}^{n}(D_{i}-P)/\sqrt{n} \\
&=&\sum_{i=1}^{n}\{D_{i}G_{i}(q_{\lambda },x)-E[D_{i}G_{i}(q_{\lambda
},x)]\}/\sqrt{n}+o_{p}(1) \\
&&+O(\sqrt{n}h^{2})-\lambda \sum_{i=1}^{n}(D_{i}-P)/\sqrt{n} \\
&=&\sum_{i=1}^{n}D_{i}[G_{i}(q_{\lambda },x)-\lambda ]/\sqrt{n}%
+o_{p}(1)=O_{p}(1).
\end{eqnarray*}

Next, note that from standard uniform convergence of kernel density results, 
$\hat{G}_{1}^{\prime }(y,x)$ converges uniformly in probability to $%
G_{1}^{\prime }(y,x),$ where the "prime" superscript denotes the partial
derivative with respect to $y$. Therefore, for $\bar{q}_{\lambda }\overset{p}%
{\longrightarrow }q_{\lambda },$ $\hat{G}_{1}^{\prime }(\bar{q}_{\lambda },x)%
\overset{p}{\longrightarrow }G_{1}^{\prime }(q_{\lambda },x) >0$, and
hence $\hat{G}_{1}^{\prime }(\bar{q}_{\lambda },x)^{-1}=O_{p}(1)$. An
expansion then gives $\lambda =\hat{G}_{1}(\hat{q}_{\lambda },x)=\hat{G}%
_{1}(q_{\lambda },x)+\hat{G}_{1}^{\prime }(\bar{q}_{\lambda },x)(\hat{q}%
_{\lambda }-q_{\lambda }).$ Solving and inverting gives%
\begin{eqnarray*}
\sqrt{n}(\hat{q}_{\lambda }-q_{\lambda }) &=&-\hat{G}_1^{\prime }(\bar{q}%
_{\lambda },x)^{-1}\sqrt{n}[\hat{G}_{1}(q_{\lambda },x)-\lambda ] \\
&=&-\hat{G}_{1}'(\bar{q}_{\lambda },x)^{-1}\left(
\sum_{i=1}^{n}D_{i}/n\right) ^{-1}\sum_{i=1}^{n}D_{i}[\bar{G}_{i}(q_{\lambda
},x)-\lambda ]/\sqrt{n} \\
&=&-G^{\prime }_1(q_{\lambda
},x)^{-1}P^{-1}\sum_{i=1}^{n}D_{i}[G_{i}(q_{\lambda },x)-\lambda ]/\sqrt{n}%
+o_{p}(1).Q.E.D.
\end{eqnarray*}

\bigskip

\textsc{Theorem A8:} \textit{If Assumptions 1, 2, and 7 are satisfied and}\ $%
E[D_{i}]>0,$ \textit{then }$\sqrt{n}(\hat{\delta}_{\lambda }-\delta
_{\lambda })\overset{d}{\longrightarrow }N(0,V_{\lambda })$ \textit{and} $%
\sum_{i=1}^{n}\hat{\psi}_{\lambda i}^{2}/n\overset{p}{\longrightarrow }%
V_{\lambda },$ \textit{where }$V_{\lambda }=E[\psi _{\lambda i}^{2}]$\textit{%
\ and} 
\begin{equation*}
\psi _{i\lambda }=- \frac{D_{i}}{P}\left\{ \frac{G_{i}(q^{a},x^{a})-\lambda }{%
G_{1}^{\prime }(q^{a},x^{a})}-\frac{G_{i}(q^{b},x^{b})-\lambda }{%
G_{1}^{\prime }(q^{b},x^{b})}\right\}
\end{equation*}%
\newline

Proof: By Lemma A7 we have%
\begin{equation*}
\sqrt{n}(\hat{\delta}_{\lambda }-\delta _{\lambda })=\sum_{i=1}^{n}\psi
_{i\lambda }/\sqrt{n}+o_{p}(1).
\end{equation*}%
The CLT gives the first conclusion. Next, note that by $\Phi (v)$ having a
bounded derivative, 
\begin{equation*}
\sum_{i=1}^{n}[\bar{G}_{i}(\hat{q}^{a},x^{a})-\bar{G}%
_{i}(q^{a},x^{a})]^{2}/n\leq Ch^{-1}(\hat{q}^{a}-q^{a})=O_{p}((h\sqrt{n}%
)^{-1})\overset{p}{\longrightarrow }0.
\end{equation*}%
Then by mean square convergence of $\bar{G}_{i}(q^{a},x^{a})$ to $%
G_{i}(q^{a},x^{a})$ and the triangle inequality we have $\sum_{i=1}^{n}[\bar{%
G}_{i}(\hat{q}^{a},x^{a})-G_{i}(q^{a},x^{a})]^{2}/n\overset{p}{%
\longrightarrow }0.$ The second conclusion then follows similarly to the
proof of Theorem A6. \textit{Q.E.D.}

\bigskip

\subsection{Proof of Theorem 1}

%\bigskip

%\textsc{Proof of Theorem 1:} Next, 
Note that $\sigma _{i}^{2}>0$ if and only if $D_{i}=1$, so that 
\begin{equation*}
\sigma _{i}^{2}=D_{i}\sigma _{i}^{2},X_{it}-\bar{X}_{i}=D_{i}(X_{it}-\bar{X}%
_{i}).
\end{equation*}%
Furthermore, since $X_{it}$ is a dummy variable, the usual difference in
means formula for the slope of a regression on a constant and dummy variable
gives%
\begin{equation*}
D_{i}\frac{\sum_{t=1}^{T}(X_{it}-\bar{X}_{i})Y_{it}}{\sum_{t=1}^{T}(X_{it}-%
\bar{X}_{i})^{2}}=D_{i}\{\bar{Y}_{i}(1)-\bar{Y}_{i}(0)\}.
\end{equation*}%
Also, by the Khintchine's weak law of large numbers (LLN), 
\begin{equation*}
n^{-1}(T-1)^{-1}\sum_{i=1}^{n}\sum_{t=1}^{T}(X_{it}-\bar{X}%
_{i})^{2}=n^{-1}\sum_{i=1}^{n}\sigma _{i}^{2}\overset{p}{\longrightarrow }%
E[\sigma _{i}^{2}]=E[D_{i}\sigma _{i}^{2}].
\end{equation*}%
Furthermore, by LLN%
\begin{eqnarray*}
n^{-1}(T-1)^{-1}\sum_{i=1}^{n}\sum_{t=1}^{T}(X_{it}-\bar{X}_{i})Y_{it}
&=&n^{-1}(T-1)^{-1}\sum_{i=1}^{n}\sum_{t=1}^{T}D_{i}(X_{it}-\bar{X}%
_{i})Y_{it} \\
&=&n^{-1}\sum_{i=1}^{n}D_{i}\sigma _{i}^{2}\{\bar{Y}_{i}(1)-\bar{Y}_{i}(0)\}
\\
&&\overset{p}{\longrightarrow }E[D_{i}\sigma _{i}^{2}\{\bar{Y}_{i}(1)-\bar{Y}%
_{i}(0)\}].
\end{eqnarray*}%
\newline
The conclusion then follows by the continuous mapping theorem. Q.E.D.

\section{Supplements to Section 4}

Here we include the proof of Theorem 2 as well as bounds that impose
monotonicity.

%Here is the proof of Theorem 2.

%\bigskip

%\textsc{Proof of Theorem 2:} 

\subsection{Proof of Theorem 2}

Let 
\begin{equation*}
\left( 
\begin{array}{c}
m_{\ell i} \\ 
m_{ui}%
\end{array}%
\right) =\left( 
\begin{array}{c}
\bar{Y}_{i}(x^{a})-\bar{Y}_{i}(x^{b})+B_{\ell
}1(T_{i}(x^{a})=0)-B_{u}1(T_{i}(x^{b})=0) \\ 
\bar{Y}_{i}(x^{a})-\bar{Y}_{i}(x^{b})+B_{u}1(T_{i}(x^{a})=0)-B_{\ell
}1(T_{i}(x^{b})=0)%
\end{array}%
\right) .
\end{equation*}%
Note that $\hat{\Delta}_{\ell }=\sum_{i=1}^{n}m_{\ell i }/n$ and $\hat{\Delta%
}_{u}=\sum_{i=1}^{n}m_{ui }/n.$ Then for $\Sigma =Var((m_{\ell i },m_{ui})),$
$\Delta _{\ell }=E[m_{\ell i }],$ and $\Delta _{u}=E[m_{ui }]$ the first and
second conclusions follow by standard arguments for a vector of sample means.

Next, note that by Lemma A4 and iterated expectations%
\begin{eqnarray}
\Delta _{\ell } &=&E[1(T_{i}(x^{a})>0)g_{i1}(x^{a})+B_{\ell
}1(T_{i}(x^{a})=0)]  \label{ATE static bound} \\
& & -E[1(T_{i}(x^{b}) >0)g_{i1}(x^{b})+B_{u}1(T_{i}(x^{b})=0)]\leq
E[g_{i1}(x^{a})]-E[g_{i1}(x^{b})]=\Delta .  \notag
\end{eqnarray}%
It follows similarly that $\Delta \leq \Delta _{u}$. To show sharpness, let $%
\tilde{\alpha}_{i}=(\alpha _{i},X_{i})$. Define%
\begin{eqnarray*}
g(x,\tilde{\alpha}_{i},\varepsilon _{it},C_{a},C_{b})
&=&1(T_{i}(x)>0)g_{0}(x,\alpha _{i},\varepsilon _{it}) \\
+1(T_{i}(x) &=&0)[C_{a}1(x=x^{a})+C_{b}1(x=x^{b})],
\end{eqnarray*}%
where $B_{\ell }\leq C_{a}\leq B_{u}$ and $B_{\ell }\leq C_{b}\leq B_{u}$.
Note that $T_{i}(X_{it})>0$ with probability one, so that $g(X_{it},\tilde{%
\alpha}_{i},\varepsilon _{it},C_{a},C_{b})=g_{0}(X_{it},\alpha
_{i},\varepsilon _{it})=Y_{it}.$ Hence the conditional distribution of $%
(Y_{i1},...,Y_{iT})^{\prime }$ given $X_{i}$ is the same for $g$ and $\tilde{%
\alpha}_{i}$ as for $g_{0}$ and $\alpha _{i}$. Also, because $(\alpha
_{i},X_{i})$ is a one-to-one function of $(\tilde{\alpha}_{i},X_{i})$ it
follows that Assumption 2 is satisfied with $\tilde{\alpha}_{i}$ replacing $%
\alpha _{i}$. When $(C_{a},C_{b})=(B_{\ell },B_{u})$ we have 
\begin{eqnarray*}
\Delta &=&E[g(x^{a},\tilde{\alpha}_{i},\varepsilon _{it},B_{\ell
},B_{u})-g(x^{b},\tilde{\alpha}_{i},\varepsilon _{it},B_{\ell },B_{u})] \\
&=&E[1(T_{i}(x^{a})>0)g_{i}(x^{a})+1(T_{i}(x^{a})=0)B_{\ell }] \\
& & -E[1(T_{i}(x^{b}) >0)g_{i}(x^{b})+1(T_{i}(x^{b})=0)B_{u}]=\Delta _{\ell },
\end{eqnarray*}%
and the lower bound is attained. Similarly the upper bound is attained when $%
(C_{a},C_{b})=(B_{u},B_{\ell })$.

Turning now to the quantile bounds, it follows as in the proof of Lemma A7
applied to $\hat{G}_{\ell }(y,x^{a})$ and to $\hat{G}_{\ell }(y,x^{b})+\bar{P%
}(x^{b})$ that 
\begin{equation*}
\hat{q}_{u}^{d}\overset{p}{\longrightarrow }q_{u}^{d},\hat{q}_{\ell }^{d}%
\overset{p}{\longrightarrow }q_{\ell }^{d},G_{\ell
}(q_{u}^{d},x^{d})=\lambda ,G_{\ell }(q_{\ell }^{d},x^{d})+\mathcal{\bar{P}}%
(x^{d})=\lambda ,d\in \{a,b\}.
\end{equation*}%
It also follows as in eq. (\ref{ATE static bound}) that $G_{\ell }(y,x)\leq
G(y,x)\leq G_{\ell }(y,x)+\mathcal{\bar{P}}(x),$ implying $\Delta _{\lambda
\ell }\leq \Delta _{\lambda }\leq \Delta _{\lambda u}$. Next, it follows as
in Lemma A7 that 
\begin{eqnarray*}
\sqrt{n}(\hat{q}_{u}^{a}-q_{u}^{a}) &=& - G_{\ell }^{\prime
}(q_{u}^{a},x^{a})^{-1}\frac{1}{\sqrt{n}}\sum_{i}\left[ G_{i}(q_{u}^{a},x)-%
\lambda \right] +o_{p}(1), \\
\sqrt{n}(\hat{q}_{\ell }^{b}-q_{\ell }^{b}) &=& - G_{\ell }^{\prime }(q_{\ell
}^{b},x^{b})^{-1}\frac{1}{\sqrt{n}}\sum_{i}\left[ G_{i}(q_{\ell
}^{b},x^{b})+1(T_{i}(x^{b})=0)-\lambda \right] +o_{p}(1).
\end{eqnarray*}%
Differencing then gives%
\begin{equation*}
\sqrt{n}(\hat{\Delta}_{u}-\Delta _{u})=  - \sum_{i=1}^{n}\frac{\Psi _{\lambda
i}^{u}}{\sqrt{n}}+o_{p}(1),\Psi _{\lambda i}^{u}= \frac{%
G_{i}(q_{u}^{a},x^{a})-\lambda }{G_{\ell }^{\prime }(q_{u}^{a},x^{a})}-\frac{%
G_{i}(q_{\ell }^{b},x^{b})+1(T_{i}(x^{b})=0)-\lambda }{G_{\ell }^{\prime
}(q_{\ell }^{b},x^{b})}.
\end{equation*}%
It follows similarly that%
\begin{equation*}
\sqrt{n}(\hat{\Delta}_{\ell }-\Delta _{\ell })=  - \sum_{i=1}^{n}\frac{\Psi
_{\lambda i}^{\ell }}{\sqrt{n}}+o_{p}(1),\Psi _{\lambda i}^{\ell }=  \frac{%
G_{i}(q_{\ell }^{a},x)+1(T_{i}(x^{a})=0)-\lambda }{G_{\ell }^{\prime
}(q_{\ell }^{a},x^{a})}-\frac{G_{i}(q_{u}^{b},x^{b})-\lambda }{G_{\ell
}^{\prime }(q_{u}^{b},x^{b})}.
\end{equation*}%
Then for $\Sigma _{\lambda }=Var(\Psi _{\lambda i}^{\ell },\Psi _{\lambda
i}^{u})$ the next conclusion follows by the CLT. It also follows by similar
arguments to the proof of Theorem A8 that $\sum_{i=1}^{n}\left( \hat{\Psi}%
_{\lambda i}^{\ell }-\Psi _{\lambda i}^{\ell }\right) ^{2}/n\overset{p}{%
\longrightarrow }0$ and $\sum_{i=1}^{n}\left( \hat{\Psi}_{\lambda
i}^{u}-\Psi _{\lambda i}^{u}\right) ^{2}/n\overset{p}{\longrightarrow }0.$
The consistency of $\hat{\Sigma}_{\lambda }$ then follows by standard
methods.

To show sharpness of the QTE bounds, define $\tilde{\alpha}_{i}$ and $g(x,%
\tilde{\alpha}_{i},\varepsilon _{it},C_{a},C_{b})$ as in the proof of the
ATE bounds, but now for any $C_{a},C_{b}\in \mathbb{R}.$ Let $%
G(y,x,C_{a},C_{b})=E[1(g(x,\tilde{\alpha}_{i},\varepsilon
_{it},C_{a},C_{b})\leq y)].$ Note that for $d\in \{a,b\},$%
\begin{equation*}
G(y,x^{d},C_{a},C_{b})=G_{\ell }(y,x^{d})+1(y\geq C_{d})\mathcal{\bar{P}}%
(x^{d}).
\end{equation*}%
Let $q(\lambda ,x,C_{a},C_{b})$ be the associated QSF. For $d\in \{a,b\},$%
\begin{equation*}
q(\lambda ,x^{d},C_{a},C_{b})=\left\{ 
\begin{array}{c}
q_{u}(\lambda ,x^{d}),\lambda <G_{\ell }(C_{d},x^{d}), \\ 
C_{d},G_{\ell }(C_{d},x^{d})\leq \lambda \leq G_{\ell }(C_{d},x^{d})+%
\mathcal{\bar{P}}(x^{d}), \\ 
q_{\ell }(\lambda ,x^{d}),\lambda >G_{\ell }(C_{d},x^{d})+\mathcal{\bar{P}}%
(x^{d}).%
\end{array}%
\right. 
\end{equation*}%
For $\lambda $ with $\mathcal{\bar{P}}(x^{d})<\lambda <1-\mathcal{\bar{P}}%
(x^{d})$ we have $q(\lambda ,x^{d},C_{a},C_{b})=q_{\ell }(\lambda ,x^{d})$
for $C_{d}$ small enough that $G_{\ell }(C_{d},x)+\mathcal{\bar{P}}%
(x^{d})<\lambda $ and $q(\lambda ,x^{d},C_{a},C_{b})=q_{u}(\lambda ,x^{d})$
for $C_{d}$ big enough. For $\lambda \leq \mathcal{\bar{P}}(x^{d})$ we have $%
q(\lambda ,x^{d},C_{a},C_{b})=q_{u}(\lambda ,x)$ for all $C_{d}$ big enough
(by $\lambda <1-\mathcal{\bar{P}}(x^{d})$) and $\lim_{C_{d}\longrightarrow
-\infty }q(\lambda ,x^{d},C_{a},C_{b})=-\infty =q_{\ell }(\lambda ,x).$ For $%
\lambda \geq 1-\mathcal{\bar{P}}(x^{d})$ we have $q(\lambda
,x^{d},C_{a},C_{b})=q_{\ell }(\lambda ,x^{d})$ for all $C_{d}$ small enough
and $\lim_{C_{d}\longrightarrow \infty }q(\lambda
,x^{d},C_{a},C_{b})=+\infty =q_{u }(\lambda ,x^{d}).$ Therefore, we have 
\begin{eqnarray*}
\lim_{C_{a}\longrightarrow -\infty ,C_{b}\longrightarrow +\infty }[q(\lambda
,x^{a},C_{a},C_{b})-q(\lambda ,x^{b},C_{a},C_{b})] &=&q_{\ell }(\lambda
,x^{a})-q_{u}(\lambda ,x^{b}), \\
\lim_{C_{a}\longrightarrow +\infty ,C_{b}\longrightarrow -\infty }[q(\lambda
,x^{a},C_{a},C_{b})-q(\lambda ,x^{b},C_{a},C_{b})] &=&q_{u}(\lambda
,x^{a})-q_{\ell }(\lambda ,x^{b}),
\end{eqnarray*}%
showing the bounds are sharp. \textit{Q.E.D.}

\bigskip

\subsection{Bounds under monotonicity}

We now turn to the bounds when $g_{0}$ is known to be monotonic, satisfying
the following condition.

\bigskip

\textsc{Assumption A1:}\textit{\ For some }$x^{a}$\textit{\ and }$x^{b},$%
\textit{\ }$g_{0}(x^{a},\alpha _{i},\varepsilon _{it})\geq
g_{0}(x^{b},\alpha _{i},\varepsilon _{it}).$

\bigskip

This condition leads to tighter bounds for the ASF and QSF. Here we will
give results showing estimable population bounds under monotonicity. We will
also briefly describe how to estimate them but for brevity do not give the
full asymptotic theory. Define $1_{i}^{a}=1(T_{i}(x^{a})>0),$ $%
1_{i}^{b}=1(T_{i}(x^{b})>0),$ $\mathcal{\bar{P}}(x^{b},x^{a})=\Pr
(T_{i}(x^{a})=T_{i}(x^{b})=0),$ and%
\begin{eqnarray*}
G_{u}^{\ast }(y,x^{a}) &=&E[G_{i}(y,x^{a})+(1-1_{i}^{a})G_{i}(y,x^{b})]+%
\mathcal{\bar{P}}(x^{b},x^{a}), \\
G_{\ell }^{\ast }(y,x^{b}) &=&E[G_{i}(y,x^{b})+(1-1_{i}^{b})G_{i}(y,x^{a})].
\end{eqnarray*}

\textsc{Theorem A9: }\textit{Suppose that Assumptions 1, 2, 5, and A1 are
satisfied. If }$E[|g_{0}(x,\alpha _{i},\varepsilon _{it})|]<\infty $\textit{%
\ for }$x\in \{x^{a},x^{b}\}$\textit{\ then }$\Delta \geq P\delta .$\textit{%
\ Also, if }$G_{u}^{\ast }(y,x^{a})$\textit{\ and }$G_{\ell }^{\ast
}(y,x^{b})$\textit{\ are continuous and strictly increasing on the interior
of their range then }$q(\lambda ,x^{a})\geq Q(\lambda ,G_{u}^{\ast }(\cdot
,x^{a}))$\textit{\ and }$q(\lambda ,x^{b})\leq Q(\lambda ,G_{\ell }^{\ast
}(\cdot ,x^{b})),$\textit{\ so that}%
\begin{equation*}
\Delta _{\lambda }\geq Q(\lambda ,G_{u}^{\ast }(\cdot ,x^{a}))-Q(\lambda
,G_{\ell }^{\ast }(\cdot ,x^{b})).
\end{equation*}

Proof: Note that $1=$ $%
1_{i}^{a}+(1-1_{i}^{a})1_{i}^{b}+(1-1_{i}^{a})(1-1_{i}^{b}).$ By Lemma A4, 
\begin{equation*}
E[1_{i}^{a}g_{i1}(x^{a})]=E[\bar{Y}_{i}(x^{a})],E[1_{i}^{b}g_{i1}(x^{b})]=E[%
\bar{Y}_{i}(x^{b})].
\end{equation*}%
Then by monotonicity
\begin{eqnarray*}
\mu (x^{a}) &=&E[g_{i1}(x^{a})]\geq
E[\{1_{i}^{a}+(1-1_{i}^{a})(1-1_{i}^{b})%
\}g_{i1}(x^{a})]+E[(1-1_{i}^{a})1_{i}^{b}g_{i1}(x^{b})] \\
&=&E[1_{i}^{a}\bar{Y}_{i}(x^{a})+(1-1_{i}^{a})1_{i}^{b}\bar{Y}%
_{i}(x^{b})+(1-1_{i}^{a})(1-1_{i}^{b})g_{i1}(x^{a})].
\end{eqnarray*}%
Similarly %
\begin{equation*}
\mu (x^{b})\leq E[1_{i}^{b}\bar{Y}_{i}(x^{b})+(1-1_{i}^{b})1_{i}^{a}\bar{Y}%
_{i}(x^{a})+(1-1_{i}^{a})(1-1_{i}^{b})g_{i1}(x^{b})].
\end{equation*}%
Subtracting this inequality from the previous one, and noting that $%
1_{i}^{a}-(1-1_{i}^{b})1_{i}^{a}=1_{i}^{b}1_{i}^{a}=D_{i}$ and $-1_{i}^{b}+$ 
$(1-1_{i}^{a})1_{i}^{b}=-D_{i},$ %
\begin{eqnarray*}
\mu (x^{a})-\mu (x^{b}) &\geq &E[D_{i}\left\{ \bar{Y}_{i}(x^{a})-\bar{Y}%
_{i}(x^{b})\right\} ]+E[(1-1_{i}^{a})(1-1_{i}^{b})\{g_{0}(x^{a},\alpha
_{i},\varepsilon _{it})-g_{0}(x^{b},\alpha _{i},\varepsilon _{it})\}] \\
&\geq &E[D_{i}\left\{ \bar{Y}_{i}(x^{a})-\bar{Y}_{i}(x^{b})\right\}
]=P\delta ,
\end{eqnarray*}%
giving the first conclusion.

Next, similarly to above,%
\begin{eqnarray*}
G(y,x^{a})
&=&E[\{1_{i}^{a}+(1-1_{i}^{a})(1-1_{i}^{b})+(1-1_{i}^{a})1_{i}^{b}%
\}1(g_{i1}(x^{a})\leq y)] \\
&\leq &E[G_{i}(y,x^{a})]+E[(1-1_{i}^{a})G_{i}(y,x^{b})]+\mathcal{\bar{P}}%
(x^{b},x^{a})=G_{u}^{\ast }(y,x^{a}). \\
G(y,x^{b}) &\geq &G_{\ell }^{\ast }(y,x^{b}).
\end{eqnarray*}%
Inverting gives the second conclusion. \textit{Q.E.D.}

\bigskip

Estimation of the bounds under monotonicity is straightforward. We can
estimate the lower bound for the ATE by $\left( \sum_{i=1}^{n}D_{i}/n\right) 
\hat{\delta}$. We can estimate the quantile bounds by inverting%
\begin{eqnarray*}
\hat{G}_{u}^{\ast }(y,x^{a}) &=&\sum_{i=1}^{n}[\bar{G}%
_{i}(y,x^{a})+(1-1_{i}^{a})\bar{G}%
_{i}(y,x^{b})+1(T_{i}(x^{b})=T_{i}(x^{a})=0)]/n, \\
\hat{G}_{\ell }^{\ast }(y,x^{b}) &=&\sum_{i=1}^{n}[\bar{G}%
_{i}(y,x^{b})+(1-1_{i}^{b})\bar{G}_{i}(y,x^{a})]/n.
\end{eqnarray*}%
Asymptotic theory for these estimators of bounds under monotonicity is
straightforward. We do not know if they are sharp.

\section{Supplements to Section 5}

Here we give the proof of Theorem 3 as well as bounds that impose
monotonicity.

% Here  is the proof of Theorem 3.

%\bigskip

%\textsc{Proof of Theorem 3: }

\subsection{Proof of Theorem 3}

We first prove the second part of Lemma A4 for the dynamic model. Let $%
d_{it}(x)=1(X_{i}\in \mathcal{X}_{t}(x)).$ By Assumption 3, $%
\sum_{t=1}^{T}d_{it}(x)=1(T_{i}(x)>0),$ and the fact that $d_{it}(x)$
depends only on $X_{it},X_{i,t-1},...,X_{i1}$ we have%
\begin{eqnarray*}
E[\hat{Y}_{i}(x)|X_{i1}]
&=&\sum_{t=1}^{T}E[d_{it}(x)Y_{it}|X_{i1}]=%
\sum_{t=1}^{T}E[d_{it}(x)E[g_{it}(x)|X_{it},...,X_{i1}]|X_{i1}] \\
&=&%
\sum_{t=1}^{T}E[d_{it}(x)E[g_{i1}(x)|X_{i1}]|X_{i1}]=E[1(T_{i}(x)>0)|X_{i1}]E[g_{i1}(x)|X_{i1}].
\end{eqnarray*}%
Let 
\begin{equation*}
\left( 
\begin{array}{c}
m_{\ell i} \\ 
m_{ui}%
\end{array}%
\right) =\left( 
\begin{array}{c}
\hat{Y}_{i}(x^{a})-\hat{Y}_{i}(x^{b})+B_{\ell
}1(T_{i}(x^{a})=0)-B_{u}1(T_{i}(x^{b})=0) \\ 
\hat{Y}_{i}(x^{a})-\hat{Y}_{i}(x^{b})+B_{u}1(T_{i}(x^{a})=0)-B_{\ell
}1(T_{i}(x^{b})=0)%
\end{array}%
\right) .
\end{equation*}%
Note that $\hat{\Delta}_{\ell }=\sum_{i=1}^{n}m_{\ell i }/n$ and $\hat{\Delta%
}_{u}=\sum_{i=1}^{n}m_{ui }/n.$Then for $\Sigma =Var((m_{\ell i },m_{ui})),$ 
$\Delta _{\ell }=E[m_{\ell i }],$ and $\Delta _{u}=E[m_{ui }]$ the first and
second conclusions follow by standard arguments for a vector of sample means.

Next, note that $E[g_{i1}(x^{a})|X_{i1}]\leq B_{u}$ by Assumption 6, so that%
\begin{equation*}
E[B_{u}1(T_{i}(x^{a})=0)|X_{i1}]\geq
E[1(T_{i}(x^{a})=0)|X_{i1}]E[g_{i1}(x^{a})|X_{i1}].
\end{equation*}%
Then by iterated expectations and $T_{i}(x^{a})\geq 0$,%
\begin{eqnarray*}
E[\hat{Y}_{i}(x^{a})+B_{u}1(T_{i}(x^{a}) =0)|X_{i1}]&\geq&
E[1(T_{i}(x^{a})>0)|X_{i1}]E[g_{i1}(x^{a})|X_{i1}] \\
& & +E[1(T_{i}(x^{a})
=0)|X_{i1}]E[g_{i1}(x^{a})|X_{i1}]=E[g_{i1}(x^{a})|X_{i1}].
\end{eqnarray*}%
Taking expectations of both sides of this inequality gives 
\begin{equation*}
E[\hat{Y}_{i}(x^{a})+B_{u}1(T_{i}(x^{a})=0)]\geq \mu (x^{a}).
\end{equation*}%
Similarly we have $E[\hat{Y}_{i}(x^{a})+B_{\ell }1(T_{i}(x^{a})=0)]\leq \mu
(x^{a}).$ Replacing $x^{a}$ by $x^{b}$ and differencing gives $\Delta _{\ell
}\leq \Delta \leq \Delta _{u}.$

Turning to the quantile bounds, we next prove the first part of Lemma A4 for
a dynamic model. Let $G_{i}(y,x)$ here, in the dynamic case, be given by%
\begin{eqnarray*}
G_{i}(y,x) &=&\sum_{t=1}^{T}d_{it}(x)1(Y_{it}\leq
y)=\sum_{t=1}^{T}d_{it}(x)1(g_{it}(x)\leq y), \\
G_{\ell }(y,x) &=&E[E[1(T_{i}(x)>0)|X_{i1}]1(g_{i1}(x)\leq y)].
\end{eqnarray*}%
\newline
Note that since $\sum_{t=1}^{T}d_{it}(x)=1(T_{i}(x)>0)$ and $d_{it}(x)$
depends only on $X_{it},X_{it-1},...,X_{i1}$, Assumption 3 implies 
\begin{eqnarray*}
E[G_{i}(y,x)] &=&E[\sum_{t=1}^{T}d_{it}(x)1(g_{it}(x)\leq
y)]=E[\sum_{t=1}^{T}d_{it}(x)E[1(g_{it}(x)\leq y)|X_{it},...,X_{i1}]] \\
&=&E[\sum_{t=1}^{T}d_{it}(x)E[1(g_{i1}(x)\leq
y)|X_{i1}]]=E[1(T_{i}(x)>0)E[1(g_{i1}(x)\leq y)|X_{i1}]] \\
&=&G_{\ell }(y,x).
\end{eqnarray*}%
Also, since $d_{it}(x)d_{is}(x)=0$ for any $s\neq t$ and $%
d_{it}(x)^{2}=d_{it}(x),$ Assumption 3 implies that%
\begin{eqnarray*}
E[\{\hat{G}_{i}(y,x)-G_{i}(y,x)\}^{2}] &=&E[\sum_{t=1}^{T}d_{it}(x)\{\Phi (%
\frac{y-g_{it}(x)}{h})-1(g_{it}(x)\leq y)\}^{2}] \\
&\leq &E[\sum_{t=1}^{T}d_{it}(x)E[\{\Phi (\frac{y-g_{it}(x)}{h}%
)-1(g_{it}(x)\leq y)\}^{2}|X_{it},...,X_{i1}]] \\
&=&E[1(T_{i}(x)>0)E[\{\Phi (\frac{y-g_{i1}(x)}{h})-1(g_{i1}(x)\leq
y)\}^{2}|X_{i1}]] \\
&=&E[E[1(T_{i}(x)>0)|X_{i1}]\{\Phi (\frac{y-g_{i1}(x)}{h})-1(g_{i1}(x)\leq
y)\}^{2}]
\end{eqnarray*}%
By Assumption 7 with $X_{i1}$ replacing $X_i$ it follows that $g_{i1}(x)$ is
continuously distributed for the probability measure weighted by $%
E[1(T_{i}(x)>0)|X_{i1}]$. Therefore it follows similarly to the proof of
Lemma A7 that $E[\{\hat{G}_{i}(y,x)-G_{i}(y,x)\}^{2}]\longrightarrow 0$ as $%
h\longrightarrow 0$. It also follows similarly to the proof of Lemma A7 
\begin{equation*}
E[\hat{G}_{i}(y,x)]=E[G_{i}(y,x)]+O(h^{2}).
\end{equation*}%
The conclusion now follows exactly like the proof of Theorem 2. \textit{%
Q.E.D.}

\subsection{Bounds under monotonicity}

We now turn to the bounds when $g_{0}$ is known to be monotonic, satisfying
Assumption A1, in the dynamic model. This condition leads to tighter bounds
for the ASF and QSF. Here we will give results showing estimable population
bounds under monotonicity. We will also briefly describe how to estimate
them but for brevity do not give the full asymptotic theory. For $d \in
\{a,b\},$ define $1_{it}^{d}=1(X_{i} \in \mathcal{X}_t(x^d)),$ $t = 1, ..., T
$, $\bar 1_i^d = 1(X_{i} \in \bar{\mathcal{X}}(x^d)),$ and $\tilde 1_{iT}^d
= 1(X_{iT} = x^d).$  Let 
\begin{eqnarray*}
G_{u}^{\ast }(y,x^{a}) &=&E[G_{i}(y,x^{a})+ \bar 1_{i}^{a} \{ \tilde
1_{iT}^b 1(Y_{iT} \leq y) + (1 - \tilde 1_{iT}^b) \}], \\
G_{\ell }^{\ast }(y,x^{b}) &=&E[G_{i}(y,x^{b})+ \bar 1_{i}^{b} \tilde
1_{iT}^a 1(Y_{iT} \leq y)].
\end{eqnarray*}

\textsc{Theorem A10: }\textit{Suppose that Assumptions 1, 3, 5, and A1 are
satisfied. If }$E[|g_{0}(x,\alpha _{i},\varepsilon _{it})|]<\infty $\textit{%
\ for }$x\in \{x^{a},x^{b}\}$\textit{\ then } 
\begin{equation*}
\Delta \geq E[\hat Y_{i}(x^a) - \hat Y_{i}(x^b)] + E[\bar 1_i^a (\tilde
1_{iT}^bY_{iT} + (1 - \tilde 1_{iT}^b)B_{\ell})] - E[\bar 1_i^b (\tilde
1_{iT}^aY_{iT} + (1 - \tilde 1_{iT}^a)B_{u})].
\end{equation*}
\textit{\ Also, if }$G_{u}^{\ast }(y,x^{a})$\textit{\ and }$G_{\ell }^{\ast
}(y,x^{b})$\textit{\ are continuous and strictly increasing on the interior
of their range then }$q(\lambda ,x^{a})\geq Q(\lambda ,G_{u}^{\ast }(\cdot
,x^{a}))$\textit{\ and }$q(\lambda ,x^{b})\leq Q(\lambda ,G_{\ell }^{\ast
}(\cdot ,x^{b})),$\textit{\ so that}%
\begin{equation*}
\Delta _{\lambda }\geq Q(\lambda ,G_{u}^{\ast }(\cdot ,x^{a}))-Q(\lambda
,G_{\ell }^{\ast }(\cdot ,x^{b})).
\end{equation*}

Proof: Note that $1=$ $\sum_{t=1}^T 1_{it}^{a} + \bar 1_{i}^{a} \tilde
1_{iT}^{b} + \bar 1_{i}^{a} (1- \tilde 1_{iT}^{b}).$ By Lemma A4, 
\begin{equation*}
\sum_{t=1}^T E[1_{it}^{a} g_{it}(x^{a})]=E[\hat{Y}_{i}(x^{a})], \sum_{t=1}^T
E[1_{it}^{b}g_{it}(x^{b})]=E[\hat{Y}_{i}(x^{b})].
\end{equation*}%
Then by Assumption 3, monotonicity and $g_{iT}(x^{a}) \geq B_{\ell}$, 
\begin{eqnarray*}
\mu (x^{a}) &=& \sum_{t=1}^T E[1_{it}^{a} g_{it}(x^{a})] + E[\bar 1_{i}^{a}
g_{iT}(x^{a})] \\
&\geq& \sum_{t=1}^T E[1_{it}^{a} g_{it}(x^{a})] +E[\bar 1_{i}^{a} \tilde
1_{iT}^{b} g_{iT}(x^{b})] + E[\bar 1_{i}^{a} (1-\tilde 1_{iT}^{b}) ]B_{\ell}
\\
&=& E[\hat{Y}_{i}(x^{a})]+ E[\bar 1_{i}^{a} \tilde 1_{iT}^{b} Y_{iT}] +
E[\bar 1_{i}^{a}(1-\tilde 1_{iT}^{b})]B_{\ell}.
\end{eqnarray*}%
Similarly we have%
\begin{equation*}
\mu (x^{b})\leq E[\hat{Y}_{i}(x^{b})]+ E[ \bar 1_{i}^{b} \tilde 1_{iT}^{a}
Y_{iT}] + E[\bar 1_{i}^{b} (1-\tilde 1_{iT}^{a})]B_u.
\end{equation*}%
Subtracting this inequality from the previous one gives the first conclusion.

Next, similarly to above,%
\begin{eqnarray*}
G(y,x^{a}) &=& \sum_{t=1}^T E[1_{it}^{a} 1(g_{it}(x^{a})\leq y)] + E[\bar
1_{i}^{a} 1(g_{iT}(x^{a})\leq y)] \\
&\leq &E[G_{i}(y,x^{a})]+E[\bar 1_{i}^{a} \tilde 1_{iT}^b 1(Y_{iT} \leq
y)]+E[\bar 1_{i}^{a} (1-\tilde 1_{iT}^b)] =G_{u}^{\ast }(y,x^{a}). \\
G(y,x^{b}) &\geq &G_{\ell }^{\ast }(y,x^{b}).
\end{eqnarray*}%
Inverting gives the second conclusion. \textit{Q.E.D.}

\bigskip

If $X_{it} \in \{0,1\},$ $x^b = 0$, and $x^a = 1,$ then $\bar 1_{i}^{b}
(1-\tilde 1_{iT}^{a}) = \bar 1_{i}^{a} (1-\tilde 1_{iT}^{b}) = 0$ and the
lower bound for $\Delta$ does not depend on $B_{\ell}$ and $B_u.$

Estimation of the bounds under monotonicity is straightforward. We can
estimate the lower bound for the ATE by 
\begin{equation*}
\sum_{i=1}^{n}[\hat{Y}_{i}(x^{a})-\hat{Y}_{i}(x^{b})+\bar{1}_{i}^{a}(\tilde{1%
}_{iT}^{b}Y_{iT}+(1-\tilde{1}_{iT}^{b})B_{\ell })-\bar{1}_{i}^{b}(\tilde{1}%
_{iT}^{a}Y_{iT}+(1-\tilde{1}_{iT}^{a})B_{u})]/n.
\end{equation*}%
We can estimate the quantile bounds by inverting%
\begin{eqnarray*}
\hat{G}_{u}^{\ast }(y,x^{a}) &=&\sum_{i=1}^{n}[\hat{G}_{i}(y,x^{a})+\bar{1}%
_{i}^{a}\{\tilde{1}_{iT}^{b}1(Y_{iT}\leq y)+(1-\tilde{1}_{iT}^{a})\}]/n, \\
\hat{G}_{\ell }^{\ast }(y,x^{b}) &=&\sum_{i=1}^{n}[\hat{G}_{i}(y,x^{b})+\bar{%
1}_{i}^{b}\tilde{1}_{iT}^{a}1(Y_{iT}\leq y)]/n.
\end{eqnarray*}%
Asymptotic theory for these estimators of bounds under monotonicity is
straightforward. We do not know if they are sharp.

\section{Supplements to Section 6}

In addition to the proofs of the rate results of Section 6, we here give
necessary and sufficient conditions for identification as $T\longrightarrow
\infty $ and extend the identification and rate results to the QTE.

\subsection{Identification as $T \to \infty$}

We begin with the identification result. The necessary and sufficient
condition for identification of $\Delta $ as $T$ grows is

\bigskip

\textsc{Assumption A2:} $\Pr (\Pr \left( X_{it}=x|\alpha _{i}\right) >0)=1$ 
\textit{for }$x\in \{x^{a},x^{b}\}$ \textit{ and some } $t \in \{1, \ldots, T\}$.

\bigskip

If this condition does not hold for both $x^{b}$ and $x^{a}$ then some
individuals, as represented by $\alpha _{i}$, will never reach either $x^{b}$
or $x^{a}$, so we cannot nonparametrically identify the treatment effect for
those individuals, and hence the overall treatment effect is not identified.
A related condition was formulated in Chamberlain (1982, p. 17) but was used
for a different purpose, as a sufficient condition for a least squares
estimate for a single individual to converge to that individual's
coefficient.

The following result shows the key role of Assumption A2 in achieving
identification as $T\longrightarrow \infty $.

\bigskip

\textsc{Theorem A11: }\textit{Suppose that Assumptions 1 and 5 are
satisfied. If Assumption A2 is not satisfied then }$\mathcal{\bar{P}}(x)$%
\textit{\ is bounded away from zero uniformly in }$T$ \textit{for }$x=x^{a}$ 
\textit{or }$x=x^{b},$\textit{\ so that if Assumption 6 is satisfied, }$%
\Delta _{u}-\Delta _{\ell }$\textit{\ does not converge to zero as }$T$ 
\textit{grows. Suppose also that }$(X_{i1},X_{i2},...)$\textit{\ is
stationary and ergodic conditional on }$\alpha _{i}.$ \textit{If Assumptions
2 and A2 are satisfied and }$E[|g_{0}(x,\alpha _{i},\varepsilon
_{i1})|]<\infty $ \textit{for }$x=x^{a}$ \textit{and }$x=x^{b},$ \textit{%
then \ }$\delta \longrightarrow $ $\Delta $\textit{\ as }$T\longrightarrow
\infty $.\textit{\ If Assumptions 3, 6, and A2 are satisfied then }$\Delta
_{u}-\Delta _{\ell }\longrightarrow 0$\textit{\ as }$T\longrightarrow \infty 
$\textit{.}

\bigskip

Proof: First, note that if Assumption A2 is not satisfied then for some $%
x^{d}\in \{x^{a},x^{b}\}$ there is a set $\mathcal{A}$ with $\Pr (\mathcal{A)%
}>0$ such that $\Pr (X_{it}=x^{d}|\alpha _{i})=0$ for all $t$ and $\alpha
_{i}\in \mathcal{A}$. Then 
\begin{equation*}
E[T_{i}(x^{d})|\alpha _{i} \in \mathcal{A}]=\sum_{t=1}^{T}E[1(X_{it}=x^{d})|%
\alpha _{i}\in \mathcal{A}]=0.
\end{equation*}%
Since $T_{i}(x^{d})$ is a nonnegative random variable, this implies that $%
\Pr (T_{i}(x^{d})=0|\alpha _{i})=1$ for all $T$ and $\alpha _{i}\in \mathcal{%
A}$. Therefore%
\begin{equation*}
\mathcal{\bar{P}}(x^{d})=E[\Pr (T_{i}(x^{d})=0|\alpha _{i})]\geq E[1(%
\mathcal{A})\Pr (T_{i}(x^{d})=0|\alpha _{i})]=\Pr (\mathcal{A)}>0.
\end{equation*}%
Thus $\mathcal{\bar{P}}(x^{d})$ is bounded away from zero for all $T,$ and
hence under Assumption 6, $(B_{u}-B_{\ell })[\mathcal{\bar{P}}(x^{a})+%
\mathcal{\bar{P}}(x^{b})]\geq (B_{u}-B_{\ell })\mathcal{\bar{P}}(x^{d})$
does not converge to zero.

Next suppose that Assumptions 2 and A2 are satisfied, $(X_{i1},X_{i2},...)$ is
stationary and ergodic conditional on $\alpha _{i},$ and that $x\in
\{x^{a},x^{b}\}$. Recall that $T_{i}(x)=\sum_{t=1}^{T}1(X_{it}=x).$\ By the
ergodic theorem, there is a set of $\alpha _{i}$ having probability one such
that%
\begin{equation*}
T_{i}(x)/T\overset{a.s.}{\longrightarrow }E[1(X_{it}=x)|\alpha _{i}]=\Pr
(X_{it}=x\mid \alpha _{i}).
\end{equation*}%
Under Assumption A2 $\Pr (X_{it}=x\mid \alpha _{i})>0$ on a set of $\alpha
_{i}$ with probability one (a.s. $\alpha _{i}$ henceforth). Therefore $%
1(T_{i}(x)>0)\overset{a.s.}{\longrightarrow }1$ a.s. $\alpha _{i}$. Since
this holds for both $x^{a}$ and $x^{b},$ it follows that 
\begin{equation*}
D_{i}=1(T_{i}(x^{a})>0)1(T_{i}(x^{b})>0)\overset{a.s.}{\longrightarrow }1
\end{equation*}%
a.s. $\alpha _{i}$. Let $\Delta _{i}=g_{i1}(x^{a})-g_{i1}(x^{b})$. Note that 
$|D_{i}\Delta _{i}|\leq |\Delta _{i}|$ and $E[|\Delta _{i}||\alpha
_{i}]<\infty $ a.s. $\alpha _{i}$. Then by the dominated convergence theorem
(DCT henceforth), 
\begin{equation*}
E[D_{i}\Delta _{i}|\alpha _{i}]\longrightarrow E[\Delta _{i}|\alpha
_{i}],E[D_{i}|\alpha _{i}]\longrightarrow 1\text{ a.s. }\alpha _{i}\text{.}
\end{equation*}%
Then by the applying the DCT again, 
\begin{equation*}
E[D_{i}\Delta _{i}]\longrightarrow E[\Delta _{i}]=\Delta
,E[D_{i}]\longrightarrow 1,
\end{equation*}%
giving the first conclusion.

Suppose next that Assumptions 3 and 6 are satisfied, and $(X_{i1},X_{i2},...)$ is
stationary and ergodic conditional on $\alpha _{i}.$ Recall that $\Delta
_{u}-\Delta _{\ell }=(B_{u}-B_{\ell })[\mathcal{\bar{P}}(x^{a})+\mathcal{%
\bar{P}}(x^{b})]$. If Assumption A2 is satisfied then since $%
1(T_{i}(x^{a})>0)\geq D_{i}$ we have 
\begin{equation*}
\mathcal{\bar{P}}(x^{a})=E[1(T_{i}(x^{a})=0)]\leq 1-E[D_{i}]\longrightarrow 0%
\text{ }
\end{equation*}%
Similarly we have $\mathcal{\bar{P}}(x^{b})\longrightarrow 0$ so the second
conclusion holds. Q.E.D.

\bigskip

%Next, we will prove the results given in Section 6.

%\bigskip

%\textsc{Proof of Theorem 4:} 

\subsection{Proof of Theorem 4}

Let $\Pi _{t=1}^{T}1(X_{it}\neq x)$ be the indicator function for the event
that none of the elements of $X_{i}$ is equal to $x$ so that $\mathcal{\bar{P%
}}(x)=E[\Pi _{t=1}^{T}1(X_{it}\neq x)].$ By iterated expectations, for $T>J$%
, 
\begin{eqnarray*}
\mathcal{\bar{P}}(x) &=&E[\Pi _{t=1}^{T-1}1(X_{it}\neq x)E[1(X_{iT}\neq
x)|X_{i,T-1},...,X_{i1},\alpha _{i}]] \\
&=&E[\{\Pi _{t=1}^{T-1}1(X_{it}\neq x)\}\Pr (X_{iT}\neq
x|X_{i,T-1},...,X_{i,T-J},\alpha _{i})]\leq (1-\varepsilon )E[\Pi
_{t=1}^{T-1}1(X_{it}\neq x)].
\end{eqnarray*}%
Repeating the argument for $T-1,...,J$ gives%
\begin{equation*}
\mathcal{\bar{P}}(x)\leq (1-\varepsilon )^{T-J}E[\Pi
_{t=1}^{J-1}1(X_{it}\neq x)]\leq (1-\varepsilon )^{T-J},
\end{equation*}%
giving the first conclusion.

For the second conclusion, note that the conditional i.i.d. assumption and
the bound implies that for $P_{i}=\Pr (X_{it}\neq x|\alpha _{i})$ we have $%
\mathcal{\bar{P}}(x)=E[P_{i}^{T}]$ being no greater than a constant times
the $T^{th}$ raw moment of a Beta distribution with parameters $\gamma $ and 
$v.$ Also, it is well known that $T^{v}\Gamma (T+\gamma )/\Gamma (T+\gamma
+v)\longrightarrow 1$ as $T\longrightarrow \infty $. Therefore, we have%
\begin{eqnarray*}
E[P_{i}^{T}] &\leq &C[\Gamma (\gamma +v)/\Gamma (\gamma )\Gamma
(v)]\int_{0}^{1}p^{T+\gamma -1}(1-p)^{v-1}dp \\
&\leq &C[\Gamma (\gamma +v)/\Gamma (\gamma )\Gamma (v)][\Gamma (T+\gamma
)\Gamma (v)/\Gamma (T+\gamma +v)] \\
&=&C\Gamma (T+\gamma )/\Gamma (T+\gamma +v)\leq CT^{-v}.\text{ \ }Q.E.D.
\end{eqnarray*}

\bigskip

%\textsc{Proof of Theorem 5:} 

\subsection{Proof of Theorem 5}

Note that $\Pr (Y_{it}=0|Y_{i,t-1}=0,\alpha _{i})=1-H(\alpha _{1i})$%
\begin{eqnarray*}
\mathcal{\bar{P}}(1) &=&E[\Pr (Y_{i,T-1}=Y_{i,T-2}=...=Y_{i0}=0|\alpha _{i})]
\\
&=&E[\Pi _{t=1}^{T-1}\Pr (Y_{it}=0|Y_{i,t-1}=0,\alpha _{i})\Pr
(Y_{i0}=0|\alpha _{i})] \\
&\leq &E[\{1-H(\alpha _{i1})\}^{T-1}].
\end{eqnarray*}%
By a change of variables we find that the pdf $f(p)$ of $1-H(\alpha _{i1})$
is 
\begin{equation*}
f(p)=f_{1}(H^{-1}(1-p))/f_{\varepsilon }(H^{-1}(1-p))\leq
C(1-p)^{v-1}p^{v-1}.
\end{equation*}%
Thus, the pdf of $1-H(\alpha _{i1})$ is bounded above by a Beta pdf with
parameters $v,v$. It then follows as in the proof of Theorem 4 that $%
\mathcal{\bar{P}}(1)\leq C(T-1)^{-v}\leq CT^{-v}.$ It follows similarly that 
$\mathcal{\bar{P}}(0)\leq CT^{-v}$. \textit{Q.E.D.}

\bigskip

\subsection{Identification rates for QTE}

Finally, we show that the nonparametric rates and nonidentification results
apply to the QTE. We do this by giving Lemmas for quantile bounds that apply
to both static and dynamic models. The first Lemma shows that the
identification rate is at least as fast as the rate at which $\mathcal{\bar{P%
}}(x)$ decreases.

\bigskip

\textsc{Lemma A12:} \textit{Suppose that }$G(y)$\textit{\ is a CDF that is
strictly increasing and continuously differentiable on }$\{y:0<G(y)<1\}$%
\textit{\ and that }$G_{T}(y)$\textit{\ is a continuous function and }$\mathcal{\bar{P%
}}_{T}$\textit{\ a nonnegative constant satisfying}%
\begin{equation*}
G_{T}(y)\leq G(y)\leq G_{T}(y)+\mathcal{\bar{P}}_{T},G_{T}(-\infty
)=0,G_{T}(\infty )+\mathcal{\bar{P}}_{T}=1.
\end{equation*}%
\textit{If }$\mathcal{\bar{P}}_{T}\longrightarrow 0$\textit{\ as }$T\longrightarrow
\infty $\textit{\ then for }$0<\lambda <1$\textit{\ and large enough }$T$%
\textit{\ there are }$q_{\ell T}\leq q\leq q_{uT}$\textit{\ satisfying} 
\begin{equation*}
\lambda =G_{T}(q_{uT})=G(q)=G_{T}(q_{\ell T})+\mathcal{\bar{P}}_{T}.
\end{equation*}%
\textit{Also, any such }$q_{uT}$\textit{\ and }$q_{\ell T}$\textit{\ satisfy:%
}  $q_{uT}-q_{\ell T}=O(\mathcal{\bar{P}}_{T})$\textit{.}

\bigskip

Proof: Choose $T$ large enough that $\mathcal{\bar{P}}_{T}<\min (\lambda
,1-\lambda).$ Then $G_{T}(\infty )=1-\mathcal{\bar{P}}_{T}>\lambda $ and $%
G_{T}(-\infty )+\mathcal{\bar{P}}_{T}=\mathcal{\bar{P}}_{T}<\lambda .$
Therefore by continuity of $G_{T}(y)$ there exist $q_{uT}$ such that $%
\lambda =G_{T}(q_{uT})$ and $q_{\ell T}$ such that $\lambda =G_{T}(q_{\ell
T})+\mathcal{\bar{P}}_{T}$. Also, by $G(y)$ being a strictly increasing CDF
there is a unique $q$ with $\lambda =G(q).$ Note $G(q)=G_{T}(q_{uT})\leq
G(q_{uT})$ so that $q_{uT}\geq q$ by $G(q)$ strictly monotonic. It follow
similarly that $q_{\ell T}\leq q.$ Also, for any $\varepsilon >0$ we have $%
G(q-\varepsilon )<G(q)$, so that for large enough $T$ it follow 
\begin{equation*}
G(q-\varepsilon )<G(q)-\mathcal{\bar{P}}_{T}=G_{T}(q_{\ell T})\leq G(q_{\ell
T}).
\end{equation*}%
By strict monotonicity of $G(q)$ it follows that $q_{\ell T}>q-\varepsilon $
for large enough $T$. Since $\varepsilon $ is arbitrary we have $q_{\ell
T}\longrightarrow q$. It follow similarly that $q_{uT}\longrightarrow q$.

Next, choose $\varepsilon $ small enough that $\partial G(\tilde{q}%
)/\partial q\geq C>0$ for $\tilde{q}\in \mathcal{I=}[q-\varepsilon
,q+\varepsilon ]$. Note that for $T$ large enough, $q_{\ell T},q_{uT}\in 
\mathcal{I}$. Also we have%
\begin{equation*}
G(q_{\ell T})+2\mathcal{\bar{P}}_{T}\geq G_{T}(q_{\ell T})+2\mathcal{\bar{P}}%
_{T}=G(q)+\mathcal{\bar{P}}_{T}=G_{T}(q_{uT})+\mathcal{\bar{P}}_{T}\geq
G(q_{uT}).
\end{equation*}%
Subtracting $G(q_{\ell T})$ from both sides and expanding gives 
\begin{equation*}
2\mathcal{\bar{P}}_{T}\geq G(q_{uT})-G(q_{\ell T})=\frac{\partial G(\bar{q}%
_{T})}{\partial q}(q_{uT}-q_{\ell T})\geq C(q_{uT}-q_{\ell T}).
\end{equation*}%
Dividing through by $C$ gives $q_{uT}-q_{\ell T}\leq C\mathcal{\bar{P}}_{T},$
implying the conclusion. \textit{Q.E.D.}

\bigskip

The next result gives conditions under which the identification rate is no
faster than the rate at which $\mathcal{\bar{P}}(x)$ decreases. This result
will also show that quantile effects are not identified as $T\longrightarrow
\infty $ if $\mathcal{\bar{P}}(x)$ does not go to zero.

\bigskip

\textsc{Lemma A13:} \textit{If the conditions of Lemma A12 are satisfied and 
}$G_{T}(y)$\textit{\ is continuously differentiable with }$|\partial
G_{T}(y)/\partial y|\leq C$\textit{\ for all }$y$ \textit{and }$T$ \textit{%
then there is }$C$ such that for $\mathcal{\bar{P}}_{T}>0$, \textit{\ }%
\begin{equation*}
q_{uT}-q_{\ell T}\geq C\mathcal{\bar{P}}_{T}.
\end{equation*}%
Proof: As in the proof of Lemma A12 we have $G_{T}(q_{uT})=G_{T}(q_{\ell T})+%
\mathcal{\bar{P}}_{T}$. By the intermediate value theorem it follows that
for some $q_{\ell T}\leq \bar{q}\leq q_{uT}$%
\begin{equation*}
\frac{\partial G_{T}(\bar{q})}{\partial q}(q_{uT}-q_{\ell T})=\mathcal{\bar{P%
}}_{T}\text{.}
\end{equation*}%
For $\mathcal{\bar{P}}_{T}>0$ we must have $\partial G_{T}(\bar{q})/\partial
q\neq 0$, so that%
\begin{equation*}
q_{uT}-q_{\ell T}=\left[ \frac{\partial G_{T}(\bar{q})}{\partial q}\right]
^{-1}\mathcal{\bar{P}}_{T}\geq C^{-1}\mathcal{\bar{P}}_{T}\text{. \ }Q.E.D.
\end{equation*}

Taken together these two results show that the identification rate for the
QTE is the same as the rate at which $\mathcal{\bar{P}}(x)$ decreases.
Together they also show that if $\mathcal{\bar{P}}(x)$ does not go to zero
the bounds do not shrink to a point. It is straightforward to check that the
conditions of these results are satisfied.

\section{Supplements to Section 7}

We now turn to the results of Section 7 and to one additional result on the
consistency of non-linear fixed effects estimators for the identified ATE.

% We will first prove the identification rate result from Theorem 6.

%\bigskip 

%\textsc{Proof of Theorem 6:} 

\subsection{Proof of Theorem 6}

Consider first the static case where $X_{it}\in \{0,1\}.$ We show the result
for $X^{k}=(0,...,0)^{\prime }.$ The result for $X^{k}=(1,...,1)^{\prime }$
will follow similarly. Note that $\beta ^{\ast }$ is identified for logit so 
$B=\{\beta ^{\ast }\}$. Let $Z=H(\alpha )$ and let $G(z)$ be the CDF of $Z$
when $F\in \mathcal{F}_{k}=\mathcal{F}_{k}(\beta ^{\ast },\mathcal{P})$ is
the CDF of $\alpha $. By $(Y_{i1},...,Y_{iT})$ mutually independent
conditional on $\alpha $ we have%
\begin{equation*}
M_{t}=\Pr (Y_{it}=1,...,Y_{i1}=1|X_{i}\in X^{k})=\int H(\alpha
)^{t}dF(\alpha )=\int Z^{t}dG(Z),
\end{equation*}%
so that $M_{t}$ is identified for $t=1,...,T$. Now consider a $T^{th}$ order
polynomial $P(z,T)=b_{0}+b_{1}z+...+b_{T}z^{T}$ in $z.$ Note that%
\begin{equation*}
\int P(Z,T)dG(Z)=b_{0}+\sum_{t=1}^{T}b_{t}M_{t}
\end{equation*}%
does not depend on $F\in \mathcal{F}_{k}$. As a special case, $\int
ZdG(Z)=M_{1}$ also does not depend on $F\in \mathcal{F}_{k}.$ Define the
function $h(z)=$ $H(\beta ^{\ast }+H^{-1}(z))=$ $ze^{\beta ^{\ast
}}/1-(1-e^{\beta ^{\ast }})z.$ Note $\Delta ^{k}=\int [h(Z)-Z]dG(Z)$ for all 
$F\in \mathcal{F}_{k}.$ For any polynomial $P(z,t)$ let $R(z,t)=h(z)-P(z,t)$
be the remainder. Then we have%
\begin{eqnarray}
\Delta _{u}^{k}-\Delta _{\ell }^{k} &=&\sup_{F\in \mathcal{F}_{k}}\int
[h(Z)-Z]dG(Z)-\inf_{F\in \mathcal{F}_{k}}\int [h(Z)-Z]dG(Z)  \label{rem} \\
&=&\sup_{F\in \mathcal{F}_{k}}\int [P(Z,T)+R(Z,T)]dG(Z)-\inf_{F\in \mathcal{F%
}_{k}}\int [P(Z,T)+R(Z,T)]dG(Z)  \notag \\
&=&\sup_{F\in \mathcal{F}_{k}}\int R(Z,T)dG(Z)-\inf_{F\in \mathcal{F}%
_{k}}\int R(Z,T)dG(Z)\leq 2\sup_{0\leq z\leq 1}|R(z,T)|.  \notag
\end{eqnarray}%
The function $h(z)$ is continuously differentiable of order $r$ for every $r$
with 
\begin{equation*}
\left\vert \frac{d^{r}h(z)}{dz^{r}}\right\vert \leq r!e^{|\beta ^{\ast
}|}(e^{|\beta ^{\ast }|}-1|)^{r-1}.
\end{equation*}%
Then by Jackson's Theorem (e.g. Judd (1998) Chap. 3) there exists $P(z,T)$
such that for $\gamma =\pi (e^{|\beta ^{\ast }|}-1|)/4$ 
\begin{eqnarray*}
\sup_{0\leq z\leq 1}\left\vert R(z,T)\right\vert &\leq &\frac{(T-r)!}{T!}%
\left( \frac{\pi }{4}\right) ^{r}\sup_{0\leq z\leq 1}\left\vert \frac{%
d^{r}h(z)}{dz^{r}}\right\vert \\
&\leq &\frac{(T-r)!r!}{T!}\left( \frac{\pi }{4}\right) ^{r}e^{|\beta ^{\ast
}|}(e^{|\beta ^{\ast }|}-1|)^{r-1}\leq C\left( \frac{r\gamma }{T}\right)
^{r}.
\end{eqnarray*}%
This inequality continues to hold if $\gamma $ is replaced by $\max \{\gamma
,1\}$, so we can assume $\gamma >1.$ Then choose $r$ equal to $T/\gamma e$,
so that 
\begin{equation*}
\sup_{0\leq z\leq 1}\left\vert R(z,T)\right\vert \leq Ce^{-T/\gamma e}.
\end{equation*}%
The conclusion then follows by eq. (\ref{rem}).

Next consider the dynamic binary logit model where $X_{it}=Y_{i,t-1}.$ It is
known from Cox (1958) and Chamberlain (1985) that $\beta ^{\ast }$
identified for $T$ large enough. We show the result for $\Delta ^{1}$ where $%
\mathcal{X}^{1}=\{X_{i}:X_{i1}=0\}.$ The result for the ATE conditional on $%
X_{i1}=1$ will follow analogously. 
%Next, consider the dynamic model and $\Delta ^{1}$ for $X^{1}=\{X_{i}:X_{i1}=0\}.$
Then%
\begin{equation*}
\Pr (Y_{it}=0,...,Y_{i1}=0|X_{i1}=0)=\int [1-H(\alpha )]^{t}dF(\alpha )
\end{equation*}%
is identified for $t=1,...,T$. It follows by a standard argument that $%
M_{t}=\int H(\alpha )^{t}dF(\alpha )$ is identified for $t=1,...,T$. The
proof then proceeds exactly as for the static case. \ \textit{Q.E.D.}

\bigskip

\subsection{Consistency of fixed effects for identified ATE}

We now consider the fixed effects estimator in a binary choice model with a
binary regressor and $T=2.$ In some models fixed effect (FE) estimators of
the ATE appear to have small biases; e.g. see Hahn and Newey (2004) and
Fern\'andez-Val (2009). Here we show consistency of FE for $\delta $. To
describe this result, note that the FE estimator of the ASF conditional on $%
X_{i}=X^{k}$ is%
\begin{eqnarray*}
\hat{\mu}_{k}^{FE}(x) &=&\sum_{i=1}^{n}1(X_{i}=X^{k})H(x\hat{\beta}_{FE}+%
\hat{\alpha}_{i})/\sum_{i=1}^{n}1(X_{i}=X^{k}), \\
\hat{\beta}_{FE},\hat{\alpha}_{1},...,\hat{\alpha}_{n} &=&\arg \max_{\beta
,\alpha _{1},...,\alpha _{n}}\sum_{i,t}\ln \{H(X_{it}\beta +\alpha
_{i})^{Y_{it}}[1-H(X_{it}\beta +\alpha _{i})]^{1-Y_{it}}\}.
\end{eqnarray*}%
Let $\beta _{T}$ denote the limit of $\hat{\beta}_{FE}$. In the multinomial
choice model $\hat{\alpha}_{i}$ will have a limit distribution conditional
on $X_{i}=X^{k}$ that is discrete with $J$ support points $\alpha
_{j}^{k}(\beta _{T})$ and $\Pr (\alpha =\alpha _{j}^{k}(\beta _{T}))=$ $%
\mathcal{P}_{j}^{k}$, $(j=1,...,J)$. These limits will satisfy%
\begin{eqnarray}
\beta _{T} &=&\text{argmax}_{\beta }\sum_{k=1}^{K}\mathcal{P}%
^{k}\sum_{j=1}^{J}\mathcal{P}_{j}^{k}\log \mathcal{L}_{j}^{k}\left( \alpha
_{j}^{k}(\beta ),\beta \right) , \\
\alpha _{j}^{k}(\beta ) &=&\text{argmax}_{\alpha }\mathcal{L}_{j}^{k}\left(
\alpha ,\beta \right) ,(j=1,...,J;k=1,...,K),  \notag
\end{eqnarray}%
where $\mathcal{P}^{k}=E[1(X_{i}=X^{k})]$. The corresponding limit of $\hat{%
\mu}_{k}^{FE}(x)$ is then given by 
\begin{equation*}
\mu _{k}^{T}(x)=\sum_{j=1}^{J}\mathcal{P}_{j}^{k}H(x^{\prime }\beta
_{T}+\alpha _{j}^{k}(\beta _{T})).
\end{equation*}%
Note that with binary $X_{it}$ and $T=2$ we have $K=4.$ Let $X^{1}=(0,0)$, $%
X^{2}=(0,1)$, $X^{3}=(1,0)$, and $X^{4}=(1,1)$, so that the identified
effect equals $\delta =\sum_{k=2}^{3}\mathcal{P}^{k}\Delta
^{k}/\sum_{k=2}^{3}\mathcal{P}^{k}$.

\bigskip

\textsc{Theorem A14: }\textit{If} $H^{\prime }(x)>0,$ $H(-x)=1-H(x),$ $%
X_{it}\in \{0,1\}$, $T=2$ \textit{and }$\mathcal{P}_{2}+\mathcal{P}_{3}$ $>0$
\textit{then}%
\begin{equation*}
\sum_{k=2}^{3}\mathcal{P}^{k}[\mu _{k}^{T}(1)-\mu _{k}^{T}(0)]/\sum_{k=2}^{3}%
\mathcal{P}^{k}=\delta .
\end{equation*}

Proof: Let $Y^{1}=(0,0)^{\prime },Y^{2}=(0,1)^{\prime },Y^{3}=(1,0)^{\prime
},Y^{4}=(1,1)^{\prime }$ and $X^{1}=(0,0)^{\prime },$ $X^{2}=(0,1)^{\prime
}, $ $X^{3}=(1,0)^{\prime },$ $X^{4}=(1,1)^{\prime }$. The identified effect
is 
\begin{eqnarray*}
\delta &=&\left\{ \mathcal{P}^{2}E[Y_{i2}-Y_{i1}|X_{i}=X^{2}]+\mathcal{P}%
^{3}E[Y_{i1}-Y_{i2}|X_{i}=X^{3}]\right\} /(\mathcal{P}^{2}+\mathcal{P}^{3})
\\
&=&\left[ \mathcal{P}^{2}(\mathcal{P}_{2}^{2}-\mathcal{P}_{3}^{2})+\mathcal{P%
}^{3}(\mathcal{P}_{3}^{3}-\mathcal{P}_{2}^{3})\right] /(\mathcal{P}^{2}+%
\mathcal{P}^{3}).
\end{eqnarray*}%
Next, the symmetry$H(-x)=1-H(x)$ implies that $\alpha _{j}^{k}(\beta )$ take
the form 
\begin{equation*}
\alpha _{j}^{k}(\beta )=\left\{ 
\begin{array}{ll}
-\infty , & j=1, \\ 
-\beta (X_{1}^{k}+X_{2}^{k})/2, & j=2,3, \\ 
\infty , & j=4.%
\end{array}%
\right.
\end{equation*}%
Note that for $k=2$ or $k=3$ we have $X_{1}^{k}+X_{2}^{k}=1,$ so that $%
\alpha _{j}^{k}(\beta )=-\tilde{\beta}$ for $\tilde{\beta}=\beta /2$. Thus,%
\begin{equation*}
H(\beta +\alpha _{j}^{k}(\beta ))-H(\alpha _{j}^{k}(\beta ))=H(\tilde{\beta}%
)-H(-\tilde{\beta})=2H(\tilde{\beta})-1.
\end{equation*}%
Therefore the limit of the fixed effects estimator of the identified effect
is%
\begin{equation*}
A[2H(\tilde{\beta})-1],A=\left[ \mathcal{P}^{2}(\mathcal{P}_{2}^{2}+\mathcal{%
P}_{3}^{2})+\mathcal{P}^{3}(\mathcal{P}_{2}^{3}+\mathcal{P}_{3}^{3})\right]
/(\mathcal{P}^{2}+\mathcal{P}^{3}).
\end{equation*}%
Next, the limit of the concentrated log likelihood is 
\begin{equation*}
2\mathcal{P}^{2}[\mathcal{P}_{2}^{2}\ln H(\tilde{\beta})+\mathcal{P}%
_{3}^{2}\ln H(-\tilde{\beta})]+2\mathcal{P}^{3}[\mathcal{P}_{2}^{3}\ln H(-%
\tilde{\beta})+\mathcal{P}_{3}^{3}\ln H(\tilde{\beta})].
\end{equation*}%
The first-order conditions for maximization of this object are%
\begin{equation*}
0=2\mathcal{P}^{2}[\mathcal{P}_{2}^{2}\lambda (\tilde{\beta})-\mathcal{P}%
_{3}^{2}\lambda (-\tilde{\beta})]+2\mathcal{P}^{3}[-\mathcal{P}%
_{2}^{3}\lambda (-\tilde{\beta})+\mathcal{P}_{3}^{3}\lambda (\tilde{\beta})],
\end{equation*}%
where $\lambda (x)=H^{\prime }(x)/H(x).$ By symmetry, $H^{\prime }(-\tilde{%
\beta})=H^{\prime }(\tilde{\beta}).$ Divide the first order conditions by $%
H^{\prime }(\tilde{\beta})$ and multiply by $H(\tilde{\beta})H(-\tilde{\beta}%
)$ to obtain

\begin{eqnarray*}
0 &=&2\mathcal{P}^{2}[\mathcal{P}_{2}^{2}H(-\tilde{\beta})-\mathcal{P}%
_{3}^{2}H(\tilde{\beta})]+2\mathcal{P}^{3}[-\mathcal{P}_{2}^{3}H(\tilde{\beta%
})+\mathcal{P}_{3}^{3}H(-\tilde{\beta})] \\
&=&2(\mathcal{P}^{2}+\mathcal{P}^{3})[\delta -A(2H(\tilde{\beta})-1)].\text{
\ }Q.E.D.
\end{eqnarray*}

In numerical examples this same result continues to hold for $T=3$ and $T=4.$
It would be interesting to extend this result to larger $T$ but it is beyond
the scope of this paper to do so. Unfortunately this result does not extend
to the overall ATE.

\section{Supplements to Section 8}

Here we give the proofs of Section 8 and additional numerical results for
the logit model.

%\textsc{Proof of Lemma 7:} 

\subsection{Proof of Lemma 7}

Let the vector of model probabilities for $(Y^{1},....,Y^{J})$ be%
\begin{equation*}
\mathcal{L}^{k}\left( \alpha ,\beta \right) \equiv \left( \mathcal{L}%
_{1}^{k}\left( \alpha ,\beta \right) ,...,\mathcal{L}_{J}^{k}\left( \alpha
,\beta \right) \right) ^{\prime }.
\end{equation*}%
Let $\Gamma _{k}(\beta )\equiv \left\{ \mathcal{L}^{k}\left( \alpha ,\beta
\right) :\alpha \in \Upsilon \right\} $ and $\breve{\Gamma}_{k}(\beta )$ be
the convex hull of $\Gamma _{k}(\beta )$. By Lemma 3 of Chamberlain (1987), $%
\breve{\Gamma}_{k}(\beta )=\{\int \mathcal{L}^{k}\left( \alpha ,\beta
\right) dF(\alpha ):F$ is a CDF on $\Upsilon \}.$ Therefore, $\int \mathcal{L%
}^{k}\left( \alpha ,\beta \right) dF_{k}(\alpha )\in \breve{\Gamma}%
_{k}\left( \beta \right) .$ Note that $\Gamma _{k}(\beta )$ is contained in
the unit simplex and so has dimension $J-1.$ By the Carath\'{e}odory Theorem
there exist $J$ vectors $\mathcal{L}^{k}\left( \alpha _{m}^{k},\beta \right)
,(m=1,...,J)$ and $0\leq \pi _{m}^{k}\leq 1$ with $\sum_{m=1}^{J}\pi
_{m}^{k}=1$ such that 
\begin{equation*}
\int \mathcal{L}^{k}\left( \alpha ,\beta \right) dF_{k}(\alpha
)=\sum_{m=1}^{J}\pi _{m}^{k}\mathcal{L}^{k}\left( \alpha _{m}^{k},\beta
\right) ,
\end{equation*}%
giving the conclusion for the discrete distribution $F_{k}^{J}$ with $J$
support points at $(\alpha _{1}^{k},...,\alpha _{J}^{k})$ and probabilities $%
(\pi _{1}^{k},...,\pi _{J}^{k}).$

Next, for any $\epsilon >0$ let $\beta \in B$ and $F_{k\beta }\in \mathcal{F}%
_{k}(\beta ,\mathcal{P})$ satisfy%
\begin{equation*}
\Delta _{u}^{k}-\epsilon <\int \Delta (\alpha ,\beta )dF_{k\beta }\left(
\alpha \right) \equiv \bar{\Delta}(\beta ).
\end{equation*}%
Similarly to the previous paragraph, let $\Gamma _{k}^{\Delta }(\beta
)\equiv \left\{ (\mathcal{L}^{k}\left( \alpha ,\beta \right) ^{\prime
},\Delta (\alpha ,\beta ))^{\prime }:\alpha \in \Upsilon \right\} $ and $%
\breve{\Gamma}_{k}^{\Delta }(\beta )$ be the convex hull of $\Gamma
_{k}^{\Delta }(\beta )$. Then $(\mathcal{P}_{1}^{k},...,\mathcal{P}_{J}^{k},%
\bar{\Delta}(\beta ))^{\prime }\in \breve{\Gamma}_{k}^{\Delta }\left( \beta
\right) ,$ so by Caratheodory's Theorem there exists a discrete distribution 
$F_{k\beta }^{J+1}$ with $J+1$ support points $(\alpha _{1}^{k},...,\alpha
_{J+1}^{k})$ and probabilities $\pi _{1}^{k},...,\pi _{J+1}^{k}$ such that $%
F_{k\beta }^{J+1}\in \mathcal{F}_{k}(\beta ,\mathcal{P})$ and $\int \Delta
(\alpha ,\beta )dF_{k\beta }^{J+1}\left( \alpha \right) =\bar{\Delta}(\beta
).$

We now show that it suffices to have mass over just $J$ points. Consider the
problem of allocating $\pi _{1}^{k},...,\pi _{J+1}^{k}$ among $\left( \alpha
_{1}^{k},...,\alpha _{J+1}^{k}\right) $ in order to solve 
\begin{eqnarray*}
&&\max_{\left( \pi _{1}^{k},...,\pi _{J+1}^{k}\right)
}\sum_{m=1}^{J+1}\Delta (\alpha _{m}^{k},\beta )\pi _{m}^{k},s.t. \\
\sum_{m=1}^{J+1}\pi _{m}^{k}\mathcal{L}_{j}^{k}\left( \alpha _{m}^{k},\beta
\right)  &=&\mathcal{P}_{j}^{k},\sum_{m=1}^{J+1}\pi _{m}^{k}=1,\pi
_{m}^{k}\geq 0,(m=1,...,J+1).
\end{eqnarray*}%
This is a linear program of the form 
\begin{equation*}
\max_{\pi ^{k}\in \mathbb{R}^{J+1}}c^{\prime }\pi ^{k}\quad \text{ such that 
}\quad \pi ^{k}\geq 0,\quad A\pi ^{k}=b,\quad 1^{\prime }\pi ^{k}=1,
\end{equation*}%
and any basic feasible solution to this program has $J+1$ active
constraints, of which at most $rank\left( A\right) +1$ can be equality
constraints. This means that at least $J+1-rank([A^{\prime },1]^{\prime })$
of active constraints are of the form $\pi _{m}^{k}=0$, see, e.g., Theorem
2.3 and Definition 2.9 (ii) in Bertsimas and Tsitsiklis (1997). Since each
column of $A$ sums to $1$, $rank([A^{\prime },1]^{\prime })\leq J$ and a
basic solution to this linear programming problem will have at least one
zero. Thus, there are at most $J$ strictly positive $\pi _{m}^{k}$'s.%
\footnote{%
Note that $rank([A^{\prime },1]^{\prime })\leq J$, since $\sum_{j=1}^{J}%
\mathcal{L}_{j}^{k}\left( \alpha ,\beta \right) =1$. The exact rank of $%
[A^{\prime },1]^{\prime }$ depends on the sequence $X^{k}$, the parameter $%
\beta $, the form of $\mathcal{L}_{j}^{k}\left( \alpha ,\beta \right) $, and 
$T$. For example in the model of equation (8) of the main text with $T=2$
and $X$ binary, $rank(A)=J-2=2$ when $x_{1}=x_{2}$, $\beta =0$, or $H$ is
the logistic distribution; whereas $rank(A)=J-1=3$ for $X_{1}^{k}\neq
X_{2}^{k}$, $\beta \neq 0$, and $H$ is any continuous distribution different
from the logistic.} Therefore, we have shown that there exists a
distribution $F_{k\beta }^{J}\in \mathcal{F}_{k}(\beta ,\mathcal{P})$ with
just $J$ points of support such that 
\begin{equation*}
\Delta _{u}^{k}-\epsilon <\int \Delta (\alpha ,\beta )dF_{k\beta
}^{J+1}\left( \alpha \right) \leq \int \Delta (\alpha ,\beta )dF_{k\beta
}^{J}\left( \alpha \right) .
\end{equation*}%
This construction works for every $\epsilon >0$. $Q.E.D.$

\bigskip

\subsection{Numerical results for logit model}

We carry out some additional numerical calculations for the logit model
where 
\begin{equation*}
Y_{it}=1(\beta ^{\ast }X_{it}+\alpha _{i}\geq \varepsilon _{it}),\varepsilon
_{it}\sim L(0,1),X_{it}=1(\alpha _{i}\geq \eta _{it}),\eta _{it}\sim
N(0,1),\alpha _{i}\sim N(0,1),
\end{equation*}%
where $L(0,1)$ denotes the standard logistic distribution normalized to have
zero mean and unit variance. We consider different DGPs indexed by $\beta
^{\ast }\in \lbrack -2,2]$ and $T\in \{2,3\}$. Figures 1 and 2 show
nonparametric bounds for ATEs and semiparametric bounds for $\beta ^{\ast }$
and ATEs for $T=2$ and $T=3$, respectively. The semiparametric bounds are
obtained using the computational algorithm described in Section 8 of the paper with $M=100$ and $%
\lambda _{M}=1.3\times 10^{-8}$. The elements of the fixed grid $\Upsilon
_{M}$ are located at the percentiles of the standard normal distribution. As
is well-known, we find that $\beta ^{\ast }$ is identified for $T\geq 2$.
The nonparametric bounds for the ATEs (NP-bounds) can be very wide, even
when we impose monotonicity (NPM-bounds). The semiparametric bounds for the
ATEs (SP-bounds) are tighter than the nonparametric bounds and shrink
exponentially fast with $T$, as shown in Theorem 6.

\bigskip

%\textsc{Proof of Lemma 8:} 

\subsection{Proof of Lemma 8}

Consider the set $\bar{\Re}=(-\infty ,+\infty )\cup \{-\infty ,+\infty \}.$
By assumption $H(v)$ is strictly monotonic and continuous on $\bar{\Re}$
with $H(-\infty )=0$ and $H(+\infty )=1.$ Let $H^{-1}(u)$ be the inverse
function defined on $[0,1].$ Let $\bar{v}=\max_{X^{k}\in
\{X^{1},...,X^{K}\},\beta \in B}|X_{t}^{k\prime }\beta |$ and define the
function%
\begin{equation*}
T(u)=\left\{ 
\begin{array}{l}
\bar{v}+H^{-1}(u),\;\frac{3}{4}\leq u\leq 1 \\ 
(4u-2)\left[ \bar{v}+H^{-1}(\frac{3}{4})\right] ,\;\frac{1}{4}<u<\frac{3}{4}
\\ 
-\bar{v}+H^{-1}(u),\;0\leq u\leq \frac{1}{4}.%
\end{array}%
\right. 
\end{equation*}%
This function is continuous and differentiable except at $u=\frac{1}{4}$ and 
$u=\frac{3}{4}.$ \ At $u=\frac{1}{4}$ the left derivative is $\left[
h(H^{-1}\left( \frac{1}{4}\right) )\right] ^{-1}$ and the right derivative
is $4\left[ \bar{v}+H^{-1}\left( \frac{3}{4}\right) \right] .$

Consider the function $H(v+T(u)).$ By the chain rule, $H(v+T(u))$ is
differentiable everywhere on $\left[ -\bar{v},\bar{v}\right] \times \left( 
\frac{1}{4},\frac{3}{4}\right) $ and right differentiable at $\left( v,\frac{%
1}{4}\right) $ and left differentiable at $\left( v,\frac{3}{4}\right) $
with derivative (right or left) equal to%
\begin{equation*}
h(v+T(u))4\left[ \bar{v}+H^{-1}(\frac{3}{4})\right] .
\end{equation*}%
This derivative is uniformly bounded on $\left[ -\bar{v},\bar{v}\right]
\times \left( \frac{1}{4},\frac{3}{4}\right) $ by $h$ uniformly bounded. \
Also $H(v+T(u))$ is differentiable everywhere on $\left[ -\bar{v},\bar{v}%
\right] \times \left\{ \left( \frac{3}{4},\infty \right) \cup \left( -\infty
,\frac{1}{4}\right) \right\} ,$ right differentiable at $\left[ -\bar{v},%
\bar{v}\right] \times \left\{ \frac{3}{4}\right\} $ and left differentiable
at $\left[ -\bar{v},\bar{v}\right] \times \left\{ \frac{1}{4}\right\} .$ For 
$u\in \lbrack 3/4,1]$ the (right) derivative is 
\begin{equation*}
\frac{\partial }{\partial u}H(v+T(u))=H^{\prime }(v+T(u))T^{\prime }(u)=%
\frac{h(v+\bar{v}+H^{-1}(u))}{h(H^{-1}(u))}\leq \frac{h(H^{-1}(u))}{%
h(H^{-1}(u))}=1
\end{equation*}%
where the inequality holds by $\bar{v}+v\geq 0$ (implied by $v\geq -\bar{v})$
and by $H^{-1}(u)>0.$ It follows similarly that $\partial H(v+T(u))/\partial
u$ is uniformly bounded by $1$ on $\left[ -\bar{v},\bar{v}\right] \times
\lbrack 0,\frac{1}{4}].$ It follows that there is a constant $C$ such that
for all $v\in \lbrack -\bar{v},\bar{v}]$ and $u,\tilde{u}\in \lbrack 0,1],$%
\begin{equation*}
|H(v+T(\tilde{u}))-H(v+T(u))|\leq C|\tilde{u}-u|.
\end{equation*}

Note that $T^{-1}(\alpha )$ is a strictly monotonic increasing function on $%
\bar{\Re}$. Define $d(\tilde{\alpha},\alpha )=|T^{-1}(\tilde{\alpha}%
)-T^{-1}(\alpha )|$. Note that $d(\tilde{\alpha},\alpha )\geq 0$ with
equality if and only if $\tilde{\alpha}=\alpha ,$ and for any three points $%
\bar{\alpha},$ $\tilde{\alpha},$ and $\alpha ,$ the triangle inequality
implies%
\begin{equation*}
d(\tilde{\alpha},\alpha )=|T^{-1}(\tilde{\alpha})-T^{-1}(\alpha )|\leq
|T^{-1}(\tilde{\alpha})-T^{-1}(\bar{\alpha})|+|T^{-1}(\bar{\alpha}%
)-T^{-1}(\alpha )|=d(\tilde{\alpha},\bar{\alpha})+d(\bar{\alpha},\alpha ).
\end{equation*}%
Therefore $d(\tilde{\alpha},\alpha )$ is a metric. Also, for $\tilde{u}%
=T^{-1}(\tilde{\alpha})$ and $u=T^{-1}(\alpha ),$ we have 
\begin{equation*}
\sup_{v\in \lbrack -\bar{v},\bar{v}]}|H(v+\tilde{\alpha})-H(v+\alpha )|\leq
C|T^{-1}(\tilde{\alpha})-T^{-1}(\alpha )|=Cd(\tilde{\alpha},\alpha ).
\end{equation*}%
Also, by $|X_{t}^{k\prime }\beta |\leq \bar{v},$ and $0\leq H(X_{t}^{k\prime
}\beta +\alpha )\leq 1,$ for all $t,$ $k,$ and $\beta \in \mathbb{B}$,

\begin{eqnarray*}
\left\vert \mathcal{L}_{j}^{k}\left( \tilde{\alpha},\tilde{\beta}\right) -%
\mathcal{L}_{j}^{k}\left( \alpha ,\beta \right) \right\vert &\leq
&\left\vert \mathcal{L}_{j}^{k}\left( \tilde{\alpha},\tilde{\beta}\right) -%
\mathcal{L}_{j}^{k}\left( \alpha ,\tilde{\beta}\right) \right\vert
+\left\vert \mathcal{L}_{j}^{k}\left( \alpha ,\tilde{\beta}\right) -\mathcal{%
L}_{j}^{k}\left( \alpha ,\beta \right) \right\vert \\
&\leq &Cd(\tilde{\alpha},\alpha )+\sup_{\alpha ,t,k}|H(X_{t}^{k\prime }%
\tilde{\beta}+\alpha )-H(X_{t}^{k\prime }\beta +\alpha )| \\
&\leq &Cd(\tilde{\alpha},\alpha )+\sup_{v}h(v)\sup_{t,k}\left\Vert
X_{t}^{k}\right\Vert \left\Vert \tilde{\beta}-\beta \right\Vert \\
&\leq &C[d(\tilde{\alpha},\alpha )+\left\Vert \tilde{\beta}-\beta
\right\Vert ].
\end{eqnarray*}

Finally, for every $M$ let $\bar{\alpha}_{mM}=T((m-1)/(M-1)),(m=1,...,M).$
Then 
\begin{equation*}
\eta (M)=\sup_{\alpha \in \bar{\Re}}\min_{\tilde{\alpha}\in \Upsilon
_{M}}d(\alpha ,\tilde{\alpha})=\sup_{u\in \lbrack 0,1]}\min_{\tilde{u}\in
\{0,1/(M-1),2/(M-1),...,1\}}|u-\tilde{u}|=1/(M-1).\text{ \ }Q.E.D.
\end{equation*}

\bigskip

%\textsc{Proof of Theorem 9:} 

\subsection{Proof of Theorem 9}

This proof is omitted because it is very similar (but easier) than the proof
of Theorem 10 to follow.

\section{Supplements to Section 9}

Here we describe the estimation algorithm, give the proofs of Theorems 10
and 11, and present an alternative inference method based on projection.

%Here  is the proof of Theorem 10.

\subsection{Estimation: Implementation Details}

To implement the estimation method, we also start from simpler estimates of
the bounds corresponding to those described in the computation section.
Specifically, for $\hat{\pi}(\beta )\in \arg \min_{\pi \in S_{M}^{K}}\hat{T}%
_{\lambda }(\beta ,\pi )$ let $\hat{S}^{k}(\beta )=\{\pi
^{k}:P_{j}^{k}(\beta ,\pi ,\hat{M})=$ $P_{j}^{k}(\beta ,\hat{\pi}(\beta ),%
\hat{M}),$ $j=1,...,J\}$ and let 
\begin{equation*}
\check{\Delta}_{\ell }^{k}=\min_{\beta \in \hat{B},\pi ^{k}\in \hat{S}%
^{k}(\beta )}\sum_{m=1}^{M}\pi _{m}^{k}\Delta (\bar{\alpha}_{mM},\beta ),%
\text{ }\check{\Delta}_{u}^{k}=\max_{\beta \in \hat{B},\pi ^{k}\in \hat{S}%
^{k}(\beta )}\sum_{m=1}^{M}\pi _{m}^{k}\Delta (\bar{\alpha}_{mM},\beta ).
\end{equation*}%
We use these estimated bounds as starting values and then search over other
possible values of $\pi ,$ similar to the computational approach.

The choice of $\hat{M}$ is important for this estimator. In our empirical
examples we have proceeded by starting with a small $\hat{M},$ and stopping
when the change in the estimated sets is small. We have found that quite
small $\hat{M}$ often suffices. The choice of weights $\hat{w}_{j}^{k}$ is
also important. The optimal choice, corresponding to minimum chi-square
would be $\hat{w}_{j}^{k}=\mathcal{P}^{k}/\mathcal{P}_{j}^{k}$. Using sample
frequencies in place of population frequencies does not work well due to
small cell sizes. One could use a two-step procedure where one first
computes the identified set for weights like $\hat{w}_{j}^{k}=\hat{P}^{k}$
and then reestimates the identified set using weights $\hat{w}_{j}^{k}=\hat{P%
}_{k}/P_{j}^{k}(\beta ,\hat{\pi}(\beta ),\hat{M})$ for some $\beta \in \hat{B%
}$.

%\textsc{Proof of Theorem 10: }

\subsection{Proof of Theorem 10}

For notational convenience we here denote the probabilities associated with
the fixed grid $\{\bar{\alpha}_{1M},...,\bar{\alpha}_{MM}\}$ by $\bar{\pi}%
^{k}.$ Let $\bar{\pi}=(\bar{\pi}^{1\prime },...,\bar{\pi}^{K\prime
})^{\prime }$ be a $KM\times 1$ vector with each $\bar{\pi}^{k}$ in the $M$%
-dimensional unit simplex $\mathcal{S}_{M}.$ Also, let the probabilities
associated with a variable grid $\{\alpha _{1}^{k},...,\alpha _{J+1}^{k}\}$
be $\pi ^{k}$ so that $\pi =(\pi ^{1\prime },...,\pi ^{K\prime })^{\prime }$
is a $[(J+1)K]\times 1$ vector of probabilities with each $\pi ^{k}$ in the $%
J+1$-dimensional unit simplex $\mathcal{S}_{J+1}.$ Let $\alpha ^{k}=(\alpha
_{1}^{k},...,\alpha _{J+1}^{k})^{\prime },$ $\alpha =(\alpha ^{1\prime
},...,\alpha ^{K\prime })^{\prime },$ $\gamma =(\alpha ^{\prime },\pi
^{\prime })^{\prime }$, $\theta =(\beta ^{\prime },\gamma ^{\prime
})^{\prime },$ $\tilde{P}_{j}^{k}(\theta )=\sum_{\ell =1}^{J+1}\mathcal{L}%
_{j}^{k}\left( \alpha _{\ell }^{k},\beta \right) \pi _{\ell }^{k}$, $\Delta
^{k}(\theta )=\sum_{\ell =1}^{J+1}\Delta \left( \alpha _{\ell }^{k},\beta
\right) \pi _{\ell }^{k},$ $\Theta =\mathbb{B\times }\Upsilon
^{(J+1)K}\times \mathcal{S}_{J+1}^{K},$ and 
\begin{equation*}
\hat{Q}(\theta )=\sum_{j,k}\hat{w}_{j}^{k}\left[ \hat{P}_{j}^{k}-\tilde{P}%
_{j}^{k}(\theta )\right] ^{2},Q(\theta )=\sum_{j,k}w_{j}^{k}\left[ \mathcal{P%
}_{j}^{k}-\tilde{P}_{j}^{k}(\theta )\right] ^{2}.
\end{equation*}%
By applying the Caratheodory Theorem as in the proof of Lemma 12, for every $%
\bar{\pi}$ there is $\theta (\bar{\pi},\beta )=(\beta ^{\prime },\gamma (%
\bar{\pi},\beta )^{\prime })^{\prime }$ with%
\begin{equation*}
\Delta ^{k}(\theta (\bar{\pi},\beta ))=\sum_{m=1}^{M}\Delta (\bar{\alpha}%
_{mM},\beta )\bar{\pi}_{m}^{k},\tilde{P}_{j}^{k}(\theta (\bar{\pi},\beta
))=P_{j}^{k}(\beta ,\bar{\pi},M),(j=1,...,J;k=1,...,K).
\end{equation*}%
Let $\Theta _{I}=\{\theta :Q(\theta )=0\},$%
\begin{equation*}
\tilde{\Theta}=\{\theta (\bar{\pi},\beta ):\hat{Q}(\theta (\bar{\pi},\beta
))+\lambda _{n}\bar{\pi}^{\prime }\bar{\pi}\leq \epsilon _{n}\},\Theta
_{M}=\{\theta (\bar{\pi},\beta ):\bar{\pi}\in \mathcal{S}_{M}^{K},\beta \in 
\mathbb{B}\}.
\end{equation*}%
By construction the projection of $\tilde{\Theta}$ on $\mathbb{B}$ coincides
with $\hat{B}$ and the projection of $\Theta _{I}$ on $\mathbb{B}$ coincides
with $B$. Also the identified set of marginal effects is $\{\Delta
^{k}(\theta ):\theta \in \Theta _{I}\},$ $\Delta ^{k}(\theta )$ is a
continuous function of $\theta ,$ and $\hat{D}^{k}=\{\Delta ^{k}(\theta
):\theta \in \tilde{\Theta}\}$. Since the minimum and maximum of a set are
continuous in the Hausdorff metric, it suffices to show that $d_{H}(\tilde{%
\Theta},\Theta _{I})\overset{p}{\longrightarrow }0$.

Let $d(\theta ,\tilde{\theta})=\max_{j,k}\max \{d(\alpha _{j}^{k},\tilde{%
\alpha}_{j}^{k}),|\pi _{j}^{k}-\tilde{\pi}_{j}^{k}|,\left\Vert \beta -\tilde{%
\beta}\right\Vert \}$. From Assumption 9 and $\hat{M}\overset{p}{%
\longrightarrow }\infty $ we have 
\begin{equation*}
\sup_{\alpha \in \Upsilon }\min_{\tilde{\alpha}\in \Upsilon _{\hat{M}%
}}d(\alpha ,\tilde{\alpha}\mathcal{)}\leq \eta (\hat{M})\overset{p}{%
\longrightarrow }0.
\end{equation*}%
Therefore for every $\alpha \in \Upsilon $ there is $\bar{\alpha}_{m(\alpha
),\hat{M}}$ with $d(\alpha ,\bar{\alpha}_{m(\alpha ),\hat{M}})\leq \eta (%
\hat{M}),$ so that for any $\theta \in \Theta $ there are $\bar{\alpha}%
_{m(\alpha _{\ell }^{k}),\hat{M}}$ with $\max_{1\leq \ell \leq
J+1,k}\{d(\alpha _{\ell }^{k},\bar{\alpha}_{m(\alpha _{\ell }^{k}),\hat{M}%
})\}\leq \eta (\hat{M}).$ Let $\alpha ^{k}(\theta )=(\bar{\alpha}_{m(\alpha
_{1}^{k}),\hat{M}},...,\bar{\alpha}_{m(\alpha _{J+1}^{k}),\hat{M}})^{\prime
} $, $\alpha (\theta )=(\alpha ^{1}(\theta )^{\prime },...,\alpha
^{K}(\theta )^{\prime })^{\prime },$ and $\bar{\theta}(\theta )=(\beta
^{\prime },\alpha (\theta )^{\prime },\pi ^{\prime })^{\prime }.$ By
construction, $\bar{\theta}(\theta )\in \Theta _{M}$ and $d(\bar{\theta}%
(\theta ),\theta )\leq \eta (\hat{M})$. Thus, 
\begin{equation*}
\sup_{\theta \in \Theta }\inf_{\tilde{\theta}\in \Theta _{\hat{M}}}d(\theta ,%
\tilde{\theta})\leq \eta (\hat{M}).
\end{equation*}%
Also, by Assumption 9,%
\begin{equation*}
|\tilde{P}_{j}^{k}(\theta )-\tilde{P}_{j}^{k}(\tilde{\theta})|\leq
\sum_{\ell =1}^{J}\left\vert \mathcal{L}_{j}^{k}\left( \alpha _{\ell
}^{k},\beta \right) \pi _{\ell }^{k}-\mathcal{L}_{j}^{k}\left( \tilde{\alpha}%
_{\ell }^{k},\tilde{\beta}\right) \tilde{\pi}_{\ell }^{k}\right\vert \leq
Cd(\theta ,\tilde{\theta}).
\end{equation*}%
It then follows by standard calculations that there is $\hat{C}=O_{p}(1)$
such that%
\begin{equation*}
|\hat{Q}(\theta )-\hat{Q}(\tilde{\theta})|\leq \hat{C}d(\theta ,\tilde{\theta%
})\text{ for all }\theta ,\tilde{\theta}\in \Theta .
\end{equation*}%
Therefore we have%
\begin{equation*}
\sup_{\theta \in \Theta }\inf_{\tilde{\theta}\in \Theta _{\hat{M}}}|\hat{Q}%
(\theta )-\hat{Q}(\tilde{\theta})|\leq \hat{C}\eta (\hat{M}).
\end{equation*}%
Also note that 
\begin{equation*}
\sup_{\theta \in \Theta _{I}}\hat{Q}(\theta )=\sum_{j,k}\hat{w}_{j}^{k}[\hat{%
P}_{j}^{k}-\mathcal{P}_{j}^{k}]^{2}=O_{p}(n^{-1}).
\end{equation*}

Next let $\delta >0$ be any positive constant and define the events 
\begin{equation*}
\mathcal{E}_{1}=\left\{ \eta (\hat{M})<\delta \right\} ,\mathcal{E}%
_{2}=\left\{ \hat{C}\eta (\hat{M})<\frac{\epsilon _{n}}{3}\right\} ,\mathcal{%
E}_{3}=\left\{ \sup_{\theta \in \Theta _{I}}\hat{Q}(\theta )<\frac{\epsilon
_{n}}{3}\right\} ,\mathcal{E}_{4}=\sup_{\bar{\pi}\in \mathcal{S}%
_{M}^{K}}\lambda _{n}\bar{\pi}^{\prime }\bar{\pi}<\frac{\epsilon _{n}}{3}.
\end{equation*}%
By $(n^{-1}+\eta (\hat{M})+\lambda _{n})/\epsilon _{n}\overset{p}{%
\longrightarrow }0$ it follows that%
\begin{eqnarray*}
\Pr (\mathcal{E}_{1}) &\longrightarrow &1,\Pr (\mathcal{E}_{2})=\Pr \left( 
\hat{C}<\frac{\eta (\hat{M})^{-1}\epsilon _{n}}{3}\right) \longrightarrow 1,
\\
\Pr (\mathcal{E}_{3}) &=&\Pr \left( n\sup_{\theta \in \Theta _{I}}\hat{Q}%
(\theta )<\frac{n\epsilon _{n}}{3}\right) \longrightarrow 1,\Pr (\mathcal{E}%
_{4})\geq \Pr (\lambda _{n}K\leq \frac{\epsilon _{n}}{3})\longrightarrow 1.
\end{eqnarray*}%
It follows that $\Pr (\cap _{r=1}^{4}\mathcal{E}_{r})\longrightarrow 1.$
When $\cap _{r=1}^{4}\mathcal{E}_{r}$ occurs then for every $\theta \in
\Theta _{I}$ there is $\bar{\pi}$ with $\theta _{M}=\theta (\bar{\pi},\beta
)\in \Theta _{M}$ such that $d(\theta ,\bar{\theta})<\delta $ and 
\begin{eqnarray*}
\hat{Q}(\bar{\theta})+\lambda _{n}\bar{\pi}^{\prime }\bar{\pi} &\leq &\hat{Q}%
(\bar{\theta})+\frac{\epsilon _{n}}{3}\leq \hat{Q}(\theta )+\hat{Q}(\bar{%
\theta})-\hat{Q}(\theta )+\frac{\epsilon _{n}}{3} \\
&\leq &\sup_{\theta \in \Theta _{I}}\hat{Q}(\theta )+\hat{C}\hat{\eta}(M)+%
\frac{\epsilon _{n}}{3}\leq \epsilon _{n},
\end{eqnarray*}%
i.e. $\bar{\theta}\in \tilde{\Theta}.$ Thus, with probability approaching
one, 
\begin{equation*}
\sup_{\theta \in \Theta _{I}}\inf_{\tilde{\theta}\in \tilde{\Theta}}d(\theta
,\tilde{\theta})\leq \delta .
\end{equation*}

Next, note that $\hat{Q}(\theta )\overset{p}{\longrightarrow }Q(\theta )$ so
it follows by Theorem 2.1$\ $of Newey (1991) that $\sup_{\theta \in \Theta
}\left\vert \hat{Q}(\theta )-Q(\theta )\right\vert \overset{p}{%
\longrightarrow }0.$ Define $\Theta _{I}^{\delta }=\left\{ \theta :\inf_{%
\tilde{\theta}\in \Theta _{I}}d(\theta ,\tilde{\theta})<\delta \right\} .$
Note that $\Theta _{I}^{\delta }$ is open so that $\Theta \backslash \Theta
_{I}^{\delta }$ is compact, so by continuity of $Q(\theta ),$ $%
\inf\limits_{\Theta \backslash \Theta _{I}^{\delta }}Q(\theta )=\rho >0.$ It
follows by uniform convergence that$\;\inf\limits_{\Theta \backslash \Theta
_{I}^{\delta }}\hat{Q}(\theta )>\frac{\rho }{2}$ with probability
approaching 1 (w.p.a. 1). \ By $\epsilon _{n}\rightarrow 0,$ 
\begin{equation*}
\sup\limits_{\theta \in \tilde{\Theta}}\hat{Q}(\theta )\leq \sup_{\bar{\pi}%
}\{\hat{Q}(\theta (\bar{\pi},\beta ))+\lambda _{n}\bar{\pi}^{\prime }\bar{\pi%
}\leq \epsilon _{n}\}<\rho /2,
\end{equation*}%
so that $\tilde{\Theta}\subseteq \Theta _{I}^{\delta }$. Therefore w.p.a.1
for all $\tilde{\theta}\in \tilde{\Theta}$ there exists $\theta \in \Theta
_{I}$ such that $d(\tilde{\theta},\theta )<\delta ,$ i.e. $\sup_{\tilde{%
\theta}\in \tilde{\Theta}}\inf_{\theta \in \Theta _{I}}d(\theta ,\tilde{%
\theta})\leq \delta .$ It follows that with w.p.a.1, $d_{H}(\tilde{\Theta}%
,\Theta _{I})\leq \delta $. Since $\delta >0$ is arbitrary, it follows that $%
d_{H}(\tilde{\Theta},\Theta _{I})\overset{p}{\longrightarrow }0.$ $\ Q.E.D.$

\bigskip

%\textsc{Proof of Theorem 11. } 

\subsection{Proof of Theorem 11}

We have that for $S_{n}(\mathcal{P})=\hat{\theta}-\theta ^{\ast }=\hat{\theta%
}-\theta ^{\ast }(\mathcal{P})$ 
\begin{eqnarray*}
&&\text{Pr}_{\Pi }\{\theta ^{\ast }\not\in \left[ \underline{\theta },%
\overline{\theta }\right] \}=\text{Pr}_{\Pi }\{S_{n}(\mathcal{P})\not\in
\lbrack \underline{G}_{n}^{-1}(\alpha _{2},\mathcal{P}),\overline{G}%
_{n}^{-1}(1-\alpha _{1},\mathcal{P})]\} \\
&\leq &\text{Pr}_{\Pi }[\{S_{n}(\mathcal{P})\not\in \lbrack \underline{G}%
_{n}^{-1}(\alpha _{2},\mathcal{P}),\overline{G}_{n}^{-1}(1-\alpha _{1},%
\mathcal{P})]\}\cap \{\mathcal{P}\in \text{CR}_{1-\gamma }(\mathcal{P})\}]+%
\text{Pr}_{\Pi }\{\mathcal{P}\not\in \text{CR}_{1-\gamma }(\mathcal{P})\} \\
&\leq &\text{Pr}_{\Pi }[\{S_{n}(\mathcal{P})\not\in \lbrack {G}%
_{n}^{-1}(\alpha _{2},\mathcal{P}),{G}_{n}^{-1}(1-\alpha _{1},\mathcal{P}%
)]\}\cap \{\mathcal{P}\in \text{CR}_{1-\gamma }(\mathcal{P})\}]+\text{Pr}%
_{\Pi }\{\mathcal{P}\not\in \text{CR}_{1-\gamma }(\mathcal{P})\} \\
&\leq &\text{Pr}_{\Pi }\{S_{n}(\mathcal{P})\not\in \lbrack {G}%
_{n}^{-1}(\alpha _{2},\mathcal{P}),{G}_{n}^{-1}(1-\alpha _{1},\mathcal{P}%
)]\}+\text{Pr}_{\Pi }\{\mathcal{P}\not\in \text{CR}_{1-\gamma }(\mathcal{P}%
)\} \\
&\leq &\alpha +\text{Pr}_{\Pi }\{\mathcal{P}\not\in \text{CR}_{1-\gamma }(%
\mathcal{P})\}.
\end{eqnarray*}%
Thus if $\limsup_{n\to \infty}\text{Pr}_{\Pi }\{\mathcal{P}\not\in \text{CR}%
_{1-\gamma }(\mathcal{P})\}\leq \gamma $, we obtain that $\lim_{n}\text{Pr}%
_{\Pi }\{\theta ^{\ast }\not\in \left[ \underline{\theta },\overline{\theta }%
\right] \}\leq \alpha +\gamma ,$ which is the desired conclusion.

It now remains to show that $\limsup_{n\rightarrow \infty }\text{Pr}_{\Pi }\{%
\mathcal{P}\not\in \text{CR}_{1-\gamma }(\mathcal{P})\}\leq \gamma $. We
have that 
\begin{equation*}
\text{Pr}_{\Pi }\{\mathcal{P}\not\in \text{CR}_{1-\gamma }(\mathcal{P})\}=%
\text{Pr}_{\Pi }\{W(\mathcal{P},P)>c_{1-\gamma }(\chi _{K(J-1)}^{2})\}.
\end{equation*}%
By the uniform central limit theorem, $W(\mathcal{P},\hat{P})$ converges in
law to $\chi _{K(J-1)}^{2}$ under any sequence $\Pi $ in $\mathbb{P}$.
Therefore, 
\begin{equation*}
\lim_{n\rightarrow \infty }\text{Pr}_{\Pi }\{W(\mathcal{P},\hat{P}%
)>c_{1-\gamma }(\chi _{K(J-1)}^{2})\}=\Pr \{\chi _{K(J-1)}^{2}>c_{1-\gamma
}(\chi _{K(J-1)}^{2})\}=\gamma .
\end{equation*}%
\emph{Q.E.D.}

\subsection{Modified Projection Method}

The following method projects a confidence region for conditional choice
probabilities onto a simultaneous confidence region for all possible ATEs
and other structural parameters. In general, this method is more
conservative than the perturbed bootstrap method when a single ATE or
structural parameter is of interest. %If a single marginal
%effect is of interest, then this approach is conservative; if all (or many)
%marginal effects are of interest, then this approach is sharp (or close to
%sharp). EXPLANATION? The perturbed bootstrap method in Section 9 
%appears to be sharp, at least in large samples, when a particular single
%marginal effect is of interest. 
We include a more detailed comparison between the two methods at the end of
this section.

It is convenient to describe the modified projection method in two stages.

Stage 1. The probabilities $\mathcal{P}_{j}^{k}$ belong to the product $%
\mathcal{S}_{J}^{K}$ of $K$ unit simplexes of dimension $J.$ We can begin by
constructing a confidence region for the true choice probabilities $\mathcal{%
P}$ by collecting all probabilities $%
P=(P_{1}^{1},...,P_{J}^{1},...,P_{J}^{K})^{\prime }\in $ $\mathcal{S}_{J}^{K}$ $\ $%
that pass a goodness-of-fit test: 
\begin{equation*}
CR_{1-\alpha }(\mathcal{P})=\left\{ P\in \mathcal{S}_{J}^{K}:W(P,\hat{P})\leq
c_{1-\alpha }(\chi _{K(J-1)}^{2})\right\} ,
\end{equation*}%
where $c_{1-\alpha }(\chi _{K(J-1)}^{2})$ is the $(1-\alpha )$-quantile of
the $\chi _{K(J-1)}^{2}$ distribution and $W$ is the goodness-of-fit
statistic: 
\begin{equation*}
W(P,\hat{P})=n\sum_{j,k}\hat{P}^{k}\frac{\left( \hat{P}_{j}^{k}-P_{j}^{k}%
\right) ^{2}}{P_{j}^{k}}.
\end{equation*}

Stage 2. To construct confidence regions for marginal effects and any other
structural parameters we project each $P\in CR_{1-\alpha }(\mathcal{P})$
onto $\Xi =\{P:\exists \beta \in \mathbb{B}$ with $\mathcal{F}_{k}(\beta
,P)\neq \varnothing ,\forall k=1,...,K\}$, the space of conditional choice
probabilities that is compatible with the model. We obtain this projection $%
P^{\ast }(P)$ by solving the minimum distance problem: 
\begin{equation*}
P^{\ast }(P)=\arg \min_{\tilde{P}\in \Xi }W({\tilde{P}},P),\ \ W({\tilde{P}}%
,P)=n\sum_{j,k}\hat{P}^{k}\frac{(P_{j}^{k}-{\tilde{P}}_{j}^{k})^{2}}{\tilde{P%
}_{j}^{k}}.
\end{equation*}%
The confidence regions are then constructed from the projections of all the
choice probabilities in $CR_{1-\alpha }(\mathcal{P})$. For the identified
set of the model parameter, for example, for each $P\in CR_{1-\alpha }(%
\mathcal{P})$ we solve 
\begin{equation*}
B^{\ast }(P)=\left\{ \beta \in \mathbb{B}:\exists \tilde{P}\in P^{\ast }(P)%
\text{ with }\mathcal{F}_{k}(\beta ,\tilde{P})\neq \varnothing
,k=1,...,K\right\} .
\end{equation*}%
Denote the resulting confidence region as 
\begin{equation*}
CR_{1-\alpha }(B^{\ast })=\{B^{\ast }(P):P\in CR_{1-\alpha }(\mathcal{P})\}.
\end{equation*}%
We may interpret this set as a confidence region for the set $B^{\ast }$ of $%
\beta $ that are compatible with a best approximating model. Under correct
specification, this will be a confidence region for the identified set $B$.

If we are interested in bounds on marginal effects, for each $P\in
CR_{1-\alpha }(\mathcal{P})$ we get 
\begin{eqnarray*}
\Delta _{\ell }^{k}(P) &=&\min_{\beta \in B^{\ast }(P),F_{k}\in \mathcal{F}%
_{k}(\beta ,P^{\ast }(P))}\int \Delta (\alpha ,\beta )dF_{k}(\alpha ), \\
\Delta _{u}^{k}(P) &=&\max_{\beta \in B^{\ast }(P),F_{k}\in \mathcal{F}%
_{k}(\beta ,P^{\ast }(P))}\int \Delta (\alpha ,\beta )dF_{k}(\alpha ).
\end{eqnarray*}%
Denote the resulting confidence regions as 
\begin{equation*}
CR_{1-\alpha }[\Delta _{\ell }^{k\ast },\Delta _{u}^{k\ast }]=\{[\Delta
_{\ell }^{k}(P),\Delta _{u}^{k}(P)]:P\in CR_{1-\alpha }(\mathcal{P})\}.
\end{equation*}%
These sets are confidence regions for the sets $[\Delta _{\ell }^{k\ast
},\Delta _{u}^{k\ast }]$, where $\Delta _{\ell }^{k\ast }$ and $\Delta
_{u}^{k\ast }$ are the lower and upper bounds on the marginal effects
induced by any best approximating model. Under correct specification, these
will include the true upper and lower bounds on the marginal effect $[\Delta
_{\ell }^{k},\Delta _{u}^{k}]$ induced by any true model in $(B,\mathcal{P})$%
.

In a canonical projection method we would implement the second stage by
simply intersecting $CR_{1-\alpha }(\mathcal{P})$ with $\Xi $, but this may
give an empty intersection either in finite samples or under
misspecification. We avoid this problem by using the projection step instead
of the intersection, and also by re-targeting our confidence regions onto
the best approximating model.

\bigskip

\textsc{Theorem A15:} \textit{If Assumptions 5, 8, and 9 are satisfied then
for any sequence of data-generating process} $\Pi = \Pi_n$ \textit{satisfying Assumption 10}$,$%
%\begin{equation*}
%\lim_{n\rightarrow \infty }\text{Pr}_{\Pi }\left\{ 
%\begin{array}{lll}
%\mathcal{P} & \in  & CR_{1-\alpha }(\mathcal{P}) \\ 
%B^{\ast } & \in  & CR_{1-\alpha }(B^{\ast }) \\ 
%\lbrack \Delta _{\ell }^{k\ast },\Delta _{u}^{k\ast }] & \in  & CR_{1-\alpha
%}[\Delta _{\ell }^{k\ast },\Delta _{u}^{k\ast }],\forall k%
%\end{array}%
%\right\} =1-\alpha .
%\end{equation*}
%
\begin{equation*}
\lim_{n\rightarrow \infty }\text{Pr}_{\Pi }\left[ \{\mathcal{P} \in
CR_{1-\alpha }(\mathcal{P}) \} \cap \{B^{\ast } \in CR_{1-\alpha }(B^{\ast
})\} \cap \{\lbrack \Delta _{\ell }^{k\ast },\Delta _{u}^{k\ast }] \in
CR_{1-\alpha }[\Delta _{\ell }^{k\ast },\Delta _{u}^{k\ast }],\forall k\} %
\right] =1-\alpha.
\end{equation*}

\bigskip

Proof: By the uniform central limit theorem, $W(\mathcal{P},\hat{P})$
converges in law to $\chi _{J(K-1)}^{2}$ under any sequence of true DGPs
with $\Pi $ in $\mathbb{P}$. It follows that 
\begin{equation*}
\lim_{n\rightarrow \infty }\text{Pr}_{\Pi }\{\mathcal{P}\in CR_{1-\alpha }(%
\mathcal{P})\}=1-\alpha .
\end{equation*}%
Further, the event $\mathcal{P}\in CR_{1-\alpha }(\mathcal{P})$ implies then
the event $P^{\ast }(\mathcal{P)}\in \{P^{\ast }(P):P\in CR_{1-\alpha }(\mathcal{%
P})\}$ by construction, which in turn implies the events $B^{\ast }\in
CR_{1-\alpha }(B^{\ast })$ and $[\Delta _{\ell }^{k\ast },\Delta _{u}^{k\ast
}]\in CR_{1-\alpha }[\Delta _{\ell }^{k\ast },\Delta _{u}^{k\ast }],\forall
k $. Q.E.D.

\bigskip

%\subsection{Perturbed Bootstrap}
%
%In this section we present an approach that appears to be sharper than the
%projection method, at least in large samples, when a particular ATE is of
%interest. The estimators for parameters and ATE are obtained by nonlinear
%programming subject to data-dependent constraints that are modified to
%respect the constraints of the model. The distributions of these highly
%complex estimators are not tractable, and are also non-regular in the sense
%that the limit versions of these distributions do not vary with
%perturbations of the DGP in a continuous\ fashion. This implies that the
%usual bootstrap is not consistent. To overcome all of these difficulties we
%will rely on a variation of the bootstrap, which we call the perturbed
%bootstrap.

We conclude giving a comparison of the modified projection and perturbed
bootstrap methods. The modified projection method is well suited for
performing simultaneous inference on all possible functionals of the
parameter vector. In contrast, the perturbed bootstrap is better suited for
performing inference on a given functional of the parameter vector, such as
the average structural effect. In order to understand why the latter method
can be much sharper than the former method in the case where a single
functional is of interest, it suffices to think of how these methods perform
in the simplest situation of inference about the mean of a multinomial
distribution. In this case, the perturbed bootstrap will become
asymptotically equivalent to the usual bootstrap, since the limit
distribution is continuous with respect to the DGP in this example, and our
local perturbations of DGP converge to the true DGP (note that, more
generally, in cases with limit distributions being discontinuous with
respect to the DGP, the introduction of the local perturbations ensures that the
resulting confidence interval possesses locally uniform coverage). Therefore
in this example perturbed bootstrap inference asymptotically becomes
first-order equivalent to the t-statistic-based inference on the mean, and
is efficient. Now compare that with the Scheffe-style projection based
confidence interval, whereby one creates a confidence region for multinomial
probabilities and projects it down to the confidence interval for the mean,
a linear functional of these probabilities. It is clear that the latter is
very conservative, and is much less sharp than the t-statistic based
confidence interval. We refer the reader to Romano and Wolf (2000) for the
pertinent discussion of this example in the context of a closely related
inference method.

% FIGURE %%%%%%%%%%%%%%%%%%%%%%%%%%%%%%%%%%%%%%%%%%%%

\begin{figure}

\begin{center}

\centering\epsfig{figure=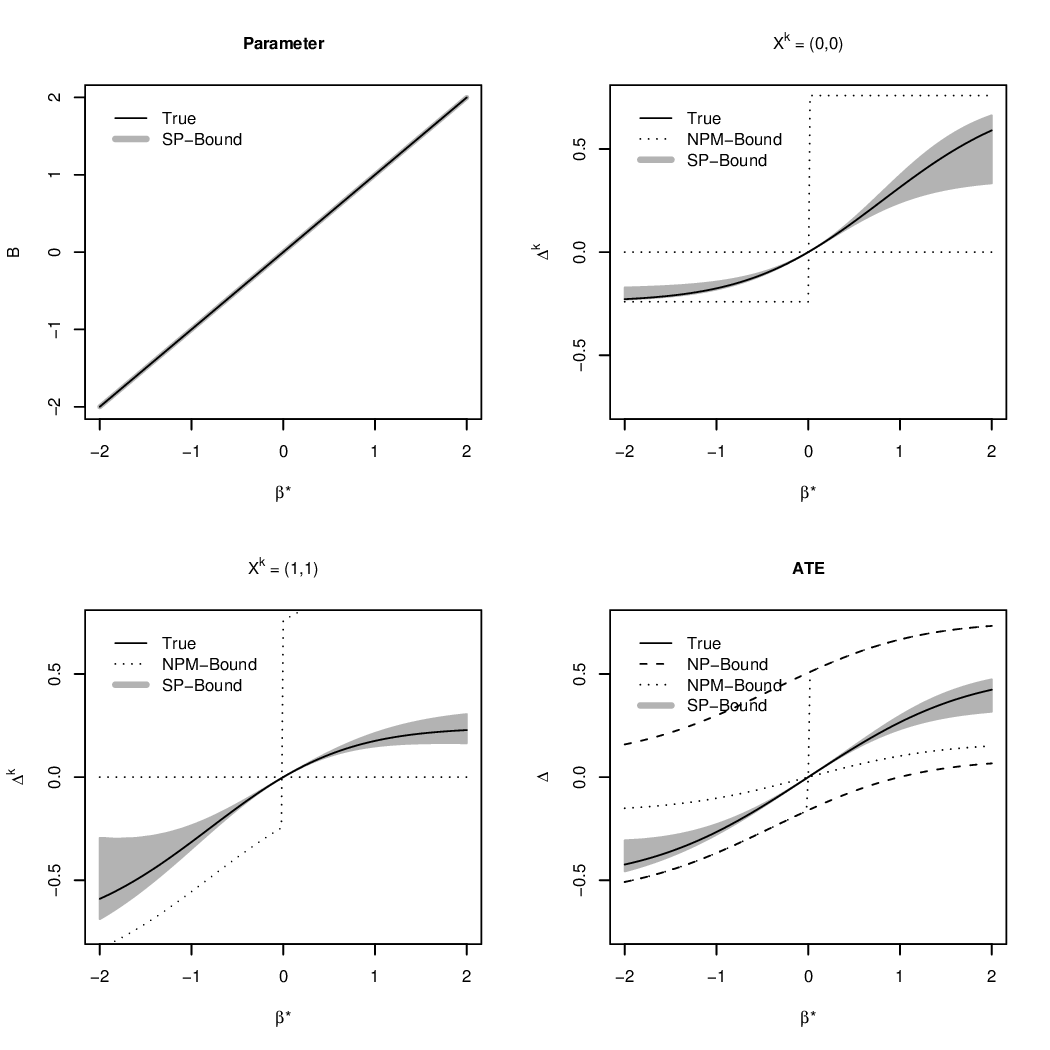,width=6.5in,height=6.5in}

\caption{\label{fig: logit-QP-T2} Identified set for parameter and
ATEs in binary choice logit models with $ Y_{it} = 1(\beta^{\ast}
X_{it} + \alpha_i \geq \varepsilon_{it})$, $\varepsilon_{it} \sim
L(0,1)$, $X_{it} = 1(\alpha_i \geq \eta_{it})$, $\eta_{it} \sim
N(0,1)$, $\alpha_i \sim N(0,1)$, $\beta^{\ast} \in [-2, 2]$, and $T
= 2$.}

\end{center}

\end{figure}

% figure %%%%%%%%%%%%%%%%%%%%%%%%%%%%%%%%%%%%%%%%%%%%

% FIGURE %%%%%%%%%%%%%%%%%%%%%%%%%%%%%%%%%%%%%%%%%%%%

\begin{figure}

\begin{center}

\centering\epsfig{figure=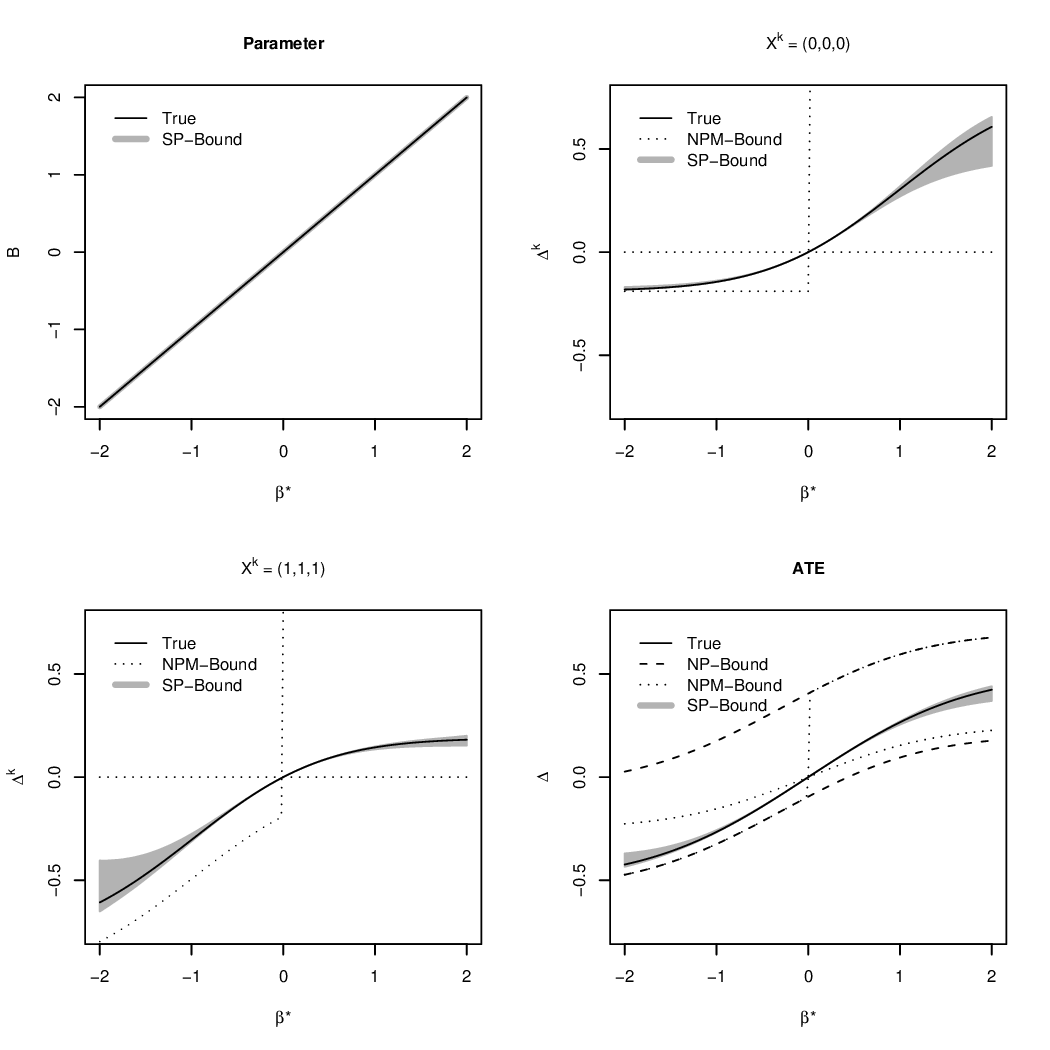,width=6.5in,height=6.5in}

\caption{\label{fig: logit-QP-T3} Identified set for parameter and
ATEs in binary choice logit models with $ Y_{it} = 1(\beta^{\ast}
X_{it} + \alpha_i \geq \varepsilon_{it})$, $\varepsilon_{it} \sim
L(0,1)$, $X_{it} = 1(\alpha_i \geq \eta_{it})$, $\eta_{it} \sim
N(0,1)$, $\alpha_i \sim N(0,1)$, $\beta^{\ast} \in [-2, 2]$, and $T
= 3$.}

\end{center}

\end{figure}

% figure %%%%%%%%%%%%%%%%%%%%%%%%%%%%%%%%%%%%%%%%%%%%

\end{document}